\documentclass[twocolumn]{aastex702}
\usepackage{booktabs}
\usepackage{multirow}
\usepackage{graphicx}
\usepackage{amsmath}
\usepackage{enumitem}
\usepackage{placeins}

\newcommand{\vsini}{\ensuremath{v\sin i_\star}} 

\newcommand{\mesa}{\textsc{mesa}}
\newcommand{\wasp}[1]{WASP-#1}

\shorttitle{The Obliquity of WASP-12 b}
\shortauthors{Zhang et al.}

\begin{document}

\title{Where Did the Angular Momentum Go? WASP-12 b May Have Circularized on a Retrograde Orbit}

\correspondingauthor{Elina Y. Zhang and Fei Dai}

\author[0009-0004-0455-2424]{Elina Y. Zhang}
\affiliation{Institute for Astronomy, University of Hawaii, 2680 Woodlawn Drive, Honolulu, HI 96822 USA}
\email[show]{yuchenzh@hawaii.edu}

\author[0000-0002-8958-0683]{Fei Dai}
\affiliation{Institute for Astronomy, University of Hawaii, 2680 Woodlawn Drive, Honolulu, HI 96822 USA}
\email[show]{fdai@hawaii.edu}

\author[0000-0003-3020-4437]{Yaguang Li}
\affiliation{Institute for Astronomy, University of Hawaii, 2680 Woodlawn Drive, Honolulu, HI 96822 USA}
\email{yaguangl@hawaii.edu}

\author[0000-0002-4284-8638]{Jennifer van Saders}
\affiliation{Institute for Astronomy, University of Hawaii, 2680 Woodlawn Drive, Honolulu, HI 96822 USA}
\email{jlvs@hawaii.edu}

\author[0000-0001-9518-9691]{Morgan Saidel}
\affiliation{Carnegie Science Observatories, 813 Santa Barbara Street, Pasadena, CA 91101, USA}
\email{msaidel@carnegiescience.edu}

\author[0000-0002-5375-4725]{Heather A. Knutson}
\affiliation{Division of Geological and Planetary Sciences, California Institute of Technology, Pasadena, CA 91125, USA}
\email{hknutso2@caltech.edu}

\author[0000-0001-8832-4488]{Daniel Huber}
\affiliation{Institute for Astronomy, University of Hawaii, 2680 Woodlawn Drive, Honolulu, HI 96822 USA}
\email{huberd@hawaii.edu}

\author[0009-0005-7924-3690]{Jingyi Zhang}
\affiliation{Institute for Astronomy, University of Hawaii, 2680 Woodlawn Drive, Honolulu, HI 96822 USA}
\email{jzhang97@hawaii.edu}

\author[0000-0001-8638-0320]{Andrew W. Howard}
\affiliation{Department of Astronomy, California Institute of Technology, Pasadena, CA 91125, USA}
\email{ahoward@caltech.edu}

\author[0000-0003-1312-9391]{Samuel P. Halverson}
\affiliation{Jet Propulsion Laboratory, California Institute of Technology, 4800 Oak Grove Drive, Pasadena, CA 91109, USA}
\email{samuel.halverson@jpl.nasa.gov}

\author[0000-0002-0531-1073]{Howard Isaacson}
\affiliation{Department of Astronomy, University of California Berkeley, Berkeley CA 94720, USA}
\email{hisaacson@berkeley.edu}

\author[0000-0003-3856-3143]{Ryan A. Rubenzahl}
\affiliation{Center for Computational Astrophysics, Flatiron Institute, 162 Fifth Avenue, New York, NY 10010, USA}
\email{rrubenzahl@flatironinstitute.org}

\author[0000-0003-3504-5316]{Benjamin J. Fulton}
\affiliation{NASA Exoplanet Science Institute/Caltech-IPAC, MC 314-6, 1200 E California Blvd, Pasadena, CA 91125, USA}
\email{bjfulton@ipac.caltech.edu}

\author[0000-0001-8342-7736]{Jack Lubin}
\affiliation{Department of Physics \& Astronomy, University of California Los Angeles, Los Angeles, CA 90095, USA}
\email{jblubin@astro.ucla.edu}

\author[0000-0002-9305-5101]{Luke B. Handley}
\altaffiliation{NSF Graduate Research Fellow}
\affiliation{Department of Astronomy, California Institute of Technology, Pasadena, CA 91125, USA}
\email{lhandley@caltech.edu}

\author[0000-0002-8965-3969]{Steven Giacalone}
\affiliation{Department of Astronomy, California Institute of Technology, Pasadena, CA 91125, USA}
\email{giacalone@astro.caltech.edu}

\author[0000-0003-0967-2893]{Erik A. Petigura}
\affiliation{Department of Physics \& Astronomy, University of California Los Angeles, Los Angeles, CA 90095, USA}
\email{petigura@astro.ucla.edu}

\author[0000-0002-2696-2406]{Jingwen Zhang}
\affiliation{Department of Physics, University of California, Santa Barbara, CA 93106, USA}
\affiliation{Institute for Astronomy, University of Hawaii, 2680 Woodlawn Drive, Honolulu, HI 96822 USA}
\email{jwzhang@ucsb.edu}

\author[0000-0002-4671-2957]{Rafael Luque}
\affiliation{Instituto de Astrofísica de Andalucía-CSIC, Glorieta de la Astronomía s/n, 18008 Granada, Spain}
\email{rluque@iaa.es}

\author[0000-0002-4290-6826]{Judah Van Zandt}
\affiliation{Department of Physics, University of California, Santa Barbara, CA 93106, USA}
\email{judahvz@ucsb.edu}

\author[0000-0002-4265-047X]{Joshua N. Winn}
\affiliation{Department of Astrophysical Sciences, Princeton University, 4 Ivy Lane, Princeton, NJ 08544, USA}
\email{jnwinn@princeton.edu}

\author[0000-0003-1762-8235]{Simon H. Albrecht}
\affiliation{Department of Physics and Astronomy, Aarhus University, Ny Munkegade 120, DK-8000 Aarhus C, Denmark}
\email{albrecht@phys.au.dk}

\author[0000-0001-7961-3907]{Samuel W. Yee}
\affiliation{Department of Physics \& Astronomy, University of California Los Angeles, Los Angeles, CA 90095, USA}
\email{syee@astro.ucla.edu}

\author[0000-0002-0376-6365]{Xian-Yu Wang}
\altaffiliation{Sullivan Prize Postdoctoral Fellow}
\affiliation{Department of Astronomy, Indiana University, Bloomington, IN 47405, USA}
\email{xwa5@iu.edu}

\author[0000-0002-7846-6981]{Songhu Wang}
\affiliation{Department of Astronomy, Indiana University, Bloomington, IN 47405, USA}
\email{sw121@iu.edu}

\author[0009-0004-4454-6053]{Steve R. Gibson}
\affil{Caltech Optical Observatories, Pasadena, CA, 91125, USA}
\email{sgibson@caltech.edu}

\author{Kodi Rider}
\affil{Space Sciences Laboratory, University of California Berkeley, Berkeley, CA 94720, USA}
\email{kodi.rider@ssl.berkeley.edu}

\author[0000-0001-8127-5775]{Arpita Roy}
\affiliation{Astrophysics \& Space Institute, Schmidt Sciences, New York, NY 10011, USA} 
\email{arpita308@gmail.com}

\author[0000-0002-6525-7013]{Ashley Baker}
\affil{Caltech Optical Observatories, Pasadena, CA, 91125, USA}
\email{abaker@caltech.edu}

\author[0009-0002-2419-8819]{Jerry Edelstein}
\affil{Space Sciences Laboratory, University of California Berkeley, Berkeley, CA 94720, USA}
\email{jerrye@ssl.berkeley.edu}

\author{Christopher L. Smith}
\affil{Space Sciences Laboratory, University of California Berkeley, Berkeley, CA 94720, USA}
\email{christopher.smith@berkeley.edu}

\author[0000-0002-6092-8295]{Josh Walawender}
\affiliation{W. M. Keck Observatory, 65-1120 Mamalahoa Hwy, Waimea, HI 96743}
\email{jwalawender@keck.hawaii.edu}

\begin{abstract}
WASP-12 b is the only exoplanet with confirmed orbital decay. Angular momentum transferred from its decaying orbit should have realigned the star's thin convective layer, yet an earlier Rossiter--McLaughlin (RM) measurement suggested a $\sim$60$^\circ$ spin--orbit misalignment. We present two new Keck Planet Finder RM measurements. We measure a sky-projected obliquity of $\lambda=-1.3\pm8.7^\circ$; and the star's projected rotation velocity is $\vsini=1.3\pm0.1~\mathrm{km\,s^{-1}}$, an order of magnitude lower than expected for a star of its mass and age. With WASP-12's low chromospheric activity, $\log R^\prime_{\rm HK}=-5.5$, our analysis favors an aligned system in which $\vsini$ reflects slow surface rotation with $P_{\rm rot}\sim60$ days, rather than a projection effect. Our evolution models favor a late-main-sequence star rather than a subgiant, with a central convective core found in 99\% of posterior samples. This internal structure suppresses the central-focusing and nonlinear wave-breaking mechanisms for inward-propagating gravity waves. We propose that WASP-12 b was initially placed on a retrograde orbit after high-eccentricity migration, when its host rotated at $\vsini\approx14$~km~s$^{-1}$ (or $P_{\rm rot}\sim$6 days). Tidal decay transferred orbital angular momentum to the convective envelope, aligning the surface. The retrograde interior acts as an angular-momentum reservoir, while interior--envelope coupling slows the surface rotation. A semi-empirical relation gives an envelope--interior angular-momentum exchange timescale of $\sim4$~Myr, similar to the observed orbital-decay timescale of $\sim3.25$~Myr. The angular-momentum budget also permits this initially retrograde evolution. In contrast, an initially prograde planet would have spun up the surface rotation to a rate substantially higher than observed.
\end{abstract}


\section{Introduction}
\label{sec:intro}

WASP-12 b is a hot Jupiter with a $1.09$~day orbital period and an inflated radius of $\simeq1.9\,R_{\rm Jup}$ \citep{Hebb2009,Li2010,Fossati2010,Haswell2012,Akinsanmi2024}. Long-baseline transit and occultation timings reveal steady orbital-period decay, $\dot P\simeq-29~\mathrm{ms\,yr^{-1}}$, implying a period-decay timescale of $\sim3.25$~Myr and an effective stellar tidal quality factor of $Q_\star'\approx1.8\times10^5$ \citep{Maciejewski2016,Patra2017,Yee2020,Wong2022,Akinsanmi2024,Biswas2026}. It remains the only exoplanet with confirmed orbital decay \citep{Yee2020,Bouma2020,Winn2025}.

Although orbital decay is well established, its efficient tidal-dissipation mechanism remains uncertain and depends sensitively on the host star's evolutionary state and internal structure \citep{Weinberg2017,Bailey2019,Yee2020,Turner2021}. \citet{Weinberg2017} showed that if WASP-12 is a subgiant without a central convective core, tidally excited internal gravity waves can reach the stellar center, where geometric focusing drives nonlinear wave breaking and dissipation consistent with the observed decay. However, \citet{Bailey2019} favor a main-sequence star retaining a convective core, preventing these waves from reaching the center. With only weak linear damping, the resulting $g$-mode resonances are too narrow to reproduce the decay without fine-tuning. Recent non-adiabatic tidal calculations likewise find that linear radiative and convective damping in main-sequence models with a convective core cannot reproduce WASP-12 b's observed decay \citep{Sun2025}. Alternative efficient-dissipation mechanisms include absorption at a corotation layer associated with a rapidly rotating interior \citep{Bailey2019} and magnetic conversion of inward-propagating internal gravity waves into outward-propagating magnetic waves that dissipate in the radiative interior \citep{Duguid2024}. The latter can operate in main-sequence stars with convective cores and may reproduce the inferred strong dissipation. Planetary obliquity tides are another possible source \citep{Millholland2018}, but recent angular-momentum and radial-velocity constraints disfavor this scenario \citep[][submitted]{Lammers2026}. Their radial-velocity analysis also hints at an observable stellar tidal bulge, suggesting equilibrium tides operate alongside other tidal effects and may transfer at least part of the planet's lost orbital angular momentum to the stellar surface layer.

Orbital decay also poses an angular momentum budget problem: the lost orbital angular momentum must go elsewhere in the system. Substantial initial deposition in the host's shallow convective envelope should tend to produce surface spin-up and orbit--surface realignment, although internal coupling may redistribute it throughout the star. Yet a previous RM observation yielded $\lambda=59^{+15}_{-20}{}^\circ$, suggesting substantial sky-projected orbit--surface misalignment \citep{Albrecht2012}. Moreover, we measure a low projected rotation velocity, $\vsini=1.3\pm0.1~\mathrm{km\,s^{-1}}$ (Section~\ref{sec:obliquity}), consistent with the spectroscopic $2.2\pm1.5~\mathrm{km\,s^{-1}}$ reported by \citet{Hebb2009} and compiled by \citet{Bonomo2017}. WASP-12 also has exceptionally weak Ca {\sc II} H\&K and Mg {\sc II} emissions \citep{Knutson2010,Haswell2012,Fossati2013}. \citet{Knutson2010} reported $\log R'_{\rm HK}=-5.50$ without uncertainty; we adopt a typical 0.05 dex. Its activity lies near the low end for similar stars (Figure~\ref{fig:activity_rossby}). The slow rotation and previously reported misalignment could arise if tidal angular momentum is deposited below the surface convective layer rather than retained in the envelope. Accurate spin--orbit and surface-rotation measurements, with improved stellar internal-structure constraints, therefore help identify the ultimate sink of the planet's lost orbital angular momentum.

We present two new Rossiter-McLaughlin \citep[RM,][]{Rossiter1924,McLaughlin1924} observations of WASP-12 b with the Keck Planet Finder \citep[KPF,][]{Gibson2016,Gibson2018,Gibson2020}, jointly analyzed with \textit{TESS} \citep{Ricker2015} photometry. The sky-projected obliquity $\lambda=-1.3\pm8.7^\circ$ and $\vsini=1.3\pm0.1\, \mathrm{km s^{-1}}$ favor alignment and anomalously slow surface rotation. Our stellar-evolution and magnetic-braking models favor a late-main-sequence host with a central convective core, a thin outer convective envelope, and much faster surface rotation than observed. We propose that orbital angular momentum was deposited partly in the surface convective layer, aligning it with the orbit. If WASP-12 b began on a retrograde orbit, a retrograde stellar interior could absorb this angular momentum by slowing its retrograde rotation rather than spinning up prograde. Interior--envelope coupling is likely fast enough for this redistribution, and the angular-momentum budget permits the required transfer.

Section~\ref{sec:obliquity} presents the stellar-obliquity measurement, and Section~\ref{sec:stellar_para} describes the stellar-evolution models. Section~\ref{sec:discussion} combines these results to explore an initially retrograde WASP-12 b; Section~\ref{sec:summary} summarizes our findings.

\section{Stellar Obliquity Measurements} \label{sec:obliquity}

\subsection{{\it TESS} Light Curves}
\label{sec:tess_lc}

The {\it Transiting Exoplanet Survey Satellite} ({\it TESS}; \citealt{Ricker2015}) observed WASP-12 in Sectors 20, 43, 44, 45, 71, and 72. We analyzed 120-s-cadence light curves from the {\it TESS} Science Processing Operations Center \citep[SPOC;][]{Jenkins2016}, publicly available through the Mikulski Archive for Space Telescopes (MAST)\footnote{\url{https://mast.stsci.edu/portal/Mashup/Clients/Mast/Portal.html}} at DOI: \dataset[10.17909/0rhv-mm50]{https://doi.org/10.17909/0rhv-mm50}.

We used the Presearch Data Conditioning Simple Aperture Photometry (PDCSAP) light curves provided by the SPOC pipeline \citep{Smith2012, Stumpe2014}. The light curves were downloaded using the \texttt{lightkurve} package \citep{lightkurve2018}. Any NaN values were removed and significant outliers were rejected using a $10\sigma$ clipping threshold.

To isolate WASP-12 b's transit signal, we masked transits within windows of 1.5 times the known duration. We detrended the masked light curves with \texttt{lightkurve}'s \texttt{flatten} routine, which uses a Savitzky--Golay filter \citep{Savitzky1964} to remove low-frequency stellar-activity variability and residual instrumental trends. The normalized light curves preserve transit profiles and photometric uncertainties for subsequent transit and obliquity analyses.

\subsection{KPF Rossiter--McLaughlin Observations}

We obtained high-resolution spectra of two WASP-12 b transits with the Keck Planet Finder \citep[KPF;][]{Gibson2016, Gibson2018,Gibson2020} on Keck~I at the W.\ M.\ Keck Observatory. KPF covers 445--870 nm at resolving power $R \sim 98{,}000$. Observations on UT 2024 March 23 and UT 2026 February 2 yielded 22 and 45 reduced spectra, respectively, with 600 s integrations. The first sequence covers the full transit and roughly 30 minutes each before ingress and after egress; the second extends approximately 140 minutes on either side. Extracted spectra typically have a signal-to-noise ratio of approximately 100 per pixel at 550 nm. Substantially poorer seeing during the latter half of the 2024 March 23 observations reduced this to $\sim50$ per pixel at 550 nm and increased RV uncertainties for several late exposures.

The spectra were reduced using the KPF Data Reduction Pipeline (DRP), version 2.12.0, applying bias subtraction, flat-field correction, order extraction, and wavelength calibration. Radial velocities were calculated using the cross-correlation function (CCF) method with numerical stellar masks adapted from those developed for {\tt ESPRESSO} \citep{Pepe2013}. Table~\ref{tab:KPF_data} lists the RV measurements used to model the RM signal. 

\begin{deluxetable}{@{}cccccc@{}}[!htbp]
\tablecaption{WASP-12 KPF RM Data \label{tab:KPF_data}}
\tabletypesize{\scriptsize}
\tablehead{\colhead{Time} & \colhead{RV} & \colhead{$\sigma_{\rm RV}$} &
\colhead{Time} & \colhead{RV} & \colhead{$\sigma_{\rm RV}$} \\
\colhead{(BJD)} & \colhead{(km/s)} & \colhead{(m/s)} & \colhead{(BJD)} & \colhead{(km/s)} & \colhead{(m/s)}
}
\startdata
3392.7443 & 19.10657 &  3.04 & 4073.7037 & 19.20251 & 2.74 \\
3392.7495 & 19.10536 &  3.13 & 4073.7116 & 19.20115 & 2.66 \\
3392.7581 & 19.09522 &  2.55 & 4073.7192 & 19.18961 & 2.67 \\
3392.7653 & 19.09374 &  2.71 & 4073.7263 & 19.18773 & 3.41 \\
3392.7733 & 19.08670 &  2.95 & 4073.7342 & 19.17513 & 2.93 \\
3392.7805 & 19.07800 &  2.77 & 4073.7419 & 19.16775 & 2.77 \\
3392.7879 & 19.06833 &  2.86 & 4073.7493 & 19.15788 & 2.67 \\
3392.7959 & 19.05687 &  3.12 & 4073.7569 & 19.15044 & 2.71 \\
3392.8032 & 19.04866 &  2.85 & 4073.7642 & 19.14286 & 2.73 \\
3392.8106 & 19.03322 &  2.82 & 4073.7718 & 19.13673 & 2.72 \\
3392.8182 & 19.01974 &  2.91 & 4073.7792 & 19.12197 & 2.98 \\
3392.8257 & 19.00782 &  2.88 & 4073.7868 & 19.11350 & 2.95 \\
3392.8329 & 18.99984 &  3.16 & 4073.7945 & 19.10544 & 2.93 \\
3392.8407 & 18.98529 &  3.53 & 4073.8020 & 19.09968 & 2.89 \\
3392.8573 & 18.96544 &  6.26 & 4073.8093 & 19.09441 & 2.75 \\
3392.8649 & 18.95242 &  6.03 & 4073.8169 & 19.08319 & 2.77 \\
3392.8728 & 18.96697 &  5.53 & 4073.8243 & 19.07640 & 2.80 \\
3392.8802 & 18.96229 &  5.18 & 4073.8319 & 19.06455 & 2.94 \\
3392.8875 & 18.94005 &  5.25 & 4073.8395 & 19.05615 & 2.80 \\
3392.8952 & 18.92928 &  5.47 & 4073.8469 & 19.04640 & 2.61 \\
3392.9026 & 18.92638 &  5.34 & 4073.8544 & 19.02916 & 2.60 \\
3392.9098 & 18.91698 &  5.86 & 4073.8619 & 19.01682 & 2.60 \\
-- & -- & -- & 4073.8694 & 19.00059 & 2.76 \\
-- & -- & -- & 4073.8771 & 18.99452 & 2.74 \\
-- & -- & -- & 4073.8843 & 18.97821 & 3.07 \\
-- & -- & -- & 4073.8918 & 18.96810 & 3.07 \\
-- & -- & -- & 4073.8994 & 18.95448 & 3.11 \\
-- & -- & -- & 4073.9072 & 18.94510 & 2.85 \\
-- & -- & -- & 4073.9144 & 18.94101 & 2.59 \\
-- & -- & -- & 4073.9219 & 18.94026 & 2.66 \\
-- & -- & -- & 4073.9294 & 18.92861 & 2.70 \\
-- & -- & -- & 4073.9370 & 18.92094 & 2.79 \\
-- & -- & -- & 4073.9446 & 18.91272 & 2.49 \\
-- & -- & -- & 4073.9594 & 18.89606 & 2.44 \\
-- & -- & -- & 4073.9670 & 18.88749 & 2.48 \\
-- & -- & -- & 4073.9744 & 18.87672 & 2.61 \\
-- & -- & -- & 4073.9820 & 18.87092 & 2.58 \\
-- & -- & -- & 4073.9895 & 18.86348 & 2.61 \\
-- & -- & -- & 4073.9970 & 18.85921 & 2.60 \\
-- & -- & -- & 4074.0044 & 18.85255 & 2.79 \\
-- & -- & -- & 4074.0119 & 18.85000 & 3.08 \\
-- & -- & -- & 4074.0195 & 18.84161 & 3.49 \\
-- & -- & -- & 4074.0269 & 18.84292 & 4.52 \\
-- & -- & -- & 4074.0345 & 18.82965 & 5.67 \\
-- & -- & -- & 4074.0420 & 18.81564 & 7.82 \\
\enddata
\tablecomments{Times are given as BJD$-$2457000. Radial velocities are reported in km s$^{-1}$, while their uncertainties are reported in m s$^{-1}$.}
\end{deluxetable}

\subsection{Joint Light Curve and Rossiter--McLaughlin Modeling}

To characterize WASP-12's spin--orbit architecture, we jointly fit the {\it TESS} transit photometry and KPF radial velocities with a modified {\tt allesfitter}\footnote{\url{https://github.com/wangxianyu7/Allesfitters},\\\url{https://doi.org/10.5281/zenodo.21792490}} package \citep{allesfitter-paper, allesfitter-code, Wang2024, wang_2026_21792490}. The modification incorporates the analytic RM prescription of \citet{Hirano2011}, extending the formulation of \citet{Ohta2005, Hirano2010}. Simultaneously modeling the transit light curve and spectroscopic RM anomaly enables self-consistent measurement of the projected stellar obliquity $\lambda$ and other transit parameters.

The {\it TESS} data contain multiple 2-min-cadence transits of WASP-12 b (Section~\ref{sec:tess_lc}), while the spectroscopic data comprise two KPF transit sequences from UT 2024 March 23 (KPF1) and UT 2026 February 2 (KPF2), treated as independent radial-velocity datasets. The transit model includes the planet-to-star radius ratio $R_{\rm p}/R_{\star}$, scaled sum of radii $(R_{\star}+R_{\rm p})/a$, orbital-inclination cosine $\cos i_{\rm p}$, and period $P$. Because WASP-12 b shows measurable transit-timing variations from orbital decay, we fitted independent transit midpoints ($T_0$) for KPF1 and KPF2. We assumed a circular orbit, justified by the planet's short period and available RV data.

We modeled the radial velocities as the RM signal plus an independent 3rd-order polynomial baseline for each KPF dataset. Because WASP-12 b has a short period ($P\approx1.09$ d), stellar radial velocity changes measurably over the $\sim3.0$ hr transit. The baselines flexibly describe low-frequency variations, including orbital motion, the tidal RV effect, and possible stellar or instrumental variability. Since the localized in-transit anomaly primarily constrains the RM signal, these low-order trends mainly absorb longer-timescale variations and are not expected to significantly affect $\lambda$. Fitting the out-of-transit RV variation with the planet's Keplerian orbit gave nearly identical obliquity constraints. The RM model adds two free parameters: projected stellar rotation $\vsini$ and projected spin--orbit angle $\lambda$.

We assumed quadratic limb darkening for the photometric and spectroscopic datasets, parameterized by the $q_1$ and $q_2$ parameters of \citet{Kipping2013}, with independent coefficients for the {\it TESS} and KPF bandpasses. To account for white noise beyond formal uncertainties, we included a photometric error-inflation term, $\ln \sigma_{\mathrm{{\it TESS}}}$, and independent radial-velocity jitter terms, $\ln \sigma_{\mathrm{jitter;KPF1}}$ and $\ln \sigma_{\mathrm{jitter;KPF2}}$, added in quadrature to the two KPF RV uncertainties.

Uniform priors, $\mathcal{U}(a,b)$, were assigned to all fitted parameters unless noted otherwise. We imposed a Gaussian prior on stellar mean density from the mass and radius in Table~\ref{tab:modeling} (Section~\ref{sec:stellar_para}). Following \citet{Hirano2011}, we also modeled the micro- and macroturbulent velocities, $V_\xi$ and $V_\zeta$, with 1 km s$^{-1}$ Gaussian priors centered on empirical calibrations. We used \citet{Bruntt2010} and \citet{Doyle2014} where applicable and otherwise the Gaia--ESO Survey calibrations of \citet{Blanco2014HighResolutionSpectralLibrary, Blanco2014iSpec, Blanco2019iSpec}.

\begin{figure*}[htbp]
    \centering
    \begin{minipage}[c]{0.65\linewidth}
        \centering
        \includegraphics[width=\linewidth]{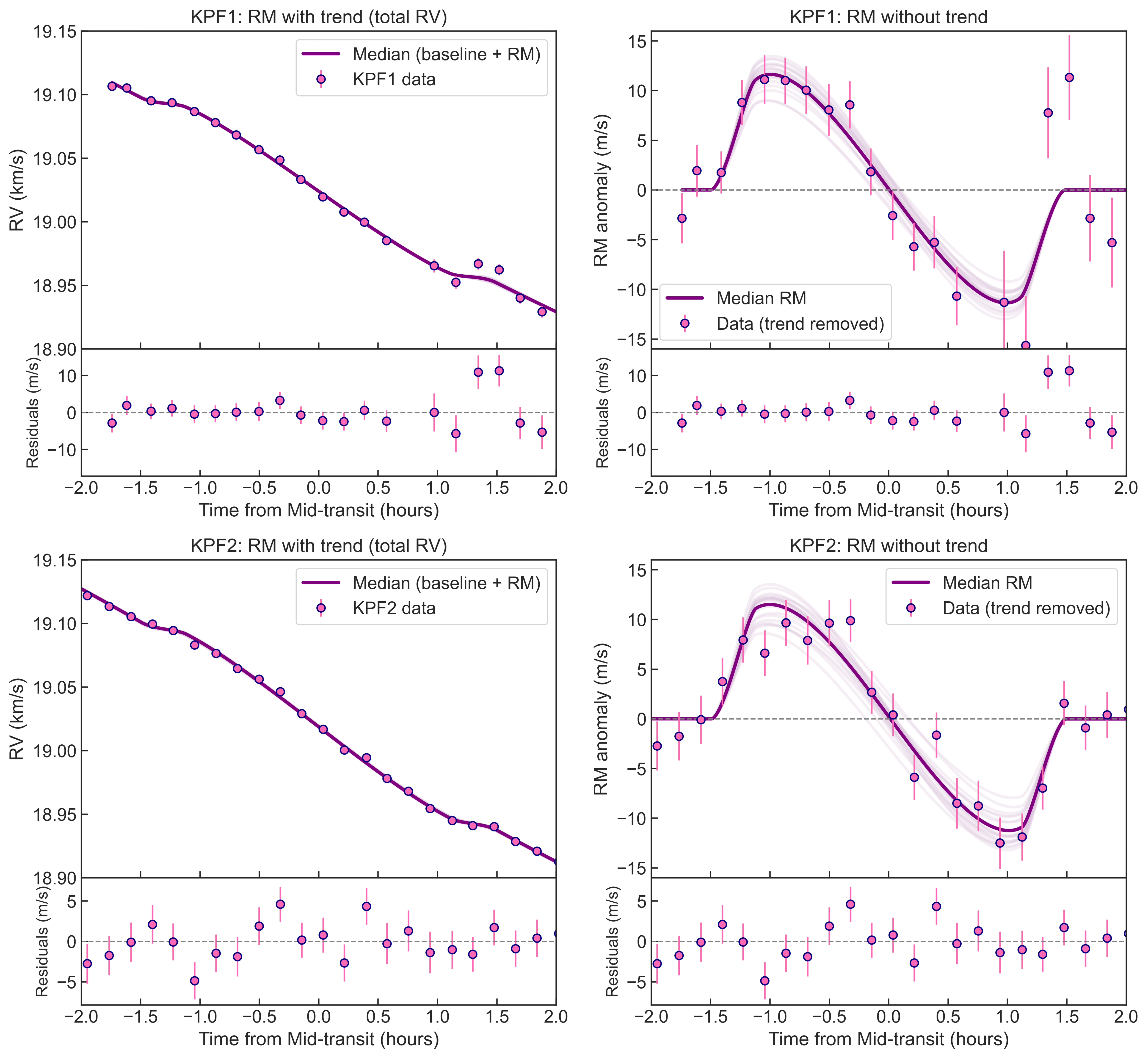}
    \end{minipage}
    \hfill
    \begin{minipage}[c]{0.33\linewidth}
        \centering
        \includegraphics[width=\linewidth]{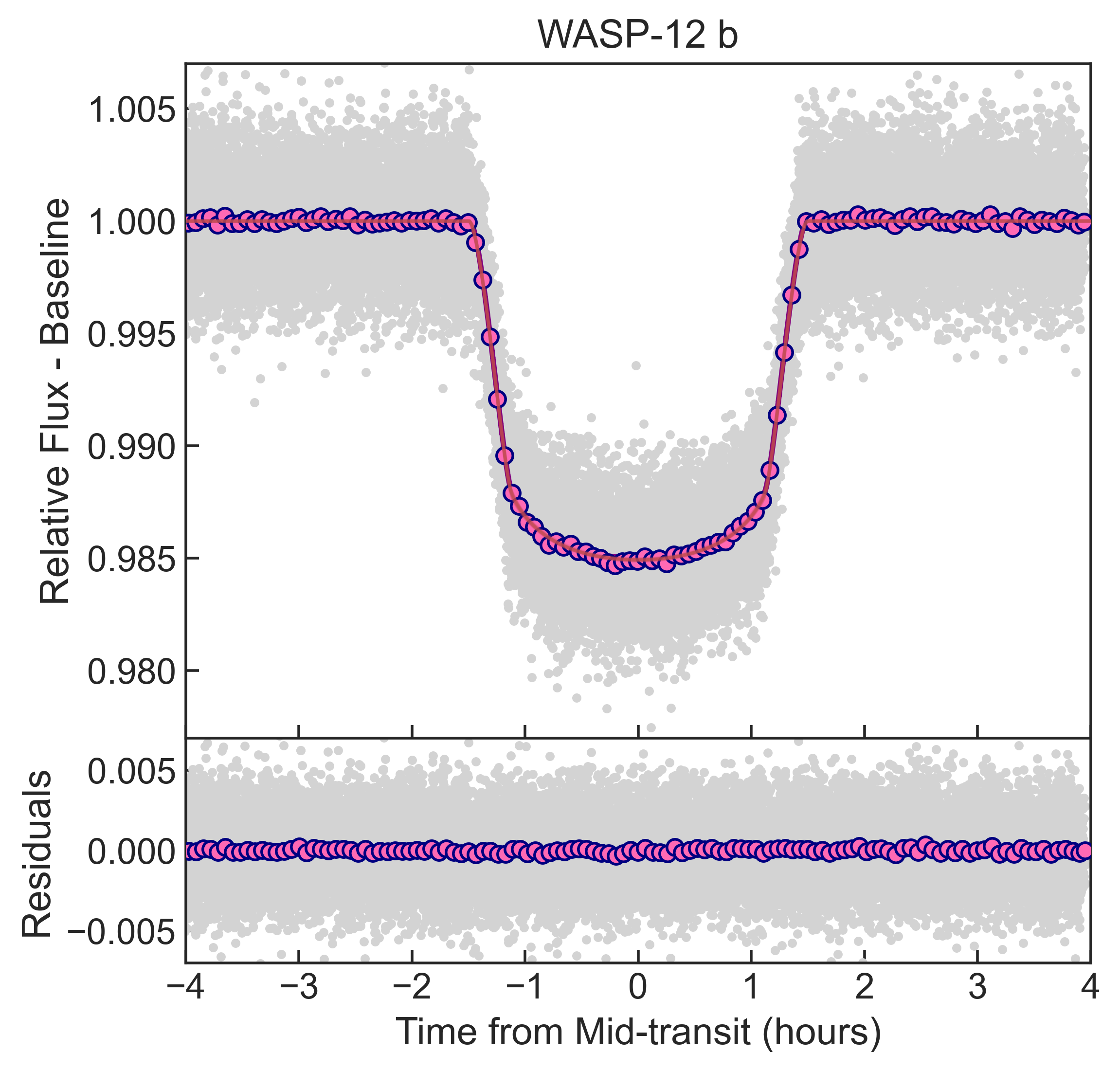}
        \vspace{0.8em}
        \includegraphics[width=\linewidth]{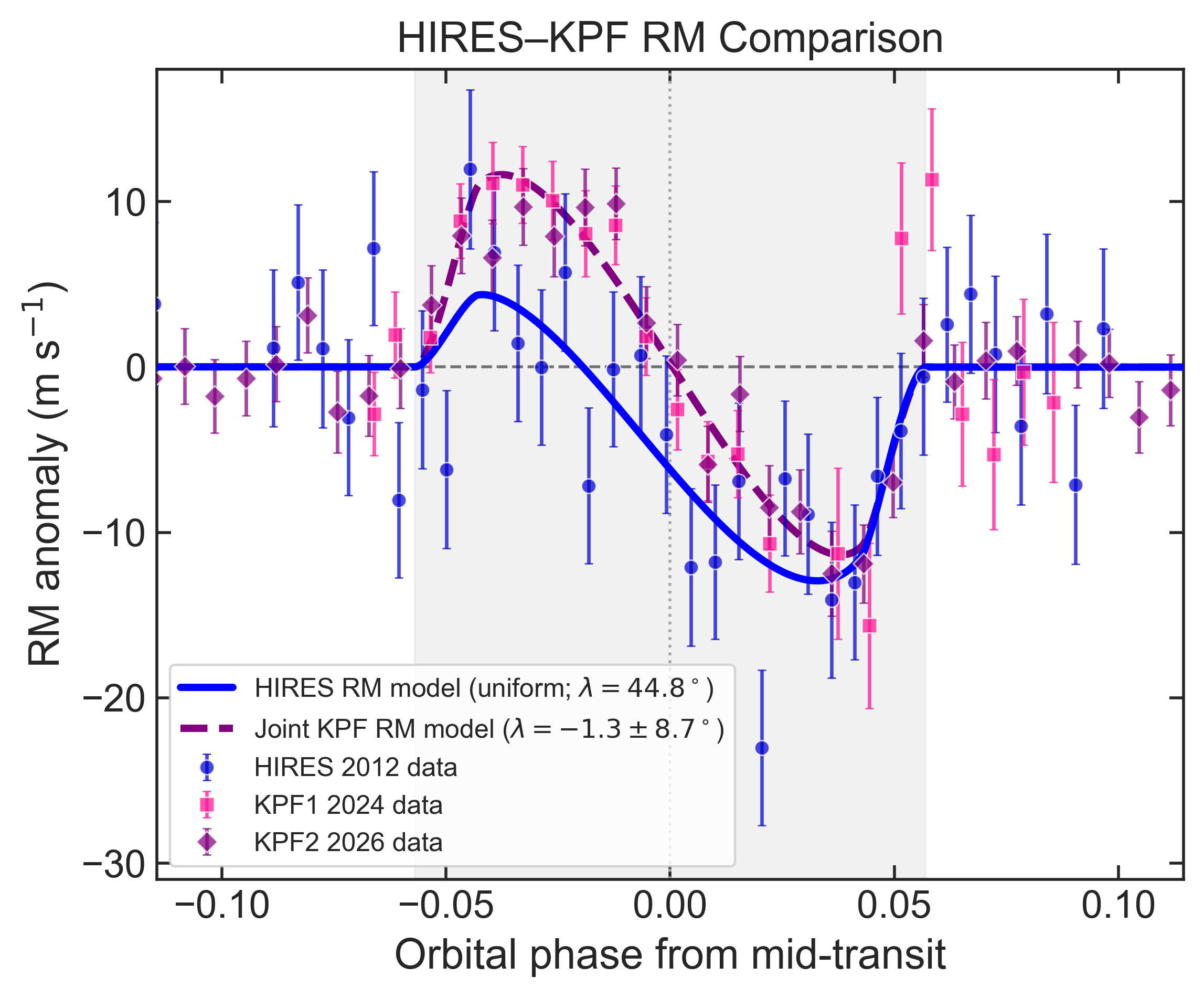}
    \end{minipage}
    \caption{
    \textbf{Left:} KPF Rossiter--McLaughlin observations obtained on UT 2024 March 23 (KPF1) and UT 2026 February 2 (KPF2). For each observing sequence, the left panels show the full radial-velocity time series including the fitted baseline and RM model, while the right panels show the isolated RM anomaly after baseline removal. The best-fit model yields a projected stellar obliquity of $\lambda=-1.3\pm8.7^\circ$. Residuals are shown in the lower panels.
    \textbf{Upper right:} Phase-folded \textit{TESS} transit light curve of WASP-12 b together with the best-fit transit model from the joint \texttt{allesfitter} analysis. The lower panel shows the residuals.
    \textbf{Lower right:} Baseline-subtracted RM measurements from HIRES (blue circles), KPF1 (pink squares), and KPF2 (purple diamonds). Error bars include the fitted RV jitter added in quadrature. The solid blue and dashed purple curves show the posterior-median HIRES model with a uniform prior on $\vsini$ and the joint KPF--\textit{TESS} model, respectively. The shaded region spans the transit window, and the vertical dotted line marks mid-transit.
    }
    \label{fig:WASP12_alles}
\end{figure*}

Posterior distributions were sampled using the nested sampling algorithm implemented in the {\tt dynesty} package \citep{Higson2019,DYNESTY}, as integrated within {\tt allesfitter}. The sampling was initialized with 500 live points and employed a random-walk ({\tt rwalk}) proposal strategy. The final posteriors contain 45,918 samples. We assessed convergence using the default criteria, posterior inspection, and the Anderson–Darling, Augmented Dickey–Fuller, Durbin–Watson, and Ljung–Box residual diagnostics provided by {\tt allesfitter}. Figure~\ref{fig:WASP12_alles} shows the best-fit RM model and detrended {\it TESS} transits; Table~\ref{tab:wasp12_all_params} summarizes median posteriors, 68\% credible intervals, and derived physical properties.

\begin{figure*}[htbp]
    \centering
    \includegraphics[width=0.90\textwidth]{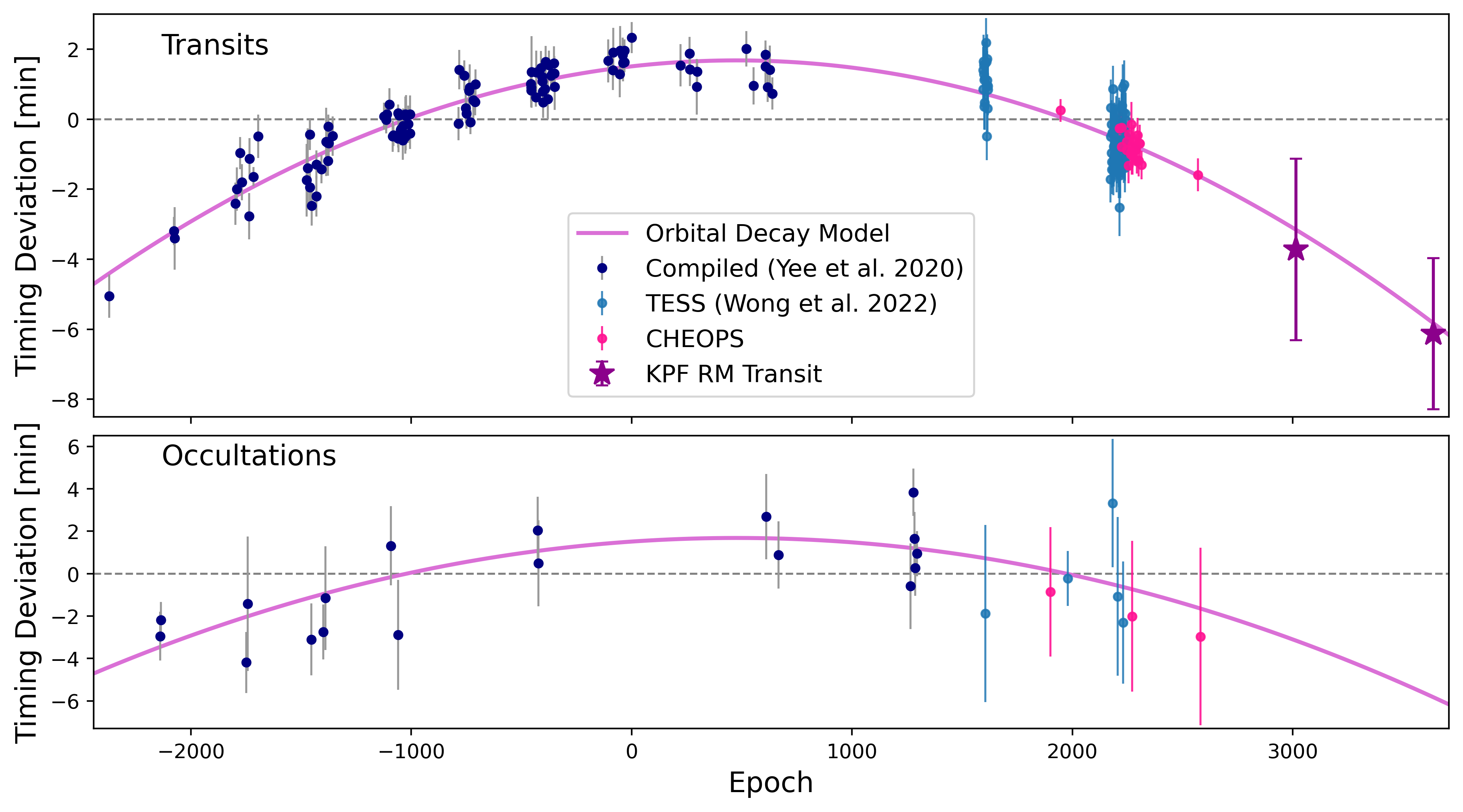}
    \caption{Transit and occultation timing deviations ($\rm O-C$) for WASP-12 b relative to the linear ephemeris. Navy points show timings compiled by \citet{Patra2017} and \citet{Yee2020}, blue points show \textit{TESS} measurements from \citet{Wong2022}, and pink points show CHEOPS measurements from \citet{Akinsanmi2024}. The orchid curve shows the quadratic orbital-decay ephemeris from \citet{Akinsanmi2024}. Purple stars mark the KPF RM-derived timings at $E=3014$ and $3638$. Both timings are consistent with the extrapolated decay trend within their uncertainties.}
    \label{fig:oc_diagram}
\end{figure*}

For 2024 March 23, we measure a mid-transit time of $T_0=2460392.8165\pm0.0018~{\rm BJDTDB}$, corresponding to $\rm O-C=-3.72\pm2.59$~min. For 2026 February 2, we measure $T_0=2461073.8602\pm0.0015~{\rm BJDTDB}$, corresponding to $\rm O-C=-6.13\pm2.16$~min. 
Figure~\ref{fig:oc_diagram} compares these measurements with previously published transit and occultation timings and with the quadratic orbital-decay ephemeris of \citet{Akinsanmi2024},
\begin{equation}
T(E)=T_0+PE+\frac{1}{2}\frac{dP}{dE}E^2,
\end{equation}
where $T_0=2457103.2837~{\rm BJDTDB}$, $P=1.0914~{\rm d}$, and $dP/dE=-1.043\times10^{-9}~{\rm d epoch^{-1}}$. Both KPF timings are consistent with the extrapolated quadratic ephemeris within their uncertainties. Although less precise than dedicated photometric measurements, they provide two independent checks, separated by nearly two years, of the established orbital-decay trend. We did not attempt an independent measurement of the orbital-decay rate because the RM-derived timings are not sufficiently precise.

\begin{table*}[!htbp]
\centering
\caption{Joint {\tt allesfitter} Model Parameters for WASP-12}
\label{tab:wasp12_all_params}
\footnotesize
\begin{tabular}{cccc}
\hline
Parameter & Prior & Value & Unit \\
\hline
\multicolumn{4}{c}{\textbf{Fitted Parameters}} \\
\hline
$R_b / R_\star$ & $\mathcal{U}(0,0.2)$ & $0.11618\pm0.00040$ &  \\
$(R_\star + R_b) / a_b$ & $\mathcal{U}(0,0.5)$ & $0.3689\pm0.0028$ &  \\
$\cos i_b$ & $\mathcal{U}(0,0.2)$ & $0.1243_{-0.0083}^{+0.0077}$ &  \\
$T_{0;b}-2457000\ {\ast}$ & $\mathcal{U}(2928.8,2929.1)$ & $2928.964648\pm0.000046$ & BJD \\
$P_b\ {\ast\ast}$ & $\mathcal{U}(1.0,1.2)$ & $1.09141644\pm0.00000011$ & d \\
$\sqrt{e_b}\cos\omega_b$ & fixed & 0.0 &  \\
$\sqrt{e_b}\sin\omega_b$ & fixed & 0.0 &  \\
\hline
$q_{1;\mathrm{TESS}}$ & $\mathcal{U}(0,1)$ & $0.247_{-0.038}^{+0.044}$ &  \\
$q_{2;\mathrm{TESS}}$ & $\mathcal{U}(0,1)$ & $0.236_{-0.056}^{+0.062}$ &  \\
$q_{1;\mathrm{KPF1}}$ & $\mathcal{U}(0,1)$ & $0.58_{-0.28}^{+0.26}$ &  \\
$q_{2;\mathrm{KPF1}}$ & $\mathcal{U}(0,1)$ & $0.49\pm0.32$ &  \\
$q_{1;\mathrm{KPF2}}$ & $\mathcal{U}(0,1)$ & $0.46_{-0.22}^{+0.26}$ &  \\
$q_{2;\mathrm{KPF2}}$ & $\mathcal{U}(0,1)$ & $0.64_{-0.35}^{+0.25}$ &  \\
$\ln\sigma_{\mathrm{TESS}}$ & $\mathcal{U}(-10,0)$ & $-6.1760\pm0.0041$ & ln(rel.\ flux) \\
$\ln\sigma_{\mathrm{jitter;KPF1}}$ & $\mathcal{U}(-10,0)$ & $-8.3\pm1.2$ & ln(km s$^{-1}$) \\
$\ln\sigma_{\mathrm{jitter;KPF2}}$ & $\mathcal{U}(-10,0)$ & $-8.58_{-0.96}^{+1.1}$ & ln(km s$^{-1}$) \\
$\lambda_\star$ & $\mathcal{U}(-180,180)$ & $-1.3\pm8.7$ & deg \\
$\vsini$ & $\mathcal{U}(0,10)$ & $1.29\pm0.11$ & km s$^{-1}$ \\
$V_\xi$ & $\mathcal{N}(1.0,1)$ & $1.21_{-0.73}^{+0.90}$ & km s$^{-1}$ \\
$V_\zeta$ & $\mathcal{N}(2.86,1)$ & $2.83\pm0.99$ & km s$^{-1}$ \\
\hline
\multicolumn{4}{c}{\textbf{Derived Parameters}} \\
\hline
$a_b/R_\star$ & -- & $3.026_{-0.021}^{+0.022}$ &  \\
$R_b$ & -- & $1.891\pm0.040$ & $R_{\rm Jup}$ \\
$a_b$ & -- & $0.02354\pm0.00052$ & AU \\
$i_b$ & -- & $82.86_{-0.45}^{+0.48}$ & deg \\
$b_{\rm tra;b}$ & -- & $0.376_{-0.023}^{+0.021}$ &  \\
$T_{\rm tot;b}$ & -- & $2.9815\pm0.0065$ & h \\
$T_{\rm full;b}$ & -- & $2.248_{-0.013}^{+0.014}$ & h \\
$T_{\rm eq;b}$ & -- & $2324\pm42$ & K \\
$\delta_{\rm tr;b}$ & -- & $15.067\pm0.039$ & ppt \\
\hline
$\rho_{\star;\rm combined} {\ast\ast\ast}$ & -- & $0.4399_{-0.0091}^{+0.0097}$ & g cm$^{-3}$ \\
\hline
\end{tabular}
\vspace{2mm}
\parbox{\textwidth}{
\textit{Note.}--- $\mathcal{U}(a,b)$ denotes a uniform prior between $a$ and $b$. \\
Gaussian priors $\mathcal{N}(\mu,\sigma)$ with $\sigma=1$ km s$^{-1}$ were adopted for the micro- and macroturbulent velocities. \\
KPF1 and KPF2 denote the two independent Rossiter--McLaughlin observing sequences obtained on UT 2024 March 23 and UT 2026 February 2, respectively. Separate limb-darkening coefficients and radial-velocity jitter parameters were fitted for each dataset. \\
${\ast}$ {\tt allesfitter} automatically shifts the input prior epoch to the center of the data time span. \\
${\ast\ast}$ The listed period is the best-fit linear-ephemeris period used as a reference; the system departs measurably from a strictly linear ephemeris because of orbital decay (Section~\ref{sec:obliquity}). \\
$\ast\ast\ast$ Combined host density from all orbits.
}
\end{table*}

\subsection{Reanalysis of the 2012 HIRES Observations}
\label{sec:hires_reanalysis}

To investigate the discrepancy with the earlier projected obliquity, $\lambda=59^{+15}_{-20}{}^\circ$ \citep{Albrecht2012}, we reanalyzed 38 Keck/HIRES RVs from the UT 2012 January 1-2 transit using the same modified \texttt{allesfitter} and RM prescription as for KPF. We fixed transit geometry, including impact parameter, to the joint KPF--\textit{TESS} best fit and varied the HIRES transit midpoint, quadratic limb-darkening coefficients, RV jitter, $\lambda$, and $\vsini$. As for KPF, a 3rd-order polynomial baseline described orbital and other slowly varying RV signals.

We performed three fits that differed only in the prior adopted for $\vsini$. A uniform prior, $\vsini\sim\mathcal{U}(0,10)~{\rm km\,s^{-1}}$, yielded $\lambda=44.8^{+15.1}_{-23.9}{}^\circ$ and $\vsini=1.3^{+0.5}_{-0.4}~{\rm km\,s^{-1}}$; the earlier spectroscopic constraint, $\vsini=2.2\pm1.5~{\rm km\,s^{-1}}$ \citep{Hebb2009}, yielded $\lambda=46.5^{+14.7}_{-22.8}{}^\circ$ and $\vsini=1.4^{+0.5}_{-0.4}~{\rm km\,s^{-1}}$. The informative KPF posterior prior, $\vsini=1.3\pm0.1~{\rm km\,s^{-1}}$, gave $\lambda=40.8^{+13.1}_{-16.3}{}^\circ$. Changing the prior shifts the obliquity by less than $6^\circ$, substantially below its statistical uncertainty. The uninformative fit independently favors low $\vsini$ consistent with KPF; different rotation priors therefore do not primarily explain the HIRES--KPF obliquity discrepancy.

Figure~\ref{fig:WASP12_alles} (lower right) compares baseline-subtracted HIRES, KPF1, and KPF2 observations at common orbital phases. The jointly modeled KPF nights share one RM solution. HIRES's preference for moderate misalignment appears driven mainly by several first-half-transit measurements favoring a more asymmetric RM waveform than the joint KPF model. The obliquity difference is nevertheless modest, approximately $2\sigma$. Given the lower precision of the HIRES observations, the apparent asymmetry may reflect a statistical fluctuation or an unmodeled systematic; our reanalysis does not distinguish between these possibilities.

\section{Stellar Evolution Model}
\label{sec:stellar_para}
\subsection{Non-rotating Models}
To better constrain the internal structure of WASP-12, we compute a grid of stellar evolution models using the Modules for Experiments in Stellar Astrophysics \citep[\mesa;][]{Paxton2011,Paxton2013,Paxton2015,Paxton2018,Paxton2019,Jermyn2023}, version~r25.12.1. Each model is evolved from the pre-main-sequence (with central temperature $T_c = 9.9\times10^5$ K) through the main sequence and subgiant phase, until either the surface temperature drops below $T_{\rm eff} = 5500$ K or the photospheric radius exceeds $R_\star = 3 R_\odot$.

We adopt the AGSS09 solar abundance mixture \citep{Asplund2009} with corresponding opacity tables (\texttt{kap\_file\_prefix = `a09'}), and the \texttt{pp\_and\_cno\_extras} nuclear-reaction network. Convection is treated using the \citet{Henyey1965} formulation of mixing-length theory (MLT), with the Ledoux criterion for convective stability and semiconvection and thermohaline coefficients following \citet{Choi2016}. Predictive mixing is applied at convective-core boundaries to avoid composition reversals. Exponential convective overshooting is included at both the core and envelope base. Atomic diffusion and gravitational settling of H, He, C, N, O, Ne, and Mg are enabled following \citet{Thoul1994}, with diffusion suppressed where the convective diffusion coefficient is $D_{\rm mix}>100$ cm$^2$ s$^{-1}$. The atmosphere is attached using the \citet{Eddington1926} $T$--$\tau$ relation with iterated opacities. A minimum diffusion coefficient of $D_{\rm min}=10$ cm$^2$ s$^{-1}$ is imposed throughout to account for other mixing processes.

The stellar models are constrained by four observables derived from spectroscopic or transit analyses of \wasp{12}. The effective temperature $T_{\rm eff} = 6155 \pm 100$ K and bulk metallicity $[\mathrm{M/H}] = 0.15 \pm 0.06$ come from {\tt SpecMatch-Syn} \citep{CKS1} analysis of a high-S/N iodine-free Keck/HIRES spectrum obtained UT 2012 Jan 5. Our transit analysis gives the mean stellar density $\bar{\rho} = 0.4399 \pm 0.0097$ g cm$^{-3}$. The luminosity $L = 3.75 \pm 0.13 L_\odot$ comes from \textsc{isoclassify} \citep{Huber2017}, using the \textit{Gaia} parallax, 2MASS $K_s$ magnitude with bolometric correction, and $A_{K_s} \approx 0.3$~mag extinction from the \citet{Green2019} dust map. For each grid model, $\mathcal{L} \propto \exp\!\left(-\chi^2/2\right)$, where $\chi^2 = \sum_i \left[(O_i - M_i)/\sigma_i\right]^2$ over these four observables. Marginalizing over all grid models yields posterior distributions for every output quantity.

From the constrained ensemble, we infer global and internal properties including $M_\star$, $R_\star$, age $t$, and surface gravity $\log g$. The models favor three layers: a central convective core, overlying radiative layer, and thin outer convective envelope. WASP-12 retains a convective core in 99\% of the posterior, strongly favoring a star not yet evolved to a post-main-sequence structure lacking one. Throughout, ``convective core'' denotes the central convective region, while ``stellar interior below the convective envelope'' collectively denotes the radiative layer and convective core.

Of particular interest are the convective-zone-base radius $r_{\rm bcz}$, fractional mass coordinate $m_{\rm bcz}/M_\star$, and moments of inertia of the convective envelope and underlying stellar interior. The envelope is very shallow in mass: $m_{\rm bcz}/M_\star=(99.51\pm0.61)\%$, implying $\sim0.5\%$ of the stellar mass, yet it contributes approximately 5\% of the total moment of inertia, with $\log I_{\rm env}=52.92\pm0.56$ versus $\log I_\star=54.22\pm0.07$ in cgs units. These quantities determine how strongly the observable surface responds to angular momentum transferred from the decaying orbit and redistributed within the star, and are central to the tidal angular-momentum budget in Section~\ref{sec:tides}. Table~\ref{tab:modeling} summarizes the inferred parameters and adopted uncertainties.

\begin{deluxetable}{@{}lcc@{}}[!htbp]
\tablecaption{WASP-12 Stellar Modeling Results \label{tab:modeling}}
\tabletypesize{\footnotesize}
\tablehead{\colhead{Parameter} & \colhead{Value} & \colhead{Unit}}
\startdata
\hline
\multicolumn{3}{c}{\textbf{Modeling inputs}} \\
\hline
Effective temperature $T_{\rm eff}$ & $6155 \pm 100$ & K \\
Metallicity $[\mathrm{M/H}]$ & $0.15 \pm 0.06$ & dex \\
Mean stellar density $\bar{\rho}$ & $0.4399 \pm 0.0097$ & g cm$^{-3}$ \\
Luminosity $L_\star$ & $3.75 \pm 0.13$ & $L_\odot$ \\
\hline
\multicolumn{3}{c}{\textbf{Estimated parameters}} \\
\hline
Initial helium fraction $Y_{\rm init}$ & $0.252 \pm 0.030$ &  \\
Mixing-length parameter $\alpha_{\rm MLT}$ & $2.10 \pm 0.38$ &  \\
Mass $M_\star$ & $1.353 \pm 0.059$ & $M_\odot$ \\
Radius $R_\star$ & $1.632 \pm 0.025$ & $R_\odot$ \\
Age $t$ & $3.44 \pm 0.56$ & Gyr \\
Surface gravity $\log g$ & $4.1428 \pm 0.0094$ & cm\,s$^{-2}$ \\
\textbf{Convective core prob.} & \textbf{$0.990 \pm 0.002$} &  \\
Convective envelope prob. & $1.000 \pm 0.000$ &  \\
BCZ radius $r_{\rm bcz}$ & $1.281 \pm 0.097$ & $R_\odot$ \\
Norm.\ BCZ radius $r_{\rm bcz}/R_\star$ & $0.783 \pm 0.056$ &  \\
BCZ mass coordinate $m_{\rm bcz}$ & $1.345 \pm 0.058$ & $M_\odot$ \\
Norm.\ BCZ mass coordinate $m_{\rm bcz}/M_\star$ & $0.9951 \pm 0.0061$ &  \\
ICZ radius$^1$ $r_{\rm icz}$ & $0.1072 \pm 0.0059$ & $R_\odot$ \\
Norm.\ ICZ radius$^1$ $r_{\rm icz}/R_\star$ & $0.0658 \pm 0.0033$ &  \\
ICZ mass coordinate$^1$ $m_{\rm icz}$ & $0.0973 \pm 0.0065$ & $M_\odot$ \\
Norm.\ ICZ mass coordinate$^1$ $m_{\rm icz}/M_\star$ & $0.0720 \pm 0.0040$ &  \\
Envelope moment of inertia $\log I_{\rm env}$ & $52.92 \pm 0.56$ & g\,cm$^2$ \\
Total moment of inertia $\log I_\star$ & $54.224 \pm 0.069$ & g\,cm$^2$ \\
\hline
\enddata
\tablecomments{1. Estimated based on models that possess an inner convection zone (core).}
\end{deluxetable}

The inferred evolutionary state of WASP-12 bears directly on the origin of its rapid orbital decay. \citet{Weinberg2017} showed that the observed dissipation rate can be explained by nonlinear breaking of tidally excited internal gravity waves if WASP-12 is a subgiant that has lost its central convective core. In this configuration, inward-propagating waves can reach the stellar center, where geometric focusing drives wave breaking. Within our {\it MESA}-based stellar interior model, a convective core, rather than a radiative core, is found in 99\% of posterior samples. However, unmodeled physics and systematic uncertainties in the {\it MESA} calculations could alter this conclusion and potentially favor a radiative core.

This does not imply that the orbital decay is incompatible with a main-sequence host; rather, under this interpretation, a different dissipation mechanism must operate. In main-sequence models, the convective core prevents gravity waves from reaching the stellar center, while linear radiative and convective damping appears insufficient to explain the measured decay rate without fine-tuning \citep{Bailey2019,Sun2025}. Mechanisms that operate in the presence of a convective core, such as absorption near an interior corotation layer \citep{Bailey2019} or magnetic conversion and subsequent dissipation of internal gravity waves \citep{Duguid2024}, therefore warrant particular attention. Our proposed angular-momentum-transfer scenario does not identify the mechanism responsible for the observed tidal dissipation; it requires only that a fraction of the lost orbital angular momentum is temporarily deposited in the surface layers.

\subsection{Magnetic Braking Models}
\label{sec:jen_model}

To estimate WASP-12's expected rotation rate at its current evolutionary stage, we used the \texttt{rotevol} tracer code \citep{Pinsonneault2026} on a YREC model grid \citep[described in][]{Metcalfe2020}, adopting the standard magnetic-braking law and \citet{Saunders2024} calibration based on \citet{Matt2008} and \citet{vanSaders2013}. The grid uses Sobol quasi-random sampling \citep{Sobol1967,Bellinger2016} over initial mass $M\in[0.9,1.6]M_\odot$, metallicity $[\mathrm{M/H}]\in[-0.1,0.5]$, mixing length $\alpha_{\rm MLT}\in[1.5,2.6]$, and initial helium fraction $Y_{\rm init}\in[0.22,0.32]$. We predict rotational evolution following \citet{Metcalfe2020,Metcalfe2025}\footnote{This grid has an initial solar helium abundance of $Y = 0.2716$, mixing length of $\alpha_{ml} = 1.92729$, and diffusion correction coefficient of 0.824.}, with three modifications: (1) initial rotation periods are drawn from a Gaussian centered on 8 days with a 2-day standard deviation, with disk locking released at 1 Myr; (2) mean density, but not rotation period, is included with luminosity, effective temperature, and surface metallicity as an observational constraint; and (3) systematic model errors are omitted. We obtain $v_{\rm eq}=14^{+11}_{-5}$~km s$^{-1}$ and $P_{\rm rot}=5.9^{+3.1}_{-2.5}$~d, with mass, mixing length, and age agreeing within $1\sigma$ with the MESA models above. The broad period posterior reflects weak magnetic braking from the shallow convective envelope and sensitivity to the assumed initial rotation. Nevertheless, the median equatorial velocities from both estimates are an order of magnitude above the measured $\vsini$, ruling out the observed value at $2.5\sigma$.

Turning the question around, WASP-12's thin convection zone means that we expect it to undergo very little magnetic braking on the main sequence. If we ask what initial rotation period would be required to match the observed slow surface rotation at the observed age and inferred stellar parameters, it would require a period of $\sim1000$ days for a 1 Myr disk-locking timescale and $\sim60$ days for a long, 20 Myr disk-locking timescale (an early end to disk locking results in a model with more initial angular momentum at fixed period). Rotation periods of this length have not been observed in young open clusters and associations \citep{Herbst2001,Somers2017}, although most existing efforts to measure such periods would have been insensitive to periods this long. The observed slow rotation of WASP-12 is therefore difficult to reproduce through standard rotational evolution alone and motivates the angular-momentum considerations discussed in Section~\ref{sec:discussion}.

For empirical comparison, we also use the gyrochronological model of \citet{Schlaufman2010}, whose mass--age--rotation relation follows Skumanich $P_{\rm rot}\propto t^{1/2}$ spin-down \citep{Skumanich} calibrated on young clusters. Assuming spin--orbit alignment, he found $v_{\rm eq}=\vsini=14.4\pm2.2~{\rm km\,s^{-1}}$ for WASP-12 (rotation period $\sim5.8$~d). Given the agreement between empirical and theoretical $\vsini$ estimates, we adopt fiducial values $v_{\rm eq}=14~{\rm km\,s^{-1}}$ and $P_{\rm rot}=6$ days for a WASP-12-like star hereafter.

\section{Discussion}
\label{sec:discussion}

\subsection{WASP-12 is aligned but rotates anomalously slowly}
\label{sec:slow_rotation}

Our KPF observations yield a low projected obliquity of $\lambda=-1.3\pm8.7^\circ$. However, the RM effect measures only the angle between the sky projections of the stellar spin and orbital angular-momentum vectors. The true three-dimensional obliquity is given by \citep{FabryckyWinn2009}:
\begin{equation}
    \cos\psi = \cos i_\star\cos i_b + \sin i_\star\sin i_b\cos\lambda,
    \label{eq:true_obliquity}
\end{equation}
where $i_\star$ and $i_b$ are stellar and orbital inclinations to the line of sight. Small $\lambda$ therefore does not guarantee small true obliquity $\psi$. Transiting planets have $i_b \approx 90^{\circ}$, but stellar inclination is generally unknown unless both $v_{\rm eq}$ and $\vsini$ are precisely measured \citep{Albrecht_polar}. Asteroseismology can also constrain inclination \citep{Huber2013,Ong2025}, but current photometry is unsuitable for WASP-12 (Section~\ref{sec:prospect}).

\begin{figure}[htbp]
    \centering
    \includegraphics[width=\columnwidth]{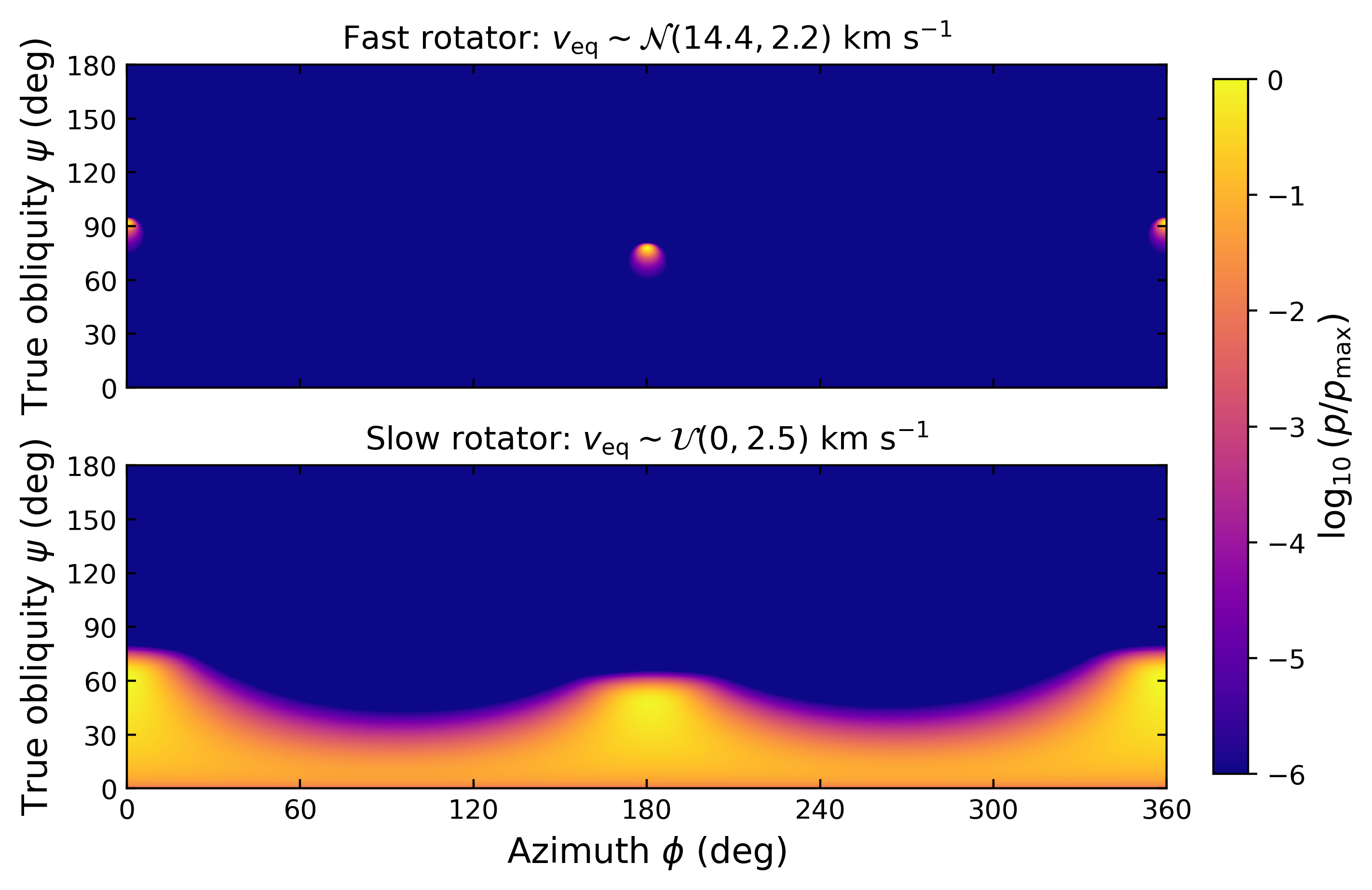}
    \par
    \noindent\makebox[\columnwidth][l]{%
    \includegraphics[width=0.9\columnwidth]{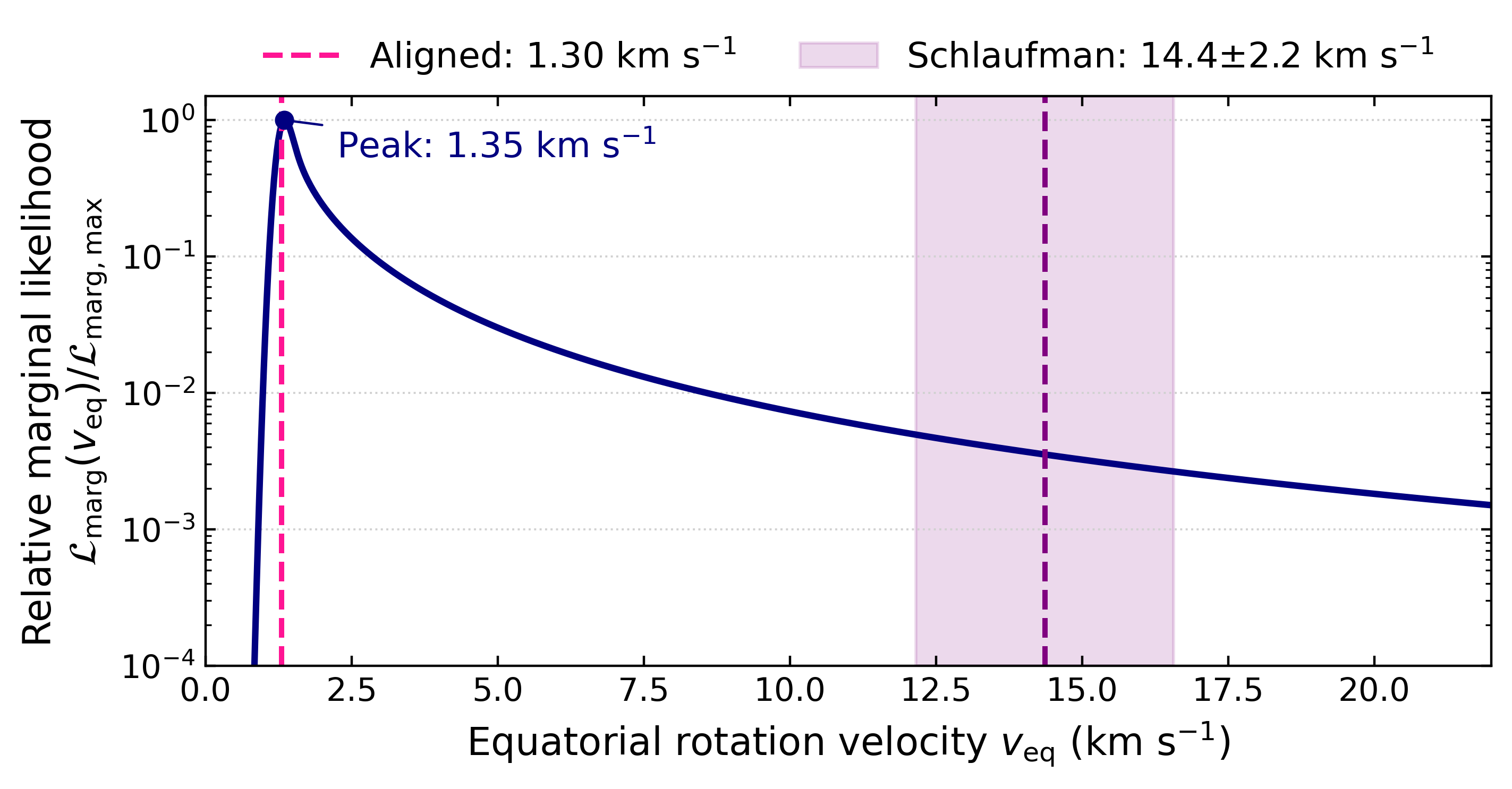}%
    }
    \par
    \includegraphics[width=\columnwidth]{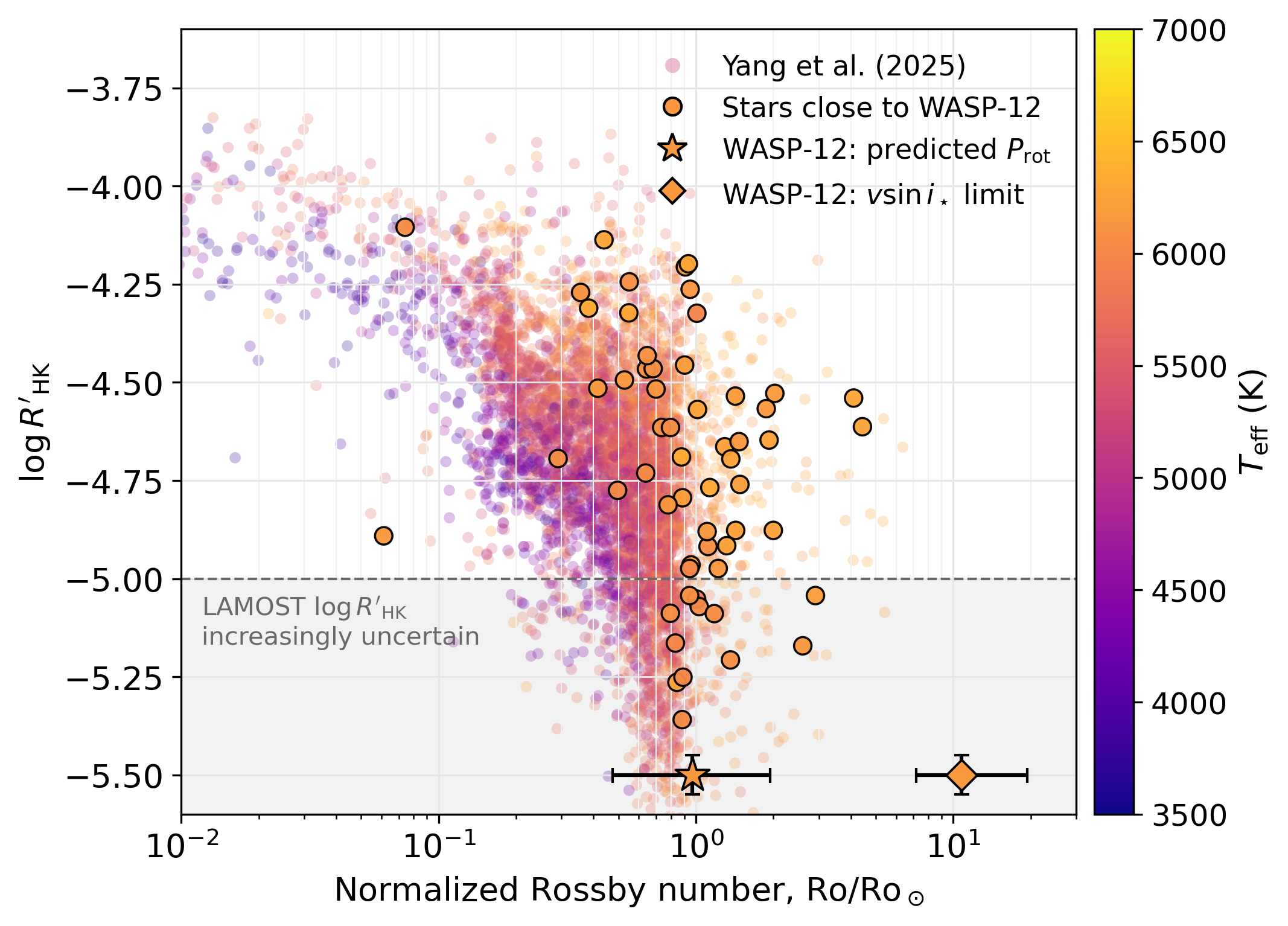}
    \caption{\textit{Top:} Joint posterior distributions of true obliquity, $\psi$, and azimuth, $\phi$, for the fast-rotation (upper) and slow-rotation (lower) models. Colors show $\log_{10}(p/p_{\rm max})$, normalized separately in each panel. \textit{Middle:} Orientation-marginalized likelihood of the equatorial rotation velocity, normalized to its maximum. The pink dashed line marks the aligned solution; the purple dashed line and shaded band show the prediction of \citet{Schlaufman2010} and its $1\sigma$ uncertainty. \textit{Bottom:} Chromospheric activity versus solar-normalized Rossby number for the LAMOST--Kepler sample of \citet{Yang2025}, colored by effective temperature. Black outlines identify stars with $|T_{\rm eff}-6155~{\rm K}|\leq200$~K and $|\log g-4.1428|\leq0.05$. The star and diamond mark WASP-12 assuming the predicted rotation period of $5.89$~d and the $\vsini$-based maximum period of $64$~d, respectively. The shaded region marks $\log R'_{\rm HK}<-5.0$, where the LAMOST activity measurements become increasingly uncertain.}
    \label{fig:veq_evidence}
    \label{fig:activity_rossby}
\end{figure}

The observed low $\lambda$ and low $\vsini$ of WASP-12 have two qualitatively different interpretations:

\begin{enumerate}
    \item WASP-12 could have a typical equatorial rotation velocity of $v_{\rm eq}\sim14~{\rm km\,s^{-1}}$ at its current evolutionary stage, but is viewed nearly pole-on, with $\sin i_\star\lesssim 0.1$, thereby producing the observed low $\vsini\simeq1.3~{\rm km\,s^{-1}}$. In this case, the true three-dimensional obliquity would be nearly polar. However, as shown shortly, the stellar-spin axis must occupy a restricted region of orientation space to reproduce both the low $\vsini$ and the measured projected obliquity $\lambda$ (Figure \ref{fig:veq_evidence}).

    \item WASP-12 could instead be intrinsically slowly rotating, with $v_{\rm eq}\simeq1.3~{\rm km\,s^{-1}}$. An approximately equator-on stellar inclination would then naturally reproduce the measured $\vsini$. Combined with the low projected obliquity, this interpretation favors a nearly aligned system.
\end{enumerate}

We compared the probabilities of these two interpretations by marginalizing over an isotropic distribution of stellar-spin orientations. With $D=\{ {\vsini}_{\rm obs}, \lambda_{\rm obs}\}$ denoting the measured quantities, we define the velocity residual as
$\Delta v={\vsini}_{\rm pred}(\psi,\phi,v_{\rm eq},i_b)-{\vsini}_{\rm obs}$
and the angular residual as
$\Delta\lambda=\lambda_{\rm pred}(\psi,\phi,i_b)-\lambda_{\rm obs}$,
wrapped onto $[-180^\circ,180^\circ)$. The corresponding chi square is
\begin{equation}
\chi^2(\psi,\phi;v_{\rm eq},i_b)=
\left(\frac{\Delta v}{\sigma_v}\right)^2 +
\left(\frac{\Delta\lambda}{\sigma_\lambda}\right)^2 .
\end{equation}

The orientation-marginalized likelihood is then
\begin{equation}
\begin{split}
\mathcal{L}_{\rm marg}(D\mid v_{\rm eq},i_b)
\propto{}&
\int_0^{2\pi}\int_0^\pi
\exp\left(-\frac{\chi^2}{2}\right)\\
&\times\frac{\sin\psi}{4\pi}\,d\psi\,d\phi .
\end{split}
\label{eq:orientation_marginal_likelihood}
\end{equation}

Here, $\psi$ is the three-dimensional angle between the stellar-spin and orbital angular-momentum vectors. The angle $\phi$ is the azimuth of the stellar-spin vector about the orbital angular-momentum axis, measured from the plane containing the orbital angular-momentum vector and the observer's line of sight. We fix the orbital inclination at $i_b=82.9^\circ$, whose uncertainty is negligible here. The predictions ${\vsini}_{\rm pred}$ and $\lambda_{\rm pred}$ depend on $(\psi,\phi,v_{\rm eq},i_b)$ and $(\psi,\phi,i_b)$, respectively. We approximate the integral on a regular $720\times720$ grid spanning $0^\circ\leq\psi\leq180^\circ$ and $0^\circ\leq\phi<360^\circ$. The factor $\sin\psi/4\pi$ is the normalized joint isotropic prior on $(\psi,\phi)$, and $\sigma_v$ and $\sigma_\lambda$ are the observational uncertainties in $\vsini$ and $\lambda$, respectively. The measured $\lambda$ constrains the joint $(\psi,\phi)$ posterior and thus remains important for inferring true obliquity.

As noted earlier, scenario~2, corresponding to the intrinsically slow-rotation interpretation, is strongly preferred. The top panels of Figure~\ref{fig:veq_evidence} illustrate the origin of this preference. Under the fast-rotation model, the solution is confined to a very limited region of the $\psi$--$\phi$ parameter space with nearly polar true obliquities and a narrow range of azimuths. The slow-rotation model, on the other hand, permits a much broader range of azimuths and shifts the posterior support toward lower true obliquities.

The middle panel of Figure~\ref{fig:veq_evidence} shows that the orientation-marginalized likelihood peaks at $v_{\rm eq}\simeq1.35~{\rm km\,s^{-1}}$. This value is close to the value implied by alignment, $v_{\rm eq,align}=\vsini/\sin i_b\simeq1.30~{\rm km\,s^{-1}}$. For a quantitative comparison of the two scenarios, we now include an ad hoc prior on the rotational velocity: a slow-rotation model with $v_{\rm eq}\mid H_{\rm slow}\sim\mathcal{U}(0,2.5~{\rm km\,s^{-1}})$ and a fast-rotation model based on the \citet{Schlaufman2010} prediction, $v_{\rm eq}\mid H_{\rm fast}\sim\mathcal{N}(14.4,2.2)~{\rm km\,s^{-1}}$. For each hypothesis $H$, we combined the marginalized likelihood with the corresponding $v_{\rm eq}$ prior:

\begin{equation}
\begin{aligned}
p(v_{\rm eq}\mid D,H)
&=
\frac{
\mathcal{L}_{\rm marg}(D\mid v_{\rm eq},i_b)
p(v_{\rm eq}\mid H)
}{
\mathcal{Z}_H
},\\
\mathcal{Z}_H
&=
\int
\mathcal{L}_{\rm marg}(D\mid v_{\rm eq},i_b)
p(v_{\rm eq}\mid H)\,dv_{\rm eq}.
\end{aligned}
\label{eq:veq_posterior}
\end{equation}

Here, $\mathcal{Z}_H$ is the Bayesian evidence for hypothesis $H$, and $B_{\rm slow,fast}=\mathcal{Z}_{\rm slow}/\mathcal{Z}_{\rm fast}$. We obtain $\ln B_{\rm slow,fast}=4.1$, corresponding to $B_{\rm slow,fast}=60.5$. Admittedly, the adopted $v_{\rm eq}$ prior is arbitrary, but our conclusion is qualitatively unchanged over the range of uniform priors tested below. As a sensitivity test, changing the upper bound of a uniform slow-rotation prior from 3 to $10~{\rm km\,s^{-1}}$ changes the Bayes factor from 55.3 to 21.2, all favoring scenario~2, or the slow-rotation interpretation. 

We therefore favor intrinsic slow rotation over a typical rapidly rotating F star viewed nearly pole-on. Under this interpretation, the small $\lambda$ is consistent with a low true obliquity. Setting $\sin i_\star=1$ in $P_{\rm rot}=2\pi R_\star\sin i_\star/\vsini$ gives a maximum rotation-period estimate of $64$~d for WASP-12.

WASP-12's low chromospheric activity provides additional context for evaluating the slow-rotation interpretation. The bottom panel of Figure~\ref{fig:activity_rossby} compares it with main-sequence LAMOST--Kepler stars \citep{Yang2025} in the activity--Rossby plane. For WASP-12, we calculate ${\rm Ro}=P_{\rm rot}/\tau_{\rm cz}$ with model convective-overturn time $\tau_{\rm cz}=2.6^{+1.3}_{-1.2}$~d (Section~\ref{sec:jen_model}), normalizing to the Sun with $\tau_{\rm cz,\odot}=10.9$~d and $P_{\rm rot,\odot}=25.4$~d. For comparison stars, we use global convective-overturn times $\tau_{\rm g}$ from \citet{Yang2025} and the corresponding solar normalization, $\tau_{\rm g,\odot}=34.9$~d. At the predicted $P_{\rm rot}=5.9$~d, WASP-12 has near-solar Rossby number but weaker observed emission than most stars at similar Rossby numbers: $\log R'_{\rm HK}=-5.50$ \citep{Knutson2010} and unusually weak Ca {\sc II} H\&K and Mg {\sc II} h\&k cores \citep{Haswell2012,Fossati2013}. The inferred maximum period of $64$~d instead places it at a much higher Rossby number, beyond this comparison sample. If intrinsic, the low chromospheric emission would be qualitatively consistent with slow rotation. However, the Ca {\sc II} H\&K and Mg {\sc II} h\&k emission cores are formed in the chromosphere and may be suppressed by absorption from circumstellar gas, possibly lost from WASP-12 b \citep{Haswell2012,Fossati2013}. We therefore do not use the activity measurement as an independent constraint on the stellar rotation or viewing geometry. We also searched for photometric variability of WASP-12 in the {\it TESS} data, and found no compelling rotational modulation. 

\subsection{Where did the orbital angular momentum go?}
\label{sec:tides}

WASP-12 b is aligned, at least with the surface convective layer, while its orbit is rapidly decaying and losing angular momentum. If the lost orbital angular momentum is transferred to the surface convective layer, the surface should have spun up, contrary to the observed slow rotation.

To quantify this discrepancy, we compare the angular-momentum reservoirs of the system. Assuming a circular orbit and $M_p\ll M_\star$, the planet's orbital angular momentum is
\begin{equation}
L_{\rm orb}\simeq M_p\sqrt{GM_\star a},
\end{equation}

We next estimate the whole-star rotational angular momentum. Using the stellar models described in Section~\ref{sec:stellar_para}, we obtain a total stellar moment of inertia of $I_{\rm tot}\simeq1.67\times10^{54}~{\rm g~cm^2}$ and a convective-envelope moment of inertia of $I_{\rm env}\simeq8.32\times10^{52}~{\rm g~cm^2}$. The outer convective envelope contributes $\sim5\%$ of the total stellar moment of inertia. We define $I_{\rm int}=I_{\rm tot}-I_{\rm env}$ as the moment of inertia of the stellar interior beneath the convective envelope, including both the radiative layer and central convective core. Table~\ref{tab:am_budget} summarizes the angular momenta and moments of inertia of the planet, surface convective layer, and stellar interior.

\begin{table*}[htbp]
\centering
\caption{
Angular-momentum reservoirs and stellar spin-reversal budget.
\textbf{Upper section:} Stellar reservoirs assume rigid rotation
with $P_{\rm rot}=64$~d (slow) or $5.9$~d (fast, corresponding to
$v_{\rm rot}=14.4~{\rm km\,s^{-1}}$). The rotation rate of the
interior below the convective envelope is observationally
unconstrained. Orbital angular momenta are calculated as
$L_{\rm orb}=M_p\sqrt{GM_\star a}$, with an effective moment of
inertia $I_{\rm orb}=M_pa^2$ such that
$L_{\rm orb}=I_{\rm orb}\Omega_{\rm orb}$ and
$\Omega_{\rm orb}=2\pi/P_{\rm orb}$. The present semimajor axis,
$a=0.02295$~AU, follows from Kepler's third law using
$M_\star=1.35\,M_\odot$ and $P_{\rm orb}=1.09$~d.
\textbf{Lower section:} Required angular momentum for evolution
from $P_{\rm rot,i}=-5.9$~d (retrograde) to
$P_{\rm rot,f}=+64$~d (prograde), calculated as
$\Delta J_{\rm rev}=I(\Omega_{\rm f}-\Omega_{\rm i})$.
The available orbital angular momentum is
$\Delta L_{\rm orb}=8.65\times10^{48}~{\rm g\,cm^2\,s^{-1}}$
for Scenario~1 and
$1.42\times10^{48}~{\rm g\,cm^2\,s^{-1}}$ for Scenario~2.
The ratio $\mathcal{R}_{\rm rev}=\Delta L_{\rm orb}/\Delta J_{\rm rev}$
indicates whether the angular-momentum budget is sufficient
($\mathcal{R}_{\rm rev}\geq1$).
}
\label{tab:am_budget}
\small
\renewcommand{\arraystretch}{1.15}
\begin{tabular}{lccc}
\hline
\multicolumn{4}{c}{\textbf{Angular-momentum reservoirs}} \\
\hline
Component
& Moment of inertia $I$
& Angular frequency $\Omega$
& Angular momentum $J=I\Omega$ \\
& (${\rm g\,cm^2}$)
& (${\rm s^{-1}}$)
& (${\rm g\,cm^2\,s^{-1}}$) \\
\hline
Whole star (slow rotator)
& $1.67\times10^{54}$ & $1.14\times10^{-6}$ & $1.90\times10^{48}$ \\
Convective envelope (slow rotator)
& $8.32\times10^{52}$ & $1.14\times10^{-6}$ & $9.45\times10^{46}$ \\
Whole star (fast rotator)
& $1.67\times10^{54}$ & $1.23\times10^{-5}$ & $2.06\times10^{49}$ \\
Convective envelope (fast rotator)
& $8.32\times10^{52}$ & $1.23\times10^{-5}$ & $1.02\times10^{48}$ \\
Stellar interior below envelope
& $1.59\times10^{54}$ & Unknown & Unknown \\
\hline
Planet orbit (current; $P_{\rm orb}=1.09$~d)
& $3.28\times10^{53}$ & $6.66\times10^{-5}$ & $2.18\times10^{49}$ \\
Planet orbit (Scenario~1; $P_{\rm orb}=2.97$~d)
& $1.25\times10^{54}$ & $2.45\times10^{-5}$ & $3.05\times10^{49}$ \\
Planet orbit (Scenario~2; $P_{\rm orb}=1.32$~d)
& $4.21\times10^{53}$ & $5.52\times10^{-5}$ & $2.33\times10^{49}$ \\
\hline\hline
\multicolumn{4}{c}{\textbf{Angular-momentum budget for stellar spin reversal}} \\
\hline
Stellar reservoir
& $\Delta J_{\rm rev}$
& Scenario~1
& Scenario~2 \\
& (${\rm g\,cm^2\,s^{-1}}$)
& Feasible? ($\mathcal{R}_{\rm rev}$)
& Feasible? ($\mathcal{R}_{\rm rev}$) \\
\hline
Convective envelope
& $1.12\times10^{48}$
& Yes ($7.7$)
& Marginally yes ($1.3$) \\
Whole star
& $2.25\times10^{49}$
& No ($0.38$)
& No ($0.06$) \\
\hline
\end{tabular}
\end{table*}

To estimate the orbital angular momentum available for stellar spin evolution, we consider two high-eccentricity migration histories for WASP-12, focusing on the orbital period just after circularization \citep[e.g.,][]{Rasio1996,Fabrycky2007,Nagasawa2008,Wu2011,Ford2006,Matsakos2016}. During circularization, tides raised primarily in the planet remove orbital energy while approximately conserving orbital angular momentum. In our simplified accounting, the angular-momentum transfer relevant to stellar spin therefore occurs during the subsequent orbital-decay stage. The post-circularization orbit is important because the difference between its angular momentum and the planet's present orbital angular momentum determines the amount available for transfer to the star. Such a migration history is plausible because WASP-12 is a hierarchical triple system: the planet-hosting primary, WASP-12 A, has a distant companion, WASP-12 BC, itself a close pair of M dwarfs. The BC pair lies at a projected separation of approximately \(1^{\prime\prime}\) from WASP-12 A, corresponding to roughly 400 AU \citep{Bergfors2012,Bechter2014}.

For a fluid planet, the Roche-limit orbital period scales approximately with the planetary mean density as \citep{Rappaport2013}
\begin{equation}
P_{\rm Roche}\simeq12.6~{\rm hr}\left(\frac{\rho_p}{1~{\rm g cm^{-3}}}\right)^{-1/2}.
\end{equation}
Because tides raised in the planet approximately conserve orbital angular momentum during circularization, the final circular semimajor axis is roughly twice the minimum pericenter distance:
\begin{equation}
a_{\rm final}\simeq2r_{\rm peri}.
\end{equation}
Traditionally, the minimum pericenter distance is assumed to be at least $2.7r_{\rm t} \equiv R_p \left( \frac{M_\star}{M_p} \right)^{1/3}$ to avoid tidal disruption \citep{Guillochon}. More recent work by \citet{Fan} revised the critical value to approximately $2r_{\rm t}$. We therefore adopt an intermediate value of 2.5$r_{\rm t}$ and assume:
\begin{equation}
a_{\rm final}\simeq5r_{\rm t}.
\end{equation}
Thus, the planet's mean density during circularization determines the orbital period at which it emerges from circularization. This post-circularization period---rather than the inflation mechanism itself---sets the orbital angular momentum subsequently available for transfer to the star. We consider two limiting possibilities:
\begin{enumerate}
\item \textbf{Scenario 1: Inflation by circularization.} The planet has already been inflated by eccentricity tides to nearly its current radius, $R_p\simeq1.89~R_{\rm Jup}$, during circularization. For $M_p\simeq1.47~M_{\rm Jup}$ \citep{ChakrabartySengupta2019}, this gives $\rho_{p,\rm infl}\simeq0.269~{\rm g~cm^{-3}}$ and $P_{\rm init}\simeq2.97$~d.
\item \textbf{Scenario 2: Inflation by irradiation.} We assume that eccentricity tidal heating is quickly shed through radiative cooling or mass loss \citep{Zanazzi}, so the planet is not inflated during circularization. We adopt a typical planetary radius of $R_p\simeq1.1~R_{\rm Jup}$ and $\rho_{p,\rm comp}\simeq1.37~{\rm g~cm^{-3}}$, giving $P_{\rm init}\simeq1.32$~d. In this case, WASP-12 b is inflated by strong irradiation or other processes after circularization.
\end{enumerate}

\begin{figure*}[htbp]
\centering
\includegraphics[width=0.65\linewidth]{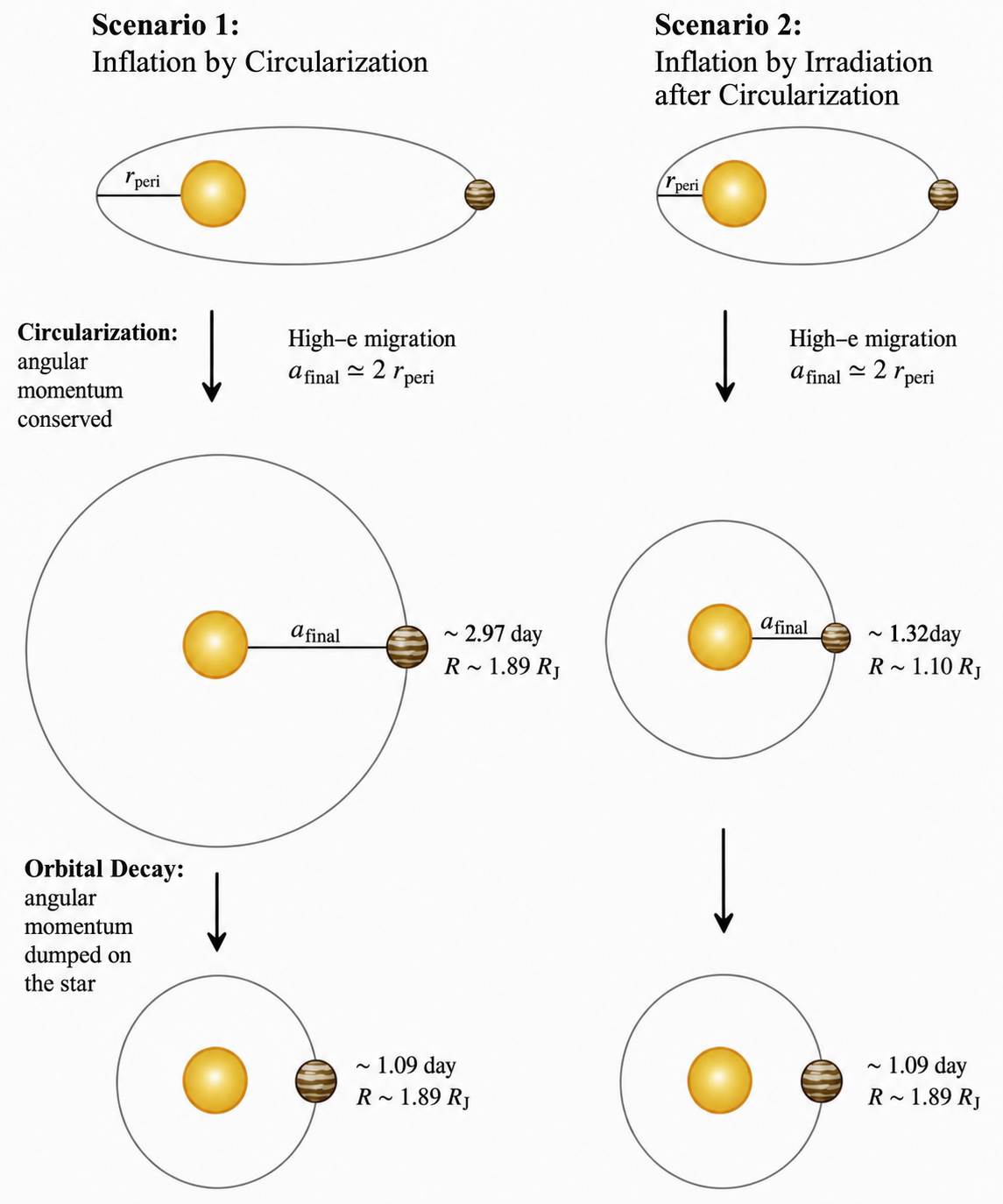}
\caption{Schematic illustration of the two post-circularization scenarios for WASP-12 b. In both scenarios, the planet undergoes high-eccentricity migration and reaches a highly eccentric orbit with pericenter distance $r_{\rm peri}$. Tidal circularization approximately conserves orbital angular momentum, producing a circular orbit with semimajor axis $a_{\rm final}\simeq2r_{\rm peri}$. \textbf{Left---Scenario 1: Inflation by circularization.} The planet has reached approximately its current radius ($R_p\simeq1.89~R_{\rm Jup}$) by circularization. Its low mean density places the Roche limit farther from the star, yielding a post-circularization period of approximately 2.97~d. \textbf{Right---Scenario 2: Inflation by irradiation after circularization.} The planet remains relatively compact ($R_p\simeq1.10~R_{\rm Jup}$) during circularization. Its higher density gives a smaller Roche limit and a post-circularization period of approximately 1.32~d. The planet subsequently inflates as tidal evolution brings it to the present-day orbital period of 1.09~d and radius of $R_p\simeq1.89~R_{\rm Jup}$. The two scenarios therefore provide substantially different amounts of orbital angular momentum for transfer during subsequent inspiral.}\label{fig:orbit_evolution}
\end{figure*}

Figure~\ref{fig:orbit_evolution} illustrates both scenarios. Subsequent tidal evolution brings the planet to its present $1.09$~d orbit, but the key difference is the total orbital angular momentum shed by WASP-12 b after circularization. In Scenario~1, the orbital angular momentum decreased from $3.05\times10^{49}$ to $2.18\times10^{49}~{\rm g~cm^2~s^{-1}}$, giving $\Delta L_{\rm orb}\simeq8.65\times10^{48}~{\rm g~cm^2~s^{-1}}$. In Scenario~2, it decreased only from $2.33\times10^{49}$ to $2.18\times10^{49}~{\rm g~cm^2~s^{-1}}$, giving $\Delta L_{\rm orb}\simeq1.42\times10^{48}~{\rm g~cm^2~s^{-1}}$. Thus, Scenario~1 provides approximately 6 times more angular momentum than Scenario~2 for modifying the stellar spin after circularization.

Adopting the empirical slow-rotation period of $P_{\rm rot}=64$~d, the present whole-star spin angular momentum is $J_\star=I_\star\Omega_\star\simeq1.90\times10^{48}~{\rm g~cm^2~s^{-1}}$, while the convective-envelope angular momentum is $J_{\rm env}\simeq9.45\times10^{46}~{\rm g~cm^2~s^{-1}}$. In both Scenarios~1 and 2, the available orbital angular momentum of WASP-12 b is comparable to that of the whole star and much larger than that in the surface convective layer. The surface rotation of WASP-12 could therefore have been significantly spun up by the decaying planet.

\subsubsection{Stellar spin-up and internal angular-momentum transport}
\label{sec:core_envelope}

We estimate the stellar response by adding the planet's lost orbital angular momentum to an initially prograde star with $P_{\rm rot,i}=5.9$~d, the YREC+\texttt{rotevol} prediction without tidal spin evolution. We calculate final periods for absorption by either the whole star or convective envelope, assuming complete transfer and fixed stellar moments of inertia.

The magnetic-braking timescale may be written as $\tau_{\rm mb}=J_\star/|\dot{J}_{\rm mb}|$ \citep{Kawaler1988}. Under the favored slow-rotation interpretation, however, WASP-12 lies in the weakened-magnetic-braking regime. In this regime, the wind torque is strongly suppressed and is set to zero in the step-function prescription of \citet{vanSaders2016}. Magnetic braking is therefore expected to be weak at the present surface rotation rate and throughout the modeled main-sequence evolution because of the star's thin surface convection zone. The post-circularization budgets below neglect wind losses.

If the transferred angular momentum is redistributed throughout the whole star, Scenario~1 changes the rotation period from $5.9$~d to $P_{\rm tot,f}\simeq4.2$~d, whereas Scenario~2 gives $P_{\rm tot,f}\simeq5.5$~d. Despite providing $\Delta L_{\rm orb}\simeq8.65\times10^{48}$ and $1.42\times10^{48}~{\rm g~cm^2~s^{-1}}$, respectively, both scenarios produce only modest whole-star spin-up because of the star's large moment of inertia.

The effect is more pronounced if transferred angular momentum remains in the thin convective envelope. Scenario~1 spins the envelope from $5.9$~d to $P_{\rm env,f}\simeq0.6$~d, while Scenario~2 gives $P_{\rm env,f}\simeq2.5$~d. Both are far shorter than the present $P_{\rm rot}\simeq64$~d inferred from $\vsini$. The observed slow photospheric rotation rules out long-term storage of transferred angular momentum solely in the convective envelope.

\begin{figure*}[htbp]
\centering
\includegraphics[width=0.92\linewidth]{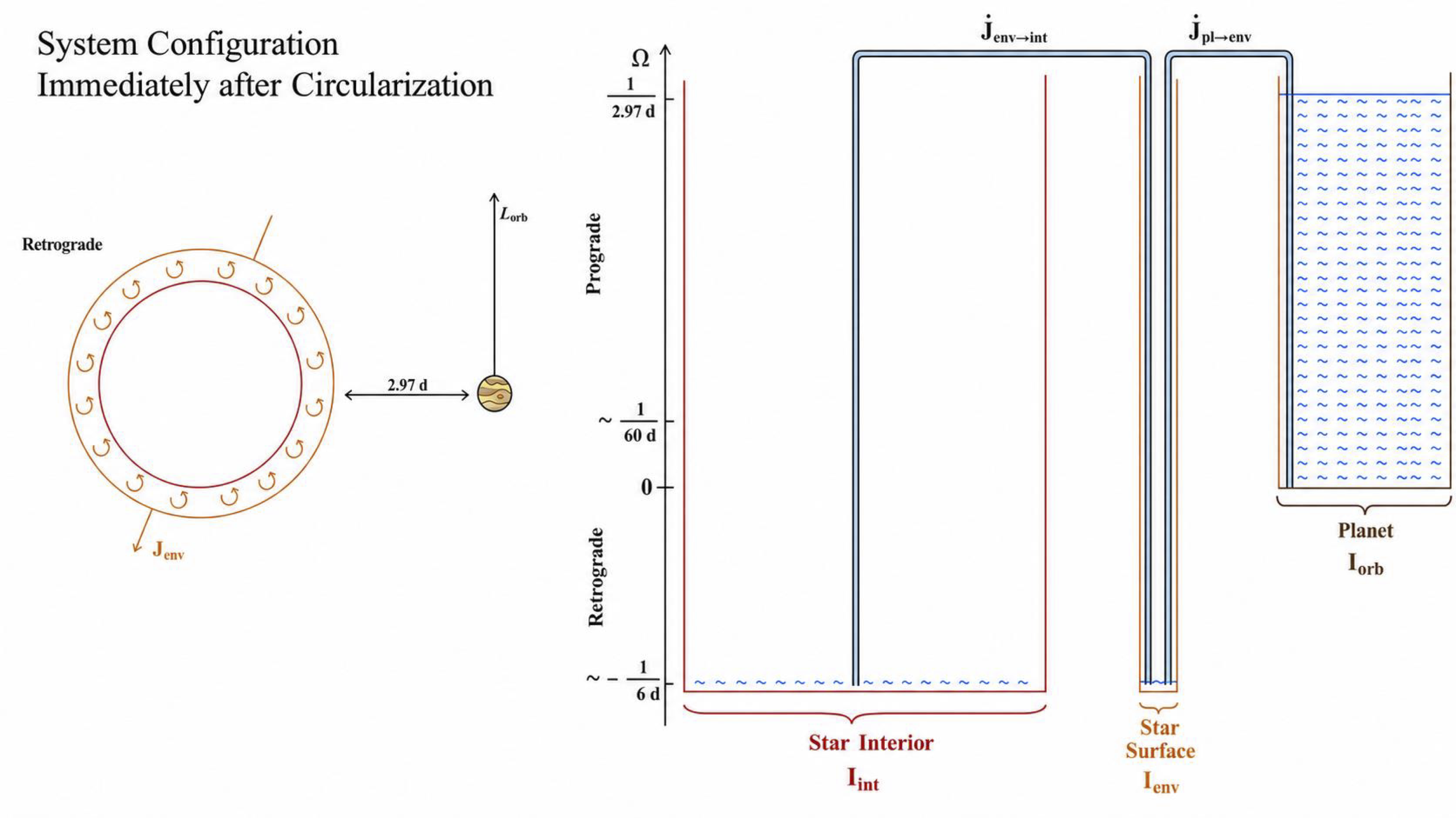}
\includegraphics[width=0.92\linewidth]{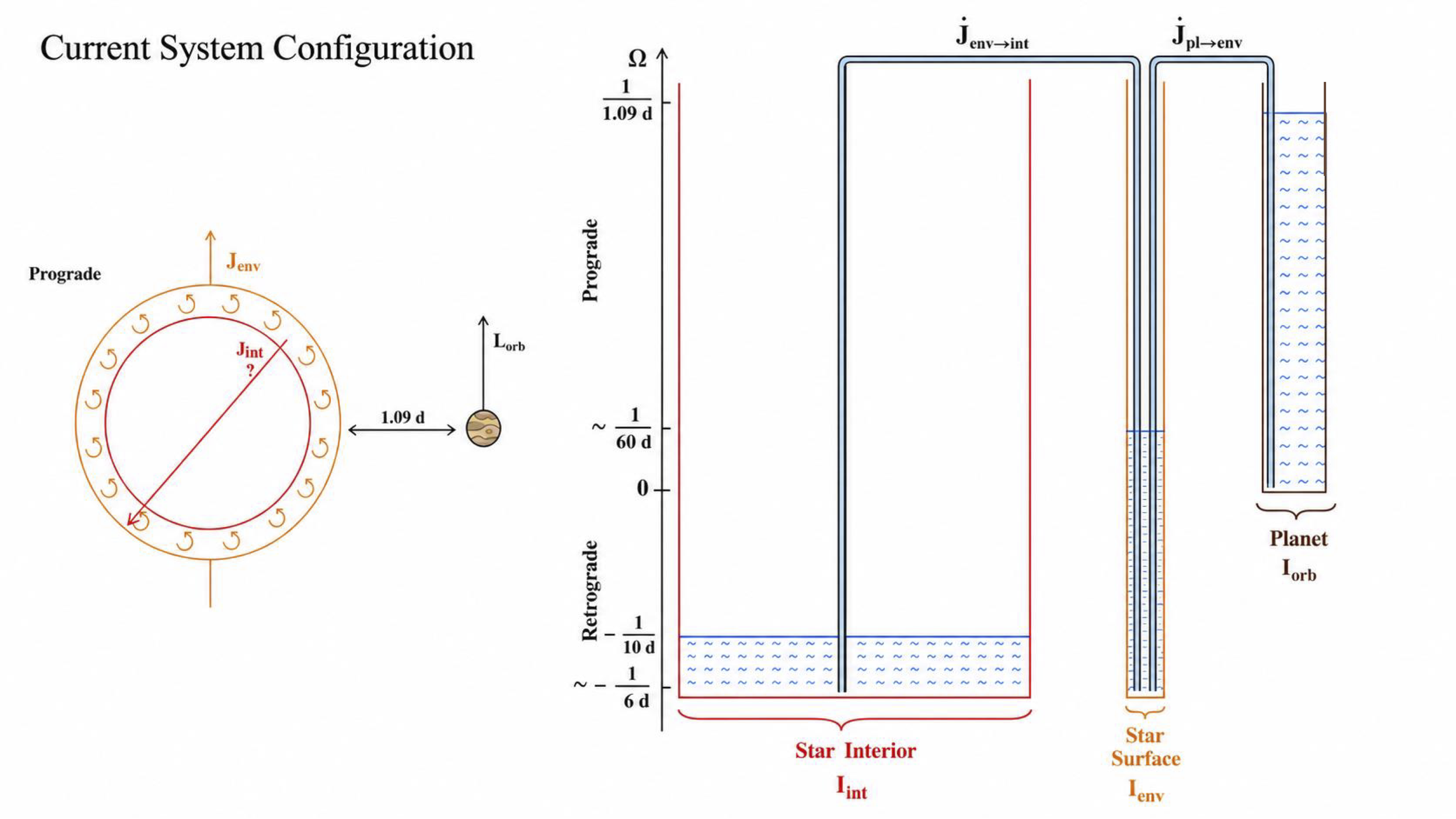}
\caption{Schematic illustration of angular-momentum transfer between WASP-12 b and its host star using a hydraulic analogy. The upper and lower panels show an illustrative initially retrograde configuration and the present-day prograde configuration, respectively. Reservoir widths represent effective moments of inertia, and fluid levels represent angular frequencies. Here, $\Omega>0$ and $\Omega<0$ denote prograde and retrograde rotation relative to the planetary orbit, respectively. The three reservoirs represent the planetary orbit ($L_{\rm orb}$), the stellar convective envelope ($J_{\rm env}$), and the stellar interior beneath the convective envelope ($J_{\rm int}$). The stellar interior and planetary orbit are much larger angular-momentum reservoirs than the convective envelope. Orbital angular momentum may be deposited into the convective envelope at a rate $\dot{J}_{\rm pl\rightarrow env}$ and subsequently redistributed into the stellar interior at a rate $\dot{J}_{\rm env\rightarrow int}$. Direct deposition into the stellar interior is also possible. In the retrograde-to-prograde scenario discussed in Section~\ref{sec:retrograde_pathway}, tidal transfer first cancels the initially retrograde stellar spin and then drives the star into the present prograde configuration while the planetary orbit loses angular momentum.}
\label{fig:am_transfer}
\end{figure*}

Figure~\ref{fig:am_transfer} summarizes angular-momentum exchange among the planetary orbit ($L_{\rm orb}$), convective envelope ($J_{\rm env}$), and stellar interior ($J_{\rm int}$). Tidal torques transfer orbital angular momentum to stellar spin, while tidal dissipation removes orbital energy. Inefficient internal transport rapidly spins up the low-inertia envelope; efficient transport redistributes angular momentum internally, allowing slow photospheric rotation. Alternatively, tidally excited internal gravity waves may transport angular momentum directly into the interior \citep{Weinberg2017,Bailey2019}. However, orbit--surface alignment suggests that some orbital angular momentum passed through the surface.

Whether the photosphere remains slowly rotating depends on how rapidly the envelope transfers angular momentum to the stellar interior. The relevant orbital angular-momentum-loss timescale is \(L_{\rm orb}/|\dot{L}_{\rm orb}| = 3P/|\dot{P}| \simeq 9.8\) Myr, broadly comparable to both the observed period-decay timescale and our illustrative core--envelope coupling timescale.

Independent evidence from stellar rotation, lithium evolution, and asteroseismology indicates that angular momentum can be exchanged efficiently between the stellar surface layers and the deeper stellar interior. The coupling generally becomes faster with increasing stellar mass \citep{Somers2016,Li2020,Cao2023}. A semi-empirical estimate follows from the two-zone rotational-evolution model of \citet{Spada2020}, which parameterizes the core--envelope coupling timescale as
\begin{equation}
\tau_{\rm ce}=\tau_{\rm ce,\odot}\left(\frac{M_\star}{M_\odot}\right)^{-\alpha},
\end{equation}
with $\tau_{\rm ce,\odot}\simeq22$~Myr and $\alpha\simeq5.6$. Extrapolating this relation to $M_\star=1.353~M_\odot$ gives
\begin{equation}
\tau_{\rm ce}\simeq22~{\rm Myr} (1.353)^{-5.6}\simeq4~{\rm Myr}.
\end{equation}

The estimated core--envelope coupling timescale is intriguingly similar to the observed period-decay timescale of $3.25\pm0.23$~Myr. Rapid envelope--interior angular-momentum transfer may therefore prevent angular momentum deposited near the surface from accumulating solely in the convective envelope. We caution that the \citet{Spada2020} relation is extrapolated beyond its calibrated mass range and was developed for magnetic-braking-dominated evolution. The effective coupling timescale may further depend on the shear or $\Delta\Omega$ between the envelope and interior. Stronger shear may further speed up angular-momentum transport. Our estimate of $\tau_{\rm ce}\simeq4$~Myr, which was calibrated on non-planet-host stars, is therefore likely an underestimate of the real coupling rate. Moreover, the core--envelope coupling timescale is also model-dependent \citep[see also][]{Gallet2013,Lanzafame2015}.

\subsection{An initially retrograde WASP-12 b}
\label{sec:retrograde_pathway}

We propose that WASP-12 b might have started on a retrograde orbit immediately following the circularization stage of its high-eccentricity migration. The envelope and interior of the host star were rotating rapidly with a period of approximately 6~days, typical for a star of its mass and evolutionary stage. The orbital angular momentum of the decaying planet quickly aligned the surface layer. However, due to rapid core--envelope angular-momentum coupling, the orbital angular momentum was eventually deposited in the deep interior of the star. The surface layer therefore currently has slow, prograde rotation. The total moment of inertia of the star is too large, and the planet probably has not yet flipped the angular-momentum axis of the whole star. Such an evolutionary pathway has also been suggested by \citet{Hallatt2026}, although Roche-lobe overflow is the mechanism for angular-momentum transfer in their analysis.

We quantitatively test an initially retrograde orbit for WASP-12 b. We adopt the YREC+\texttt{rotevol} prediction as the initial stellar rotation period $P_{\rm rot,i}=-5.9$~d, and the rotation inferred from the present $\vsini$ as the final state, $P_{\rm rot,f}=+64$~d. The signs denote retrograde and prograde rotation, respectively.

For a stellar reservoir with moment of inertia $I$, the angular momentum required to evolve between these two signed rotation states is
\begin{equation}
\Delta J_{\rm rev}
= I\left(\Omega_{\rm f}-\Omega_{\rm i}\right)
= 2\pi I\left(\frac{1}{P_{\rm rot,f}}
+\frac{1}{|P_{\rm rot,i}|}\right).
\end{equation}
This expression accounts both for canceling the initial retrograde spin and for producing the present prograde spin.

We define the available-to-required angular-momentum ratio as
\begin{equation}
\mathcal{R}_{\rm rev} = \frac{\Delta L_{\rm orb}}{\Delta J_{\rm rev}}.
\end{equation}
A value of $\mathcal{R}_{\rm rev}\geq1$ indicates that the orbital angular momentum lost after circularization is sufficient for the specified spin evolution. Table~\ref{tab:am_budget} (lower section) summarizes the results for the whole star and convective envelope.

Scenario~1 clearly distinguishes envelope from whole-star evolution. Its orbital-angular-momentum budget supplies only 38\% of that required to reverse the whole star, but $\sim7.7$ times that required to evolve the convective envelope from $-5.9$~d to $+64$~d. Scenario~1 can therefore reverse the envelope but not the whole star.

Scenario~2 supplies only 6\% of the whole-star reversal requirement but $1.3$ times the envelope requirement. It permits the signed envelope evolution in the maximum-transfer calculation, with much less margin than Scenario~1, but cannot reverse the whole star.

The signed final rotation periods show the same physical distinction. If each reservoir absorbs the full available orbital angular momentum starting from $P_{\rm rot,i}=-5.9$~d, Scenario~1 leaves the whole star and stellar interior retrograde at $-10.2$ and $-10.6$~d. Scenario~2 leaves the whole star and interior near their initial retrograde states, at $-6.3$ and $-6.4$~d, respectively. Thus, neither scenario can reverse the whole star, but both provide enough orbital angular momentum for the convective envelope to flip to prograde and reach the representative $+64$~d rotation state; Scenario~2 does so only marginally.

Our two-zone stellar interior model is of course only a toy model. We note that the strong shear between a rapidly rotating, retrograde interior and a slowly rotating, prograde envelope may be susceptible to various instabilities such as the Rayleigh criterion for centrifugal instability \citep[e.g.,][]{Barker2019} or the Goldreich--Schubert--Fricke instability \citep{GoldreichSchubert1967,Fricke1968}. We suggest that these instabilities may destroy a sharp interior--envelope boundary adopted in our toy model and instead set up a gentle transition zone with a smoother shear layer.

\subsection{Prospects for testing internal stellar rotation}\label{sec:prospect}

Our scenario raises a testable possibility: a retrograde WASP-12 interior whose strong surface--interior shear could rapidly transfer orbital angular momentum inward. Future asteroseismic rotational splittings could test differential interior--surface rotation \citep{Aerts2010,Beck2012,Deheuvels2012,Huber2013}. A related configuration may already be suggested for Kepler-56: mixed modes reveal evidence for distinct core and envelope rotation axes, with the envelope potentially closer to orbital alignment \citep{Ong2025}. Surface alignment therefore need not imply interior alignment. Although the \citet{Ong2025} mixed-mode method is not directly applicable to a late-main-sequence star, future constraints on WASP-12's interior and envelope rotation rates and orientations could test whether tidal angular-momentum transfer reversed the whole star or only its envelope. WASP-12 lies outside \textit{PLATO}'s first long-duration field \citep{Heller2026cPIC}; even with future coverage, its broad F-star mode linewidths would hinder rotation and inclination measurements unless rotational splittings were sufficiently large.

\section{Summary and Conclusions}
\label{sec:summary}

We present new KPF Rossiter--McLaughlin observations of WASP-12 b and interpret its spin--orbit architecture, orbital decay, and slow stellar rotation in a scenario in which the planet started with a retrograde orbit immediately after the circularization stage of its high-eccentricity migration. Our main conclusions are:

\begin{itemize}[topsep=0.25\baselineskip, itemsep=0.25\baselineskip, parsep=0pt, partopsep=0pt]

\item Joint modeling of two KPF RM datasets and \textit{TESS} photometry yields a low sky-projected obliquity of $\lambda=-1.3\pm8.7^\circ$. A reanalysis of the 2012 HIRES observations shows that the discrepancy is not primarily driven by the different stellar-rotation priors.

\item The measured $\vsini=1.3\pm0.1~{\rm km s^{-1}}$ is much smaller than the $\sim14~{\rm km s^{-1}}$ expected for a star of similar mass and evolutionary stage. Our Bayesian model comparison favors a slowly rotating WASP-12 with a low true obliquity $\psi$ rather than a rapidly rotating star with a high true obliquity and pathological $\sin i_\star$ values. Weak Ca and Mg chromospheric emissions of WASP-12 ($\log R^\prime_{\rm HK}=-5.5$) also favor a slowly rotating star.

\item Transit midpoints from both KPF RM observations agree with the established quadratic orbital-decay ephemeris, supporting continued rapid decay.

\item Stellar evolution modeling favors a late-main-sequence host with a central convective core, an overlying radiative layer, and a thin outer convective envelope containing only $\simeq5\%$ of the total stellar moment of inertia.

\item We calculated that the net orbital angular momentum change from the 2.97-day post-circularization orbit to the current orbit is $\Delta L_{\rm orb}\simeq8.65\times10^{48}~{\rm g~cm^2~s^{-1}}$. 

\item Our YREC+\texttt{rotevol} self-consistent stellar-evolution and magnetic-braking model predicts a rotation period of $\sim6$~d for WASP-12 in the absence of tidal spin evolution.

\item The estimated envelope--interior angular-momentum coupling timescale, $\tau_{\rm ce}\simeq4$~Myr, is similar to the observed tidal-decay timescale and may represent an upper limit because the large shear between the tidally aligned surface and initially retrograde interior could enhance angular-momentum transport. Orbital angular momentum can readily erase the surface--orbit misalignment and subsequently be redistributed into the stellar interior. This scenario could therefore reconcile the aligned orbit with the observed slow surface rotation.

\item For the 2.97-day post-circularization scenario, $\Delta L_{\rm orb}$ is 7.7 times the angular momentum required to reverse the surface convective layer from retrograde to the observed prograde rotation. However, $\Delta L_{\rm orb}$ is only 38\% of that required to reverse the whole star, and thus the stellar interior could still be rotating retrograde relative to the planet, provided that ongoing tidal forcing maintains differential rotation against envelope--interior coupling. This could be tested with future asteroseismic data.
\end{itemize}

\begin{acknowledgments}
This work made use of the Exoplanet Follow-up Observing Program for TESS (ExoFOP-TESS), available at \dataset[10.26134/ExoFOP3]{https://doi.org/10.26134/ExoFOP3}.

We thank Caleb Lammers, Tim Hallatt, Sarah Millholland, 
Jeremy Goodman, Avery Bailey, Joel Ong, and Meng Sun for thoughtful comments and helpful discussions.

R.L. is funded by the European Union (ERC, THIRSTEE, 101164189) and acknowledges financial support from the Severo Ochoa grant CEX2021-001131-S funded by MCIN/AEI/10.13039/501100011033. Views and opinions expressed are, however, those of the author(s) only and do not necessarily reflect those of the European Union or the European Research Council. Neither the European Union nor the granting authority can be held responsible for them.

This work was performed, in part, at the Jet Propulsion Laboratory, California Institute of Technology, sponsored by the United States Government under the Prime Contract 80NM0018D0004 between Caltech and NASA.
\end{acknowledgments}

\bibliography{citation}

@ARTICLE{Green2019,
       author = {{Green}, Gregory M. and {Schlafly}, Edward and {Zucker}, Catherine and {Speagle}, Joshua S. and {Finkbeiner}, Douglas},
        title = "{A 3D Dust Map Based on Gaia, Pan-STARRS 1, and 2MASS}",
      journal = {\apj},
         year = 2019,
        month = dec,
       volume = {887},
       number = {1},
          eid = {93},
        pages = {93},
          doi = {10.3847/1538-4357/ab5362},
archivePrefix = {arXiv},
       eprint = {1905.02734},
 primaryClass = {astro-ph.GA},
       adsurl = {https://ui.adsabs.harvard.edu/abs/2019ApJ...887...93G}
}

@ARTICLE{allesfitter-paper,
       author = {{G{\"u}nther}, Maximilian N. and {Daylan}, Tansu},
        title = "{Allesfitter: Flexible Star and Exoplanet Inference from Photometry and Radial Velocity}",
      journal = {\apjs},
         year = 2021,
        month = may,
       volume = {254},
       number = {1},
          eid = {13},
        pages = {13},
          doi = {10.3847/1538-4365/abe70e},
archivePrefix = {arXiv},
       eprint = {2003.14371},
 primaryClass = {astro-ph.EP},
       adsurl = {https://ui.adsabs.harvard.edu/abs/2021ApJS..254...13G}
}

@ARTICLE{Fan,
       author = {{Fan}, Qianli and {Liu}, Shang-Fei},
        title = "{Where Do Hot Jupiters Come from? Revisiting Tidal Disruption and Ejection in High-eccentricity Migration}",
      journal = {\apjl},
         year = 2026,
        month = may,
       volume = {1003},
       number = {1},
          eid = {L7},
        pages = {L7},
          doi = {10.3847/2041-8213/ae6659},
archivePrefix = {arXiv},
       eprint = {2605.14433},
 primaryClass = {astro-ph.EP},
       adsurl = {https://ui.adsabs.harvard.edu/abs/2026ApJ..1003L...7F}
}

@ARTICLE{Zanazzi,
       author = {{Zanazzi}, J.~J. and {MacLeod}, Morgan and {Bryan}, Marta L. and {Mahadevan}, Suvrath},
        title = "{Dynamical Tides during High-Eccentricity Migration produces the Hot Jupiter Pile-up, Neptune Ridge, and Neptune Desert}",
      journal = {arXiv e-prints},
         year = 2026,
        month = jun,
          eid = {arXiv:2606.20789},
        pages = {arXiv:2606.20789},
          doi = {10.48550/arXiv.2606.20789},
archivePrefix = {arXiv},
       eprint = {2606.20789},
 primaryClass = {astro-ph.EP},
       adsurl = {https://ui.adsabs.harvard.edu/abs/2026arXiv260620789Z}
}

@ARTICLE{Guillochon,
       author = {{Guillochon}, James and {Ramirez-Ruiz}, Enrico and {Lin}, Douglas},
        title = "{Consequences of the Ejection and Disruption of Giant Planets}",
      journal = {\apj},
         year = 2011,
        month = may,
       volume = {732},
       number = {2},
          eid = {74},
        pages = {74},
          doi = {10.1088/0004-637X/732/2/74},
archivePrefix = {arXiv},
       eprint = {1012.2382},
 primaryClass = {astro-ph.EP},
       adsurl = {https://ui.adsabs.harvard.edu/abs/2011ApJ...732...74G}
}

@ARTICLE{Huber2013,
       author = {{Huber}, Daniel and {Carter}, Joshua A. and {Barbieri}, Mauro and {Miglio}, Andrea and {Deck}, Katherine M. and {Fabrycky}, Daniel C. and {Montet}, Benjamin T. and {Buchhave}, Lars A. and {Chaplin}, William J. and {Hekker}, Saskia and {Montalb{\'a}n}, Josefina and {Sanchis-Ojeda}, Roberto and {Basu}, Sarbani and {Bedding}, Timothy R. and {Campante}, Tiago L. and {Christensen-Dalsgaard}, J{\o}rgen and {Elsworth}, Yvonne P. and {Stello}, Dennis and {Arentoft}, Torben and {Ford}, Eric B. and {Gilliland}, Ronald L. and {Handberg}, Rasmus and {Howard}, Andrew W. and {Isaacson}, Howard and {Johnson}, John Asher and {Karoff}, Christoffer and {Kawaler}, Steven D. and {Kjeldsen}, Hans and {Latham}, David W. and {Lund}, Mikkel N. and {Lundkvist}, Mia and {Marcy}, Geoffrey W. and {Metcalfe}, Travis S. and {Silva Aguirre}, Victor and {Winn}, Joshua N.},
        title = "{Stellar Spin-Orbit Misalignment in a Multiplanet System}",
      journal = {Science},
         year = 2013,
        month = oct,
       volume = {342},
       number = {6156},
        pages = {331-334},
          doi = {10.1126/science.1242066},
archivePrefix = {arXiv},
       eprint = {1310.4503},
 primaryClass = {astro-ph.EP},
       adsurl = {https://ui.adsabs.harvard.edu/abs/2013Sci...342..331H}
}

@ARTICLE{Albrecht_polar,
       author = {{Albrecht}, Simon H. and {Marcussen}, Marcus L. and {Winn}, Joshua N. and {Dawson}, Rebekah I. and {Knudstrup}, Emil},
        title = "{A Preponderance of Perpendicular Planets}",
      journal = {\apjl},
         year = 2021,
        month = jul,
       volume = {916},
       number = {1},
          eid = {L1},
        pages = {L1},
          doi = {10.3847/2041-8213/ac0f03},
archivePrefix = {arXiv},
       eprint = {2105.09327},
 primaryClass = {astro-ph.EP},
       adsurl = {https://ui.adsabs.harvard.edu/abs/2021ApJ...916L...1A}
}

@MISC{allesfitter-code,
 author = {{G{\"u}nther}, Maximilian~N. and {Daylan}, Tansu},
 title = "{Allesfitter: Flexible Star and Exoplanet Inference From Photometry and Radial Velocity}",
 howpublished = {Astrophysics Source Code Library},
 year = 2019,
 month = mar,
 archivePrefix = "ascl",
 eprint = {1903.003},
 adsurl = {http://adsabs.harvard.edu/abs/2019ascl.soft03003G}
}

@ARTICLE{Wang2024,
       author = {{Wang}, Xian-Yu and {Rice}, Malena and {Wang}, Songhu and {Kanodia}, Shubham and {Dai}, Fei and {Logsdon}, Sarah E. and {Schweiker}, Heidi and {Teske}, Johanna K. and {Butler}, R. Paul and {Crane}, Jeffrey D. and {Shectman}, Stephen and {Quinn}, Samuel N. and {Kostov}, Veselin and {Osborn}, Hugh P. and {Goeke}, Robert F. and {Eastman}, Jason D. and {Shporer}, Avi and {Rapetti}, David and {Collins}, Karen A. and {Watkins}, Cristilyn N. and {Relles}, Howard M. and {Ricker}, George R. and {Seager}, Sara and {Winn}, Joshua N. and {Jenkins}, Jon M.},
        title = "{Single-star Warm-Jupiter Systems Tend to Be Aligned, Even around Hot Stellar Hosts: No T $_{eff}${\textendash}{\ensuremath{\lambda}} Dependency}",
      journal = {\apjl},
         year = 2024,
        month = sep,
       volume = {973},
       number = {1},
          eid = {L21},
        pages = {L21},
          doi = {10.3847/2041-8213/ad7469},
archivePrefix = {arXiv},
       eprint = {2408.10038},
 primaryClass = {astro-ph.EP},
       adsurl = {https://ui.adsabs.harvard.edu/abs/2024ApJ...973L..21W}
}

@ARTICLE{Hirano2011,
       author = {{Hirano}, Teruyuki and {Suto}, Yasushi and {Winn}, Joshua N. and {Taruya}, Atsushi and {Narita}, Norio and {Albrecht}, Simon and {Sato}, Bun'ei},
        title = "{Improved Modeling of the Rossiter-McLaughlin Effect for Transiting Exoplanets}",
      journal = {\apj},
         year = 2011,
        month = dec,
       volume = {742},
       number = {2},
          eid = {69},
        pages = {69},
          doi = {10.1088/0004-637X/742/2/69},
archivePrefix = {arXiv},
       eprint = {1108.4430},
 primaryClass = {astro-ph.EP},
       adsurl = {https://ui.adsabs.harvard.edu/abs/2011ApJ...742...69H}
}

@ARTICLE{Ohta2005,
       author = {{Ohta}, Yasuhiro and {Taruya}, Atsushi and {Suto}, Yasushi},
        title = "{The Rossiter-McLaughlin Effect and Analytic Radial Velocity Curves for Transiting Extrasolar Planetary Systems}",
      journal = {\apj},
         year = 2005,
        month = apr,
       volume = {622},
       number = {2},
        pages = {1118-1135},
          doi = {10.1086/428344},
archivePrefix = {arXiv},
       eprint = {astro-ph/0410499},
 primaryClass = {astro-ph},
       adsurl = {https://ui.adsabs.harvard.edu/abs/2005ApJ...622.1118O}
}

@ARTICLE{Albrecht2012,
       author = {{Albrecht}, Simon and {Winn}, Joshua N. and {Johnson}, John A. and {Howard}, Andrew W. and {Marcy}, Geoffrey W. and {Butler}, R. Paul and {Arriagada}, Pamela and {Crane}, Jeffrey D. and {Shectman}, Stephen A. and {Thompson}, Ian B. and {Hirano}, Teruyuki and {Bakos}, Gaspar and {Hartman}, Joel D.},
        title = "{Obliquities of Hot Jupiter Host Stars: Evidence for Tidal Interactions and Primordial Misalignments}",
      journal = {\apj},
         year = 2012,
        month = sep,
       volume = {757},
       number = {1},
          eid = {18},
        pages = {18},
          doi = {10.1088/0004-637X/757/1/18},
archivePrefix = {arXiv},
       eprint = {1206.6105},
 primaryClass = {astro-ph.SR},
       adsurl = {https://ui.adsabs.harvard.edu/abs/2012ApJ...757...18A}
}

@ARTICLE{Weinberg2017,
       author = {{Weinberg}, Nevin N. and {Sun}, Meng and {Arras}, Phil and {Essick}, Reed},
        title = "{Tidal Dissipation in WASP-12}",
      journal = {\apjl},
         year = 2017,
        month = nov,
       volume = {849},
       number = {1},
          eid = {L11},
        pages = {L11},
          doi = {10.3847/2041-8213/aa9113},
archivePrefix = {arXiv},
       eprint = {1710.00858},
 primaryClass = {astro-ph.EP},
       adsurl = {https://ui.adsabs.harvard.edu/abs/2017ApJ...849L..11W}
}

@ARTICLE{Maciejewski2016,
       author = {{Maciejewski}, G. and {Dimitrov}, D. and {Fern{\'a}ndez}, M. and {Sota}, A. and {Nowak}, G. and {Ohlert}, J. and {Nikolov}, G. and {Bukowiecki}, {\L}. and {Hinse}, T.~C. and {Pall{\'e}}, E. and {Tingley}, B. and {Kjurkchieva}, D. and {Lee}, J.~W. and {Lee}, C.-U.},
        title = "{Departure from the constant-period ephemeris for the transiting exoplanet WASP-12}",
      journal = {\aap},
         year = 2016,
        month = apr,
       volume = {588},
          eid = {L6},
        pages = {L6},
          doi = {10.1051/0004-6361/201628312},
archivePrefix = {arXiv},
       eprint = {1602.09055},
 primaryClass = {astro-ph.EP},
       adsurl = {https://ui.adsabs.harvard.edu/abs/2016A&A...588L...6M}
}

@ARTICLE{Patra2017,
       author = {{Patra}, Kishore C. and {Winn}, Joshua N. and {Holman}, Matthew J. and {Yu}, Liang and {Deming}, Drake and {Dai}, Fei},
        title = "{The Apparently Decaying Orbit of WASP-12b}",
      journal = {\aj},
         year = 2017,
        month = jul,
       volume = {154},
       number = {1},
          eid = {4},
        pages = {4},
          doi = {10.3847/1538-3881/aa6d75},
archivePrefix = {arXiv},
       eprint = {1703.06582},
 primaryClass = {astro-ph.EP},
       adsurl = {https://ui.adsabs.harvard.edu/abs/2017AJ....154....4P}
}

@ARTICLE{Yee2020,
       author = {{Yee}, Samuel W. and {Winn}, Joshua N. and {Knutson}, Heather A. and {Patra}, Kishore C. and {Vissapragada}, Shreyas and {Zhang}, Michael M. and {Holman}, Matthew J. and {Shporer}, Avi and {Wright}, Jason T.},
        title = "{The Orbit of WASP-12b Is Decaying}",
      journal = {\apjl},
         year = 2020,
        month = jan,
       volume = {888},
       number = {1},
          eid = {L5},
        pages = {L5},
          doi = {10.3847/2041-8213/ab5c16},
archivePrefix = {arXiv},
       eprint = {1911.09131},
 primaryClass = {astro-ph.EP},
       adsurl = {https://ui.adsabs.harvard.edu/abs/2020ApJ...888L...5Y}
}

@ARTICLE{Turner2021,
       author = {{Turner}, Jake D. and {Ridden-Harper}, Andrew and {Jayawardhana}, Ray},
        title = "{Decaying Orbit of the Hot Jupiter WASP-12b: Confirmation with TESS Observations}",
      journal = {\aj},
         year = 2021,
        month = feb,
       volume = {161},
       number = {2},
          eid = {72},
        pages = {72},
          doi = {10.3847/1538-3881/abd178},
archivePrefix = {arXiv},
       eprint = {2012.02211},
 primaryClass = {astro-ph.EP},
       adsurl = {https://ui.adsabs.harvard.edu/abs/2021AJ....161...72T}
}

@INPROCEEDINGS{Jenkins2016,
       author = {{Jenkins}, Jon M. and {Twicken}, Joseph D. and {McCauliff}, Sean and {Campbell}, Jennifer and {Sanderfer}, Dwight and {Lung}, David and {Mansouri-Samani}, Masoud and {Girouard}, Forrest and {Tenenbaum}, Peter and {Klaus}, Todd and {Smith}, Jeffrey C. and {Caldwell}, Douglas A. and {Chacon}, A.~D. and {Henze}, Christopher and {Heiges}, Cory and {Latham}, David W. and {Morgan}, Edward and {Swade}, Daryl and {Rinehart}, Stephen and {Vanderspek}, Roland},
        title = "{The TESS science processing operations center}",
    booktitle = {Software and Cyberinfrastructure for Astronomy IV},
         year = 2016,
       editor = {{Chiozzi}, Gianluca and {Guzman}, Juan C.},
       series = {Society of Photo-Optical Instrumentation Engineers (SPIE) Conference Series},
       volume = {9913},
        month = aug,
          eid = {99133E},
        pages = {99133E},
          doi = {10.1117/12.2233418},
       adsurl = {https://ui.adsabs.harvard.edu/abs/2016SPIE.9913E..3EJ}
}

@article{Stumpe2014,
doi = {10.1086/674989},
url = {https://dx.doi.org/10.1086/674989},
year = {2014},
month = {jan},
publisher = {University of Chicago Press},
volume = {126},
number = {935},
pages = {100},
author = {Martin C. Stumpe and Jeffrey C. Smith and Joseph H. Catanzarite and Jeffrey E. Van Cleve and Jon M. Jenkins and Joseph D. Twicken and Forrest R. Girouard},
title = {Multiscale Systematic Error Correction via Wavelet-Based Bandsplitting in Kepler Data},
journal = {Publications of the Astronomical Society of the Pacific},
}

@misc{lightkurve2018,
       author = {{Lightkurve Collaboration} and {Cardoso}, Jos{\'e} Vin{\'\i}cius de Miranda and {Hedges}, Christina and {Gully-Santiago}, Michael and {Saunders}, Nicholas and {Cody}, Ann Marie and {Barclay}, Thomas and {Hall}, Oliver and {Sagear}, Sheila and {Turtelboom}, Emma and {Zhang}, Johnny and {Tzanidakis}, Andy and {Mighell}, Ken and {Coughlin}, Jeff and {Bell}, Keaton and {Berta-Thompson}, Zach and {Williams}, Peter and {Dotson}, Jessie and {Barentsen}, Geert},
        title = "{Lightkurve: Kepler and TESS time series analysis in Python}",
 howpublished = {Astrophysics Source Code Library, record ascl:1812.013},
         year = 2018,
        month = dec,
          eid = {ascl:1812.013},
       adsurl = {https://ui.adsabs.harvard.edu/abs/2018ascl.soft12013L}
}

@article{Smith2012,
doi = {10.1086/667697},
url = {https://dx.doi.org/10.1086/667697},
year = {2012},
month = {sep},
publisher = {University of Chicago Press},
volume = {124},
number = {919},
pages = {1000},
author = {Jeffrey C. Smith and Martin C. Stumpe and Jeffrey E. Van Cleve and Jon M. Jenkins and Thomas S. Barclay and Michael N. Fanelli and Forrest R. Girouard and Jeffery J. Kolodziejczak and Sean D. McCauliff and Robert L. Morris and Joseph D. Twicken},
title = {Kepler Presearch Data Conditioning II - A Bayesian Approach to Systematic Error Correction},
journal = {Publications of the Astronomical Society of the Pacific},
}

@ARTICLE{Savitzky1964,
       author = {{Savitzky}, A. and {Golay}, M.~J.~E.},
        title = "{Smoothing and differentiation of data by simplified least squares procedures}",
      journal = {Analytical Chemistry},
         year = 1964,
        month = jan,
       volume = {36},
        pages = {1627-1639},
          doi = {10.1021/ac60214a047},
       adsurl = {https://ui.adsabs.harvard.edu/abs/1964AnaCh..36.1627S}
}

@INPROCEEDINGS{Gibson2020,
       author = {{Gibson}, Steven R. and {Howard}, Andrew W. and {Rider}, Kodi and {Roy}, Arpita and {Edelstein}, Jerry and {Kassis}, Marc and {Grillo}, Jason and {Halverson}, Sam and {Sirk}, Martin M. and {Smith}, Christopher and {Allen}, Steve and {Baker}, Ashley and {Beichman}, Charles and {Berriman}, Bruce and {Brown}, Thomas and {Casey}, Kelleen and {Chin}, Jason and {Coutts}, David and {Cowley}, David and {Deich}, William and {Feger}, Tobias and {Fulton}, Benjamin and {Gers}, Luke and {Gurevich}, Yulia and {Ishikawa}, Yuzo and {James}, Ean and {Jelinsky}, Sharon and {Kaye}, Stephen and {Lanclos}, Kyle and {Li}, Adela and {Lilley}, Scott and {McCarney}, Ben and {Miller}, Tim and {Milner}, Steve and {O'Hanlon}, Timothy J. and {Pember}, Jacob and {Raffanti}, Mike and {Rockosi}, Constance and {Rubenzahl}, Ryan and {Rumph}, Dave and {Sandford}, Dale and {Savage}, Maureen and {Schwab}, Christian and {Seifahrt}, Andreas and {Shaum}, Abby and {Smith}, Roger and {Stuermer}, Julian and {Thorne}, Jim and {Vandenberg}, Adam and {Von Boeckmann}, Tod and {Wang}, Cindy and {Wang}, Qifan and {Weisfeiler}, Marie and {Wilcox}, Mavourneen and {Wishnow}, Edward H. and {Wizinowich}, Peter and {Wold}, Truman and {Wolfenberger}, Anna},
        title = "{Keck Planet Finder: design updates}",
    booktitle = {Ground-based and Airborne Instrumentation for Astronomy VIII},
         year = 2020,
       editor = {{Evans}, Christopher J. and {Bryant}, Julia J. and {Motohara}, Kentaro},
       series = {Society of Photo-Optical Instrumentation Engineers (SPIE) Conference Series},
       volume = {11447},
        month = dec,
          eid = {1144742},
        pages = {1144742},
          doi = {10.1117/12.2561783},
       adsurl = {https://ui.adsabs.harvard.edu/abs/2020SPIE11447E..42G}
}

@INPROCEEDINGS{Gibson2018,
       author = {{Gibson}, Steven R. and {Howard}, Andrew W. and {Roy}, Arpita and {Smith}, Christopher and {Halverson}, Sam and {Edelstein}, Jerry and {Kassis}, Marc and {Wishnow}, Edward H. and {Raffanti}, Michael and {Allen}, Steve and {Chin}, Jason and {Coutts}, David and {Cowley}, David and {Curtis}, Jim and {Deich}, William and {Feger}, Tobias and {Finstad}, Daniel and {Gurevich}, Yulia and {Ishikawa}, Yuzo and {James}, Ean and {Jhoti}, Elisha and {Lanclos}, Kyle and {Lilley}, Scott and {Miller}, Tim and {Milner}, Steve and {Payne}, Tom and {Rider}, Kodi and {Rockosi}, Constance and {Sandford}, Dale and {Schwab}, Christian and {Seifahrt}, Andreas and {Sirk}, Martin M. and {Smith}, Roger and {Stuermer}, Julian and {Weisfeiler}, Marie and {Wilcox}, Mavourneen and {Vandenberg}, Adam and {Wizinowich}, Peter},
        title = "{Keck Planet Finder: preliminary design}",
    booktitle = {Ground-based and Airborne Instrumentation for Astronomy VII},
         year = 2018,
       editor = {{Evans}, Christopher J. and {Simard}, Luc and {Takami}, Hideki},
       series = {Society of Photo-Optical Instrumentation Engineers (SPIE) Conference Series},
       volume = {10702},
        month = jul,
          eid = {107025X},
        pages = {107025X},
          doi = {10.1117/12.2311565},
       adsurl = {https://ui.adsabs.harvard.edu/abs/2018SPIE10702E..5XG}
}

@INPROCEEDINGS{Gibson2016,
       author = {{Gibson}, Steven R. and {Howard}, Andrew W. and {Marcy}, Geoffrey W. and {Edelstein}, Jerry and {Wishnow}, Edward H. and {Poppett}, Claire L.},
        title = "{KPF: Keck Planet Finder}",
    booktitle = {Ground-based and Airborne Instrumentation for Astronomy VI},
         year = 2016,
       editor = {{Evans}, Christopher J. and {Simard}, Luc and {Takami}, Hideki},
       series = {Society of Photo-Optical Instrumentation Engineers (SPIE) Conference Series},
       volume = {9908},
        month = aug,
          eid = {990870},
        pages = {990870},
          doi = {10.1117/12.2233334},
       adsurl = {https://ui.adsabs.harvard.edu/abs/2016SPIE.9908E..70G}
}

@ARTICLE{Hirano2010,
       author = {{Hirano}, Teruyuki and {Suto}, Yasushi and {Taruya}, Atsushi and {Narita}, Norio and {Sato}, Bun'ei and {Johnson}, John Asher and {Winn}, Joshua N.},
        title = "{Analytic Description of the Rossiter-Mclaughlin Effect for Transiting Exoplanets: Cross-Correlation Method and Comparison with Simulated Data}",
      journal = {\apj},
         year = 2010,
        month = jan,
       volume = {709},
       number = {1},
        pages = {458-469},
          doi = {10.1088/0004-637X/709/1/458},
archivePrefix = {arXiv},
       eprint = {0910.2365},
 primaryClass = {astro-ph.EP},
       adsurl = {https://ui.adsabs.harvard.edu/abs/2010ApJ...709..458H}
}

@ARTICLE{Pepe2013,
       author = {{Pepe}, F. and {Cristiani}, S. and {Rebolo}, R. and {Santos}, N.~C. and {Dekker}, H. and {M{\'e}gevand}, D. and {Zerbi}, F.~M. and {Cabral}, A. and {Molaro}, P. and {Di Marcantonio}, P. and {Abreu}, M. and {Affolter}, M. and {Aliverti}, M. and {Allende Prieto}, C. and {Amate}, M. and {Avila}, G. and {Baldini}, V. and {Bristow}, P. and {Broeg}, C. and {Cirami}, R. and {Coelho}, J. and {Conconi}, P. and {Coretti}, I. and {Cupani}, G. and {D'Odorico}, V. and {De Caprio}, V. and {Delabre}, B. and {Dorn}, R. and {Figueira}, P. and {Fragoso}, A. and {Galeotta}, S. and {Genolet}, L. and {Gomes}, R. and {Gonz{\'a}lez Hern{\'a}ndez}, J.~I. and {Hughes}, I. and {Iwert}, O. and {Kerber}, F. and {Landoni}, M. and {Lizon}, J.-L. and {Lovis}, C. and {Maire}, C. and {Mannetta}, M. and {Martins}, C. and {Monteiro}, M.~A. and {Oliveira}, A. and {Poretti}, E. and {Rasilla}, J.~L. and {Riva}, M. and {Santana Tschudi}, S. and {Santos}, P. and {Sosnowska}, D. and {Sousa}, S. and {Span{\`o}}, P. and {Tenegi}, F. and {Toso}, G. and {Vanzella}, E. and {Viel}, M. and {Zapatero Osorio}, M.~R.},
        title = "{ESPRESSO {\textemdash} An Echelle SPectrograph for Rocky Exoplanets Search and Stable Spectroscopic Observations}",
      journal = {The Messenger},
         year = 2013,
        month = sep,
       volume = {153},
        pages = {6-16},
       adsurl = {https://ui.adsabs.harvard.edu/abs/2013Msngr.153....6P}
}

@ARTICLE{Bailey2019,
       author = {{Bailey}, Avery and {Goodman}, Jeremy},
        title = "{Understanding WASP-12b}",
      journal = {\mnras},
         year = 2019,
        month = jan,
       volume = {482},
       number = {2},
        pages = {1872-1882},
          doi = {10.1093/mnras/sty2805},
archivePrefix = {arXiv},
       eprint = {1808.00052},
 primaryClass = {astro-ph.EP},
       adsurl = {https://ui.adsabs.harvard.edu/abs/2019MNRAS.482.1872B}
}

@ARTICLE{Haswell2012,
       author = {{Haswell}, C.~A. and {Fossati}, L. and {Ayres}, T. and {France}, K. and {Froning}, C.~S. and {Holmes}, S. and {Kolb}, U.~C. and {Busuttil}, R. and {Street}, R.~A. and {Hebb}, L. and {Collier Cameron}, A. and {Enoch}, B. and {Burwitz}, V. and {Rodriguez}, J. and {West}, R.~G. and {Pollacco}, D. and {Wheatley}, P.~J. and {Carter}, A.},
        title = "{Near-ultraviolet Absorption, Chromospheric Activity, and Star-Planet Interactions in the WASP-12 system}",
      journal = {\apj},
         year = 2012,
        month = nov,
       volume = {760},
       number = {1},
          eid = {79},
        pages = {79},
          doi = {10.1088/0004-637X/760/1/79},
archivePrefix = {arXiv},
       eprint = {1301.1860},
 primaryClass = {astro-ph.EP},
       adsurl = {https://ui.adsabs.harvard.edu/abs/2012ApJ...760...79H}
}

@ARTICLE{Fossati2010,
       author = {{Fossati}, L. and {Haswell}, C.~A. and {Froning}, C.~S. and {Hebb}, L. and {Holmes}, S. and {Kolb}, U. and {Helling}, Ch. and {Carter}, A. and {Wheatley}, P. and {Collier Cameron}, A. and {Loeillet}, B. and {Pollacco}, D. and {Street}, R. and {Stempels}, H.~C. and {Simpson}, E. and {Udry}, S. and {Joshi}, Y.~C. and {West}, R.~G. and {Skillen}, I. and {Wilson}, D.},
        title = "{Metals in the Exosphere of the Highly Irradiated Planet WASP-12b}",
      journal = {\apjl},
         year = 2010,
        month = may,
       volume = {714},
       number = {2},
        pages = {L222-L227},
          doi = {10.1088/2041-8205/714/2/L222},
archivePrefix = {arXiv},
       eprint = {1005.3656},
 primaryClass = {astro-ph.SR},
       adsurl = {https://ui.adsabs.harvard.edu/abs/2010ApJ...714L.222F}
}

@ARTICLE{Ford2006,
       author = {{Ford}, Eric B. and {Rasio}, Frederic A.},
        title = "{On the Relation between Hot Jupiters and the Roche Limit}",
      journal = {\apjl},
         year = 2006,
        month = feb,
       volume = {638},
       number = {1},
        pages = {L45-L48},
          doi = {10.1086/500734},
archivePrefix = {arXiv},
       eprint = {astro-ph/0512632},
 primaryClass = {astro-ph},
       adsurl = {https://ui.adsabs.harvard.edu/abs/2006ApJ...638L..45F}
}

@ARTICLE{Matsakos2016,
       author = {{Matsakos}, Titos and {K{\"o}nigl}, Arieh},
        title = "{On the Origin of the Sub-Jovian Desert in the Orbital-period-Planetary-mass Plane}",
      journal = {\apjl},
         year = 2016,
        month = mar,
       volume = {820},
       number = {1},
          eid = {L8},
        pages = {L8},
          doi = {10.3847/2041-8205/820/1/L8},
archivePrefix = {arXiv},
       eprint = {1603.00414},
 primaryClass = {astro-ph.EP},
       adsurl = {https://ui.adsabs.harvard.edu/abs/2016ApJ...820L...8M}
}

@ARTICLE{Rappaport2013,
       author = {{Rappaport}, Saul and {Sanchis-Ojeda}, Roberto and {Rogers}, Leslie A. and {Levine}, Alan and {Winn}, Joshua N.},
        title = "{The Roche Limit for Close-orbiting Planets: Minimum Density, Composition Constraints, and Application to the 4.2 hr Planet KOI 1843.03}",
      journal = {\apjl},
         year = 2013,
        month = aug,
       volume = {773},
       number = {1},
          eid = {L15},
        pages = {L15},
          doi = {10.1088/2041-8205/773/1/L15},
archivePrefix = {arXiv},
       eprint = {1307.4080},
 primaryClass = {astro-ph.EP},
       adsurl = {https://ui.adsabs.harvard.edu/abs/2013ApJ...773L..15R}
}

@ARTICLE{Kipping2013,
       author = {{Kipping}, David M.},
        title = "{Efficient, uninformative sampling of limb darkening coefficients for two-parameter laws}",
      journal = {\mnras},
         year = 2013,
        month = nov,
       volume = {435},
       number = {3},
        pages = {2152-2160},
          doi = {10.1093/mnras/stt1435},
archivePrefix = {arXiv},
       eprint = {1308.0009},
 primaryClass = {astro-ph.SR},
       adsurl = {https://ui.adsabs.harvard.edu/abs/2013MNRAS.435.2152K}
}

@ARTICLE{Higson2019,
       author = {{Higson}, Edward and {Handley}, Will and {Hobson}, Mike and {Lasenby}, Anthony},
        title = "{Dynamic nested sampling: an improved algorithm for parameter estimation and evidence calculation}",
      journal = {Statistics and Computing},
         year = 2019,
        month = sep,
       volume = {29},
       number = {5},
        pages = {891-913},
          doi = {10.1007/s11222-018-9844-0},
archivePrefix = {arXiv},
       eprint = {1704.03459},
 primaryClass = {stat.CO},
       adsurl = {https://ui.adsabs.harvard.edu/abs/2019S&C....29..891H}
}

@ARTICLE{Bruntt2010,
       author = {{Bruntt}, H. and {Bedding}, T.~R. and {Quirion}, P. -O. and {Lo Curto}, G. and {Carrier}, F. and {Smalley}, B. and {Dall}, T.~H. and {Arentoft}, T. and {Bazot}, M. and {Butler}, R.~P.},
        title = "{Accurate fundamental parameters for 23 bright solar-type stars}",
      journal = {\mnras},
         year = 2010,
        month = jul,
       volume = {405},
       number = {3},
        pages = {1907-1923},
          doi = {10.1111/j.1365-2966.2010.16575.x},
archivePrefix = {arXiv},
       eprint = {1002.4268},
 primaryClass = {astro-ph.SR},
       adsurl = {https://ui.adsabs.harvard.edu/abs/2010MNRAS.405.1907B}
}

@ARTICLE{Blanco2014iSpec,
       author = {{Blanco-Cuaresma}, S. and {Soubiran}, C. and {Heiter}, U. and {Jofr{\'e}}, P.},
        title = "{Determining stellar atmospheric parameters and chemical abundances of FGK stars with iSpec}",
      journal = {\aap},
         year = 2014,
        month = sep,
       volume = {569},
          eid = {A111},
        pages = {A111},
          doi = {10.1051/0004-6361/201423945},
archivePrefix = {arXiv},
       eprint = {1407.2608},
 primaryClass = {astro-ph.IM},
       adsurl = {https://ui.adsabs.harvard.edu/abs/2014A&A...569A.111B}
}

@ARTICLE{Blanco2014HighResolutionSpectralLibrary,
       author = {{Blanco-Cuaresma}, S. and {Soubiran}, C. and {Jofr{\'e}}, P. and {Heiter}, U.},
        title = "{The Gaia FGK benchmark stars. High resolution spectral library}",
      journal = {\aap},
         year = 2014,
        month = jun,
       volume = {566},
          eid = {A98},
        pages = {A98},
          doi = {10.1051/0004-6361/201323153},
archivePrefix = {arXiv},
       eprint = {1403.3090},
 primaryClass = {astro-ph.SR},
       adsurl = {https://ui.adsabs.harvard.edu/abs/2014A&A...566A..98B}
}

@ARTICLE{Blanco2019iSpec,
       author = {{Blanco-Cuaresma}, Sergi},
        title = "{Modern stellar spectroscopy caveats}",
      journal = {\mnras},
         year = 2019,
        month = jun,
       volume = {486},
       number = {2},
        pages = {2075-2101},
          doi = {10.1093/mnras/stz549},
archivePrefix = {arXiv},
       eprint = {1902.09558},
 primaryClass = {astro-ph.SR},
       adsurl = {https://ui.adsabs.harvard.edu/abs/2019MNRAS.486.2075B}
}

@ARTICLE{Doyle2014,
       author = {{Doyle}, Amanda P. and {Davies}, Guy R. and {Smalley}, Barry and {Chaplin}, William J. and {Elsworth}, Yvonne},
        title = "{Determining stellar macroturbulence using asteroseismic rotational velocities from Kepler}",
      journal = {\mnras},
         year = 2014,
        month = nov,
       volume = {444},
       number = {4},
        pages = {3592-3602},
          doi = {10.1093/mnras/stu1692},
archivePrefix = {arXiv},
       eprint = {1408.3988},
 primaryClass = {astro-ph.SR},
       adsurl = {https://ui.adsabs.harvard.edu/abs/2014MNRAS.444.3592D}
}

@BOOK{Aerts2010,
       author = {{Aerts}, Conny and {Christensen-Dalsgaard}, J{\o}rgen and {Kurtz}, Donald W.},
        title = "{Asteroseismology}",
        publisher = {Springer},
         year = 2010,
          doi = {10.1007/978-1-4020-5803-5},
       adsurl = {https://ui.adsabs.harvard.edu/abs/2010aste.book.....A}
}

@ARTICLE{Beck2012,
       author = {{Beck}, Paul G. and {Montalban}, Josefina and {Kallinger}, Thomas and {De Ridder}, Joris and {Aerts}, Conny and {Garc{\'\i}a}, Rafael A. and {Hekker}, Saskia and {Dupret}, Marc-Antoine and {Mosser}, Benoit and {Eggenberger}, Patrick and {Stello}, Dennis and {Elsworth}, Yvonne and {Frandsen}, S{\o}ren and {Carrier}, Fabien and {Hillen}, Michel and {Gruberbauer}, Michael and {Christensen-Dalsgaard}, J{\o}rgen and {Miglio}, Andrea and {Valentini}, Marica and {Bedding}, Timothy R. and {Kjeldsen}, Hans and {Girouard}, Forrest R. and {Hall}, Jennifer R. and {Ibrahim}, Khadeejah A.},
        title = "{Fast core rotation in red-giant stars as revealed by gravity-dominated mixed modes}",
      journal = {\nat},
         year = 2012,
        month = jan,
       volume = {481},
       number = {7379},
        pages = {55-57},
          doi = {10.1038/nature10612},
archivePrefix = {arXiv},
       eprint = {1112.2825},
 primaryClass = {astro-ph.SR},
       adsurl = {https://ui.adsabs.harvard.edu/abs/2012Natur.481...55B}
}

@ARTICLE{Deheuvels2012,
       author = {{Deheuvels}, S. and {Garc{\'\i}a}, R.~A. and {Chaplin}, W.~J. and {Basu}, S. and {Antia}, H.~M. and {Appourchaux}, T. and {Benomar}, O. and {Davies}, G.~R. and {Elsworth}, Y. and {Gizon}, L. and {Goupil}, M.~J. and {Reese}, D.~R. and {Regulo}, C. and {Schou}, J. and {Stahn}, T. and {Casagrande}, L. and {Christensen-Dalsgaard}, J. and {Fischer}, D. and {Hekker}, S. and {Kjeldsen}, H. and {Mathur}, S. and {Mosser}, B. and {Pinsonneault}, M. and {Valenti}, J. and {Christiansen}, J.~L. and {Kinemuchi}, K. and {Mullally}, F.},
        title = "{Seismic Evidence for a Rapidly Rotating Core in a Lower-giant-branch Star Observed with Kepler}",
      journal = {\apj},
         year = 2012,
        month = sep,
       volume = {756},
       number = {1},
          eid = {19},
        pages = {19},
          doi = {10.1088/0004-637X/756/1/19},
archivePrefix = {arXiv},
       eprint = {1206.3312},
 primaryClass = {astro-ph.SR},
       adsurl = {https://ui.adsabs.harvard.edu/abs/2012ApJ...756...19D}
}

@ARTICLE{Wong2022,
       author = {{Wong}, Ian and {Shporer}, Avi and {Vissapragada}, Shreyas and {Greklek-McKeon}, Michael and {Knutson}, Heather A. and {Winn}, Joshua N. and {Benneke}, Bj{\"o}rn},
        title = "{TESS Revisits WASP-12: Updated Orbital Decay Rate and Constraints on Atmospheric Variability}",
      journal = {\aj},
         year = 2022,
        month = apr,
       volume = {163},
       number = {4},
          eid = {175},
        pages = {175},
          doi = {10.3847/1538-3881/ac5680},
archivePrefix = {arXiv},
       eprint = {2201.08370},
 primaryClass = {astro-ph.EP},
       adsurl = {https://ui.adsabs.harvard.edu/abs/2022AJ....163..175W}
}

@ARTICLE{Akinsanmi2024,
       author = {{Akinsanmi}, B. and {Barros}, S.~C.~C. and {Lendl}, M. and {Carone}, L. and {Cubillos}, P.~E. and {Bekkelien}, A. and {Fortier}, A. and {Flor{\'e}n}, H.-G. and {Collier Cameron}, A. and {Bou{\'e}}, G. and {Bruno}, G. and {Demory}, B.-O. and {Brandeker}, A. and {Sousa}, S.~G. and {Wilson}, T.~G. and {Deline}, A. and {Bonfanti}, A. and {Scandariato}, G. and {Hooton}, M.~J. and {Correia}, A.~C.~M. and {Demangeon}, O.~D.~S. and {Smith}, A.~M.~S. and {Singh}, V. and {Alibert}, Y. and {Alonso}, R. and {Asquier}, J. and {B{\'a}rczy}, T. and {Barrado Navascues}, D. and {Baumjohann}, W. and {Beck}, M. and {Beck}, T. and {Benz}, W. and {Billot}, N. and {Bonfils}, X. and {Borsato}, L. and {Broeg}, C. and {Buder}, M. and {Charnoz}, S. and {Csizmadia}, Sz. and {Davies}, M.~B. and {Deleuil}, M. and {Delrez}, L. and {Ehrenreich}, D. and {Erikson}, A. and {Farinato}, J. and {Fossati}, L. and {Fridlund}, M. and {Gandolfi}, D. and {Gillon}, M. and {G{\"u}del}, M. and {G{\"u}nther}, M.~N. and {Heitzmann}, A. and {Helling}, Ch. and {Hoyer}, S. and {Isaak}, K.~G. and {Kiss}, L.~L. and {Lam}, K.~W.~F. and {Laskar}, J. and {Lecavelier des Etangs}, A. and {Magrin}, D. and {Maxted}, P.~F.~L. and {Mecina}, M. and {Mordasini}, C. and {Nascimbeni}, V. and {Olofsson}, G. and {Ottensamer}, R. and {Pagano}, I. and {Pall{\'e}}, E. and {Peter}, G. and {Piazza}, D. and {Piotto}, G. and {Pollacco}, D. and {Queloz}, D. and {Ragazzoni}, R. and {Rando}, N. and {Rauer}, H. and {Ribas}, I. and {Santos}, N.~C. and {S{\'e}gransan}, D. and {Simon}, A.~E. and {Stalport}, M. and {Szab{\'o}}, Gy. M. and {Thomas}, N. and {Udry}, S. and {Van Grootel}, V. and {Venturini}, J. and {Villaver}, E. and {Walton}, N.~A.},
        title = "{The tidal deformation and atmosphere of WASP-12 b from its phase curve★}",
      journal = {\aap},
         year = 2024,
        month = may,
       volume = {685},
          eid = {A63},
        pages = {A63},
          doi = {10.1051/0004-6361/202348502},
archivePrefix = {arXiv},
       eprint = {2402.10486},
 primaryClass = {astro-ph.EP},
       adsurl = {https://ui.adsabs.harvard.edu/abs/2024A&A...685A..63A}
}

@ARTICLE{Rossiter1924,
       author = {{Rossiter}, R.~A.},
        title = "{On the detection of an effect of rotation during eclipse in the velocity of the brigher component of beta Lyrae, and on the constancy of velocity of this system.}",
      journal = {\apj},
         year = 1924,
        month = jul,
       volume = {60},
        pages = {15-21},
          doi = {10.1086/142825},
       adsurl = {https://ui.adsabs.harvard.edu/abs/1924ApJ....60...15R}
}

@ARTICLE{McLaughlin1924,
       author = {{McLaughlin}, D.~B.},
        title = "{Some results of a spectrographic study of the Algol system.}",
      journal = {\apj},
         year = 1924,
        month = jul,
       volume = {60},
        pages = {22-31},
          doi = {10.1086/142826},
       adsurl = {https://ui.adsabs.harvard.edu/abs/1924ApJ....60...22M}
}

@ARTICLE{Rasio1996,
       author = {{Rasio}, Frederic A. and {Ford}, Eric B.},
        title = "{Dynamical instabilities and the formation of extrasolar planetary systems}",
      journal = {Science},
         year = 1996,
        month = nov,
       volume = {274},
        pages = {954-956},
          doi = {10.1126/science.274.5289.954},
       adsurl = {https://ui.adsabs.harvard.edu/abs/1996Sci...274..954R}
}

@ARTICLE{Fabrycky2007,
       author = {{Fabrycky}, Daniel and {Tremaine}, Scott},
        title = "{Shrinking Binary and Planetary Orbits by Kozai Cycles with Tidal Friction}",
      journal = {\apj},
         year = 2007,
        month = nov,
       volume = {669},
       number = {2},
        pages = {1298-1315},
          doi = {10.1086/521702},
archivePrefix = {arXiv},
       eprint = {0705.4285},
 primaryClass = {astro-ph},
       adsurl = {https://ui.adsabs.harvard.edu/abs/2007ApJ...669.1298F}
}

@ARTICLE{Nagasawa2008,
       author = {{Nagasawa}, M. and {Ida}, S. and {Bessho}, T.},
        title = "{Formation of Hot Planets by a Combination of Planet Scattering, Tidal Circularization, and the Kozai Mechanism}",
      journal = {\apj},
         year = 2008,
        month = may,
       volume = {678},
       number = {1},
        pages = {498-508},
          doi = {10.1086/529369},
archivePrefix = {arXiv},
       eprint = {0801.1368},
 primaryClass = {astro-ph},
       adsurl = {https://ui.adsabs.harvard.edu/abs/2008ApJ...678..498N}
}

@ARTICLE{Wu2011,
       author = {{Wu}, Yanqin and {Lithwick}, Yoram},
        title = "{Secular Chaos and the Production of Hot Jupiters}",
      journal = {\apj},
         year = 2011,
        month = jul,
       volume = {735},
       number = {2},
          eid = {109},
        pages = {109},
          doi = {10.1088/0004-637X/735/2/109},
archivePrefix = {arXiv},
       eprint = {1012.3475},
 primaryClass = {astro-ph.EP},
       adsurl = {https://ui.adsabs.harvard.edu/abs/2011ApJ...735..109W}
}

@ARTICLE{Hebb2009,
       author = {{Hebb}, L. and {Collier-Cameron}, A. and {Loeillet}, B. and {Pollacco}, D. and {H{\'e}brard}, G. and {Street}, R.~A. and {Bouchy}, F. and {Stempels}, H.~C. and {Moutou}, C. and {Simpson}, E. and {Udry}, S. and {Joshi}, Y.~C. and {West}, R.~G. and {Skillen}, I. and {Wilson}, D.~M. and {McDonald}, I. and {Gibson}, N.~P. and {Aigrain}, S. and {Anderson}, D.~R. and {Benn}, C.~R. and {Christian}, D.~J. and {Enoch}, B. and {Haswell}, C.~A. and {Hellier}, C. and {Horne}, K. and {Irwin}, J. and {Lister}, T.~A. and {Maxted}, P. and {Mayor}, M. and {Norton}, A.~J. and {Parley}, N. and {Pont}, F. and {Queloz}, D. and {Smalley}, B. and {Wheatley}, P.~J.},
        title = "{WASP-12b: The Hottest Transiting Extrasolar Planet Yet Discovered}",
      journal = {\apj},
         year = 2009,
        month = mar,
       volume = {693},
       number = {2},
        pages = {1920-1928},
          doi = {10.1088/0004-637X/693/2/1920},
archivePrefix = {arXiv},
       eprint = {0812.3240},
 primaryClass = {astro-ph},
       adsurl = {https://ui.adsabs.harvard.edu/abs/2009ApJ...693.1920H}
}

@ARTICLE{Li2010,
       author = {{Li}, Shu-Lin and {Miller}, N. and {Lin}, Douglas N.~C. and {Fortney}, Jonathan J.},
        title = "{WASP-12b as a prolate, inflated and disrupting planet from tidal dissipation}",
      journal = {\nat},
         year = 2010,
        month = feb,
       volume = {463},
       number = {7284},
        pages = {1054-1056},
          doi = {10.1038/nature08715},
archivePrefix = {arXiv},
       eprint = {1002.4608},
 primaryClass = {astro-ph.EP},
       adsurl = {https://ui.adsabs.harvard.edu/abs/2010Natur.463.1054L}
}

@ARTICLE{Heller2026cPIC,
       author = {{Heller}, Ren{\'e} and {Jiang}, Chen and {Bluhm}, Paz and {Granata}, Valentina and {Cabrera}, Juan and {Grie{\ss}bach}, Denis and {Paproth}, Carsten and {Csizmadia}, Szil{\'a}rd and {Eigm{\"u}ller}, Philipp and {Marrese}, Paola Maria and {Marinoni}, Silvia and {Samadi}, R{\'e}za and {Piotto}, Giampaolo and {Montalto}, Marco and {Sch{\"a}fer}, Martin and {Damiani}, Cilia and {Walton}, Nicholas and {Rauterberg}, Christoph and {Ammler-von Eiff}, Matthias and {Birch}, Aaron C. and {Gizon}, Laurent},
        title = "{PLATO input catalogs for technical calibration and fine guidance}",
      journal = {arXiv e-prints},
         year = 2026,
        month = apr,
          eid = {arXiv:2604.02437},
        pages = {arXiv:2604.02437},
          doi = {10.48550/arXiv.2604.02437},
archivePrefix = {arXiv},
       eprint = {2604.02437},
 primaryClass = {astro-ph.IM},
       adsurl = {https://ui.adsabs.harvard.edu/abs/2026arXiv260402437H}
}

@ARTICLE{Duguid2024,
       author = {{Duguid}, Craig D. and {de Vries}, Nils B. and {Lecoanet}, Daniel and {Barker}, Adrian J.},
        title = "{An Efficient Tidal Dissipation Mechanism via Stellar Magnetic Fields}",
      journal = {\apjl},
         year = 2024,
        month = may,
       volume = {966},
       number = {1},
          eid = {L14},
        pages = {L14},
          doi = {10.3847/2041-8213/ad3c40},
archivePrefix = {arXiv},
       eprint = {2404.07085},
 primaryClass = {astro-ph.SR},
       adsurl = {https://ui.adsabs.harvard.edu/abs/2024ApJ...966L..14D}
}

@ARTICLE{Kawaler1988,
       author = {{Kawaler}, Steven D.},
        title = "{Angular Momentum Loss in Low-Mass Stars}",
      journal = {\apj},
         year = 1988,
        month = oct,
       volume = {333},
        pages = {236},
          doi = {10.1086/166740},
       adsurl = {https://ui.adsabs.harvard.edu/abs/1988ApJ...333..236K}
}

@ARTICLE{Hallatt2026,
       author = {{Hallatt}, Tim and {Owen}, James E. and {Millholland}, Sarah},
        title = "{Revealing the Origin of Desert Dwellers via Stellar Obliquities}",
      journal = {arXiv e-prints},
         year = 2026,
        month = jun,
          eid = {arXiv:2606.11292},
        pages = {arXiv:2606.11292},
archivePrefix = {arXiv},
       eprint = {2606.11292},
 primaryClass = {astro-ph.EP},
       adsurl = {https://ui.adsabs.harvard.edu/abs/2026arXiv260611292H}
}

@ARTICLE{DYNESTY,
       author = {{Speagle}, Joshua S.},
        title = "{DYNESTY: a dynamic nested sampling package for estimating Bayesian posteriors and evidences}",
      journal = {\mnras},
         year = 2020,
        month = apr,
       volume = {493},
       number = {3},
        pages = {3132-3158},
          doi = {10.1093/mnras/staa278},
archivePrefix = {arXiv},
       eprint = {1904.02180},
 primaryClass = {astro-ph.IM},
       adsurl = {https://ui.adsabs.harvard.edu/abs/2020MNRAS.493.3132S}
}

@ARTICLE{ChakrabartySengupta2019,
       author = {{Chakrabarty}, Aritra and {Sengupta}, Sujan},
        title = "{Precise Photometric Transit Follow-up Observations of Five Close-in Exoplanets: Update on Their Physical Properties}",
      journal = {\aj},
         year = 2019,
        month = jul,
       volume = {158},
       number = {1},
          eid = {39},
        pages = {39},
          doi = {10.3847/1538-3881/ab24dd},
archivePrefix = {arXiv},
       eprint = {1905.11258},
 primaryClass = {astro-ph.EP},
       adsurl = {https://ui.adsabs.harvard.edu/abs/2019AJ....158...39C}
}

@ARTICLE{Schlaufman2010,
       author = {{Schlaufman}, Kevin C.},
        title = "{Evidence of Possible Spin-orbit Misalignment Along the Line of Sight in Transiting Exoplanet Systems}",
      journal = {\apj},
         year = 2010,
        month = aug,
       volume = {719},
       number = {1},
        pages = {602-611},
          doi = {10.1088/0004-637X/719/1/602},
archivePrefix = {arXiv},
       eprint = {1006.2851},
 primaryClass = {astro-ph.EP},
       adsurl = {https://ui.adsabs.harvard.edu/abs/2010ApJ...719..602S}
}

@ARTICLE{FabryckyWinn2009,
       author = {{Fabrycky}, Daniel C. and {Winn}, Joshua N.},
        title = "{Exoplanetary Spin-Orbit Alignment: Results from the Ensemble of Rossiter-McLaughlin Observations}",
      journal = {\apj},
         year = 2009,
        month = may,
       volume = {696},
       number = {2},
        pages = {1230-1240},
          doi = {10.1088/0004-637X/696/2/1230},
archivePrefix = {arXiv},
       eprint = {0902.0737},
 primaryClass = {astro-ph.EP},
       adsurl = {https://ui.adsabs.harvard.edu/abs/2009ApJ...696.1230F}
}

@ARTICLE{Sun2025,
       author = {{Sun}, Meng and {Townsend}, R.~H.~D. and {Xia}, Hongbo and {Liu}, Jifeng},
        title = "{Numerical Nonadiabatic Tidal Calculations with GYRE-tides: The WASP-12 Test Case}",
      journal = {\apjl},
         year = 2025,
        month = dec,
       volume = {995},
       number = {1},
          eid = {L29},
        pages = {L29},
          doi = {10.3847/2041-8213/ae23cf},
archivePrefix = {arXiv},
       eprint = {2511.18769},
 primaryClass = {astro-ph.SR},
       adsurl = {https://ui.adsabs.harvard.edu/abs/2025ApJ...995L..29S}
}

@ARTICLE{Knutson2010,
       author = {{Knutson}, Heather A. and {Howard}, Andrew W. and {Isaacson}, Howard},
        title = "{A Correlation Between Stellar Activity and Hot Jupiter Emission Spectra}",
      journal = {\apj},
         year = 2010,
        month = sep,
       volume = {720},
       number = {2},
        pages = {1569-1576},
          doi = {10.1088/0004-637X/720/2/1569},
archivePrefix = {arXiv},
       eprint = {1004.2702},
 primaryClass = {astro-ph.EP},
       adsurl = {https://ui.adsabs.harvard.edu/abs/2010ApJ...720.1569K}
}

@ARTICLE{Fossati2013,
       author = {{Fossati}, L. and {Ayres}, T.~R. and {Haswell}, C.~A. and {Bohlender}, D. and {Kochukhov}, O. and {Fl{\"o}er}, L.},
        title = "{Absorbing Gas around the WASP-12 Planetary System}",
      journal = {\apjl},
         year = 2013,
        month = apr,
       volume = {766},
       number = {2},
          eid = {L20},
        pages = {L20},
          doi = {10.1088/2041-8205/766/2/L20},
archivePrefix = {arXiv},
       eprint = {1303.3375},
 primaryClass = {astro-ph.SR},
       adsurl = {https://ui.adsabs.harvard.edu/abs/2013ApJ...766L..20F}
}

@ARTICLE{Bonomo2017,
       author = {{Bonomo}, A.~S. and {Desidera}, S. and {Benatti}, S. and {Borsa}, F. and {Crespi}, S. and {Damasso}, M. and {Lanza}, A.~F. and {Sozzetti}, A. and {Lodato}, G. and {Marzari}, F. and {Boccato}, C. and {Claudi}, R.~U. and {Cosentino}, R. and {Covino}, E. and {Gratton}, R. and {Maggio}, A. and {Micela}, G. and {Molinari}, E. and {Pagano}, I. and {Piotto}, G. and {Poretti}, E. and {Smareglia}, R. and {Affer}, L. and {Biazzo}, K. and {Bignamini}, A. and {Esposito}, M. and {Giacobbe}, P. and {H{\'e}brard}, G. and {Malavolta}, L. and {Maldonado}, J. and {Mancini}, L. and {Martinez Fiorenzano}, A. and {Masiero}, S. and {Nascimbeni}, V. and {Pedani}, M. and {Rainer}, M. and {Scandariato}, G.},
        title = "{The GAPS Programme with HARPS-N at TNG . XIV. Investigating giant planet migration history via improved eccentricity and mass determination for 231 transiting planets}",
      journal = {\aap},
         year = 2017,
        month = jun,
       volume = {602},
          eid = {A107},
        pages = {A107},
          doi = {10.1051/0004-6361/201629882},
archivePrefix = {arXiv},
       eprint = {1704.00373},
 primaryClass = {astro-ph.EP},
       adsurl = {https://ui.adsabs.harvard.edu/abs/2017A&A...602A.107B}
}

@ARTICLE{Skumanich,
       author = {{Skumanich}, A.},
        title = "{Time Scales for Ca II Emission Decay, Rotational Braking, and Lithium Depletion}",
      journal = {\apj},
         year = 1972,
        month = feb,
       volume = {171},
        pages = {565},
          doi = {10.1086/151310},
       adsurl = {https://ui.adsabs.harvard.edu/abs/1972ApJ...171..565S}
}

@ARTICLE{CKS1,
       author = {{Petigura}, Erik A. and {Howard}, Andrew W. and {Marcy}, Geoffrey W. and
         {Johnson}, John Asher and {Isaacson}, Howard and {Cargile}, Phillip A. and
         {Hebb}, Leslie and {Fulton}, Benjamin J. and {Weiss}, Lauren M. and
         {Morton}, Timothy D. and {Winn}, Joshua N. and {Rogers}, Leslie A. and
         {Sinukoff}, Evan and {Hirsch}, Lea A. and {Crossfield}, Ian J.~M.},
        title = "{The California-Kepler Survey. I. High-resolution Spectroscopy of 1305 Stars Hosting Kepler Transiting Planets}",
      journal = {\aj},
         year = 2017,
        month = sep,
       volume = {154},
       number = {3},
          eid = {107},
        pages = {107},
          doi = {10.3847/1538-3881/aa80de},
archivePrefix = {arXiv},
       eprint = {1703.10400},
 primaryClass = {astro-ph.EP},
       adsurl = {https://ui.adsabs.harvard.edu/abs/2017AJ....154..107P}
}

@ARTICLE{Millholland2018,
       author = {{Millholland}, Sarah and {Laughlin}, Gregory},
        title = "{Obliquity Tides May Drive WASP-12b{\textquoteright}s Rapid Orbital Decay}",
      journal = {\apjl},
         year = 2018,
        month = dec,
       volume = {869},
       number = {1},
          eid = {L15},
        pages = {L15},
          doi = {10.3847/2041-8213/aaedb1},
archivePrefix = {arXiv},
       eprint = {1812.01624},
 primaryClass = {astro-ph.EP},
       adsurl = {https://ui.adsabs.harvard.edu/abs/2018ApJ...869L..15M}
}

@ARTICLE{Cao2023,
       author = {{Cao}, Lyra and {Pinsonneault}, Marc H. and {van Saders}, Jennifer L.},
        title = "{Core-envelope Decoupling Drives Radial Shear Dynamos in Cool Stars}",
      journal = {\apjl},
         year = 2023,
        month = jul,
       volume = {951},
       number = {2},
          eid = {L49},
        pages = {L49},
          doi = {10.3847/2041-8213/acd780},
archivePrefix = {arXiv},
       eprint = {2301.07716},
 primaryClass = {astro-ph.SR},
       adsurl = {https://ui.adsabs.harvard.edu/abs/2023ApJ...951L..49C}
}

@ARTICLE{Somers2016,
       author = {{Somers}, Garrett and {Pinsonneault}, Marc H.},
        title = "{Lithium Depletion is a Strong Test of Core-envelope Recoupling}",
      journal = {\apj},
         year = 2016,
        month = sep,
       volume = {829},
       number = {1},
          eid = {32},
        pages = {32},
          doi = {10.3847/0004-637X/829/1/32},
archivePrefix = {arXiv},
       eprint = {1606.00004},
 primaryClass = {astro-ph.SR},
       adsurl = {https://ui.adsabs.harvard.edu/abs/2016ApJ...829...32S}
}

@ARTICLE{Winn2025,
       author = {{Winn}, Joshua N. and {Stef{\'a}nsson}, Gu{\dj}mundur},
        title = "{Orbital Decay Candidates Reconsidered: WASP-4 b Is Not Decaying and Kepler-1658 b Is Not a Planet}",
      journal = {\psj},
         year = 2025,
        month = dec,
       volume = {6},
       number = {12},
          eid = {300},
        pages = {300},
          doi = {10.3847/PSJ/ae21db},
archivePrefix = {arXiv},
       eprint = {2510.05229},
 primaryClass = {astro-ph.EP},
       adsurl = {https://ui.adsabs.harvard.edu/abs/2025PSJ.....6..300W}
}

@ARTICLE{Bouma2020,
       author = {{Bouma}, L.~G. and {Winn}, J.~N. and {Howard}, A.~W. and {Howell}, S.~B. and {Isaacson}, H. and {Knutson}, H. and {Matson}, R.~A.},
        title = "{WASP-4 Is Accelerating toward the Earth}",
      journal = {\apjl},
         year = 2020,
        month = apr,
       volume = {893},
       number = {2},
          eid = {L29},
        pages = {L29},
          doi = {10.3847/2041-8213/ab8563},
archivePrefix = {arXiv},
       eprint = {2004.00637},
 primaryClass = {astro-ph.EP},
       adsurl = {https://ui.adsabs.harvard.edu/abs/2020ApJ...893L..29B}
}

@ARTICLE{Spada2020,
       author = {{Spada}, F. and {Lanzafame}, A.~C.},
        title = "{Competing effect of wind braking and interior coupling in the rotational evolution of solar-like stars}",
      journal = {\aap},
         year = 2020,
        month = apr,
       volume = {636},
          eid = {A76},
        pages = {A76},
          doi = {10.1051/0004-6361/201936384},
archivePrefix = {arXiv},
       eprint = {1908.00345},
 primaryClass = {astro-ph.SR},
       adsurl = {https://ui.adsabs.harvard.edu/abs/2020A&A...636A..76S}
}

@ARTICLE{Gallet2013,
       author = {{Gallet}, F. and {Bouvier}, J.},
        title = "{Improved angular momentum evolution model for solar-like stars}",
      journal = {\aap},
         year = 2013,
        month = aug,
       volume = {556},
          eid = {A36},
        pages = {A36},
          doi = {10.1051/0004-6361/201321302},
archivePrefix = {arXiv},
       eprint = {1306.2130},
 primaryClass = {astro-ph.SR},
       adsurl = {https://ui.adsabs.harvard.edu/abs/2013A&A...556A..36G}
}

@ARTICLE{Lanzafame2015,
       author = {{Lanzafame}, A.~C. and {Spada}, F.},
        title = "{Rotational evolution of slow-rotator sequence stars}",
      journal = {\aap},
         year = 2015,
        month = dec,
       volume = {584},
          eid = {A30},
        pages = {A30},
          doi = {10.1051/0004-6361/201526770},
archivePrefix = {arXiv},
       eprint = {1506.05298},
 primaryClass = {astro-ph.SR},
       adsurl = {https://ui.adsabs.harvard.edu/abs/2015A&A...584A..30L}
}

@ARTICLE{Li2020,
       author = {{Li}, Gang and {Guo}, Zhao and {Fuller}, Jim and {Bedding}, Timothy R. and {Murphy}, Simon J. and {Colman}, Isabel L. and {Hey}, Daniel R.},
        title = "{The effect of tides on near-core rotation: analysis of 35 Kepler {\ensuremath{\gamma}} Doradus stars in eclipsing and spectroscopic binaries}",
      journal = {\mnras},
         year = 2020,
        month = oct,
       volume = {497},
       number = {4},
        pages = {4363-4375},
          doi = {10.1093/mnras/staa2266},
archivePrefix = {arXiv},
       eprint = {2007.14853},
 primaryClass = {astro-ph.SR},
       adsurl = {https://ui.adsabs.harvard.edu/abs/2020MNRAS.497.4363L}
}

@ARTICLE{Pinsonneault2026,
       author = {{Pinsonneault}, Marc H. and {van Saders}, Jennifer L. and {Cao}, Lyra and {Tayar}, Jamie and {Delahaye}, Franck and {Morales}, Leslie M. and {Patton}, Rachel A. and {Rendina}, Matthew C. and {Zinn}, Joel C. and {Claytor}, Zachary R. and {Ash}, Amanda L. and {Byrom}, Susan and {Cao}, Kaili and {Smedile}, Vincent A.},
        title = "{The YREC Stellar Evolution Code: Public Data Release}",
      journal = {arXiv e-prints},
         year = 2026,
        month = mar,
          eid = {arXiv:2603.25792},
        pages = {arXiv:2603.25792},
          doi = {10.48550/arXiv.2603.25792},
archivePrefix = {arXiv},
       eprint = {2603.25792},
 primaryClass = {astro-ph.SR}
}

@ARTICLE{Metcalfe2020,
       author = {{Metcalfe}, Travis S. and {van Saders}, Jennifer L. and {Basu}, Sarbani and {Buzasi}, Derek and {Chaplin}, William J. and {Christensen-Dalsgaard}, J{\o}rgen and {Creevey}, Orlagh L. and {Deheuvels}, S{\'e}bastien and {Garc{\'i}a}, Rafael A. and {Mathur}, Savita and others},
        title = "{The Evolution of Rotation and Magnetic Activity in 94 Aqr Aa from Asteroseismology with TESS}",
      journal = {\apj},
         year = 2020,
        month = sep,
       volume = {900},
       number = {2},
          eid = {154},
        pages = {154},
          doi = {10.3847/1538-4357/aba963}
}

@ARTICLE{Metcalfe2025,
       author = {{Metcalfe}, Travis S. and {Petit}, Pascal and {van Saders}, Jennifer L. and {Ayres}, Thomas R. and {Buzasi}, Derek and {Kochukhov}, Oleg and {Stassun}, Keivan G. and {Pinsonneault}, Marc H. and {Ilyin}, Ilya V. and {Strassmeier}, Klaus G. and {Finley}, Adam J. and {Garc{\'\i}a}, Rafael A. and {Huber}, Daniel and {Lu}, Yuxi (Lucy) and {See}, Victor},
        title = "{Testing the Rossby Paradigm: Weakened Magnetic Braking in Early K-type Stars}",
      journal = {\apj},
         year = 2025,
        month = jun,
       volume = {986},
       number = {2},
          eid = {120},
        pages = {120},
          doi = {10.3847/1538-4357/add40a},
archivePrefix = {arXiv},
       eprint = {2501.19169},
 primaryClass = {astro-ph.SR},
       adsurl = {https://ui.adsabs.harvard.edu/abs/2025ApJ...986..120M}
}

@ARTICLE{Saunders2024,
       author = {{Saunders}, Nicholas and {van Saders}, Jennifer L. and {Lyttle}, Alexander J. and {Metcalfe}, Travis S. and {Li}, Tanda and {Davies}, Guy R. and {Hall}, Oliver J. and {Ball}, Warrick H. and {Townsend}, Richard and {Creevey}, Orlagh and {Dodds}, Curt},
        title = "{Stellar Cruise Control: Weakened Magnetic Braking Leads to Sustained Rapid Rotation of Old Stars}",
      journal = {\apj},
         year = 2024,
        month = feb,
       volume = {962},
       number = {2},
          eid = {138},
        pages = {138},
          doi = {10.3847/1538-4357/ad1516}
}

@ARTICLE{vanSaders2013,
       author = {{van Saders}, Jennifer L. and {Pinsonneault}, Marc H.},
        title = "{Fast Star, Slow Star; Old Star, Young Star: Subgiant Rotation as a Population and Stellar Physics Diagnostic}",
      journal = {\apj},
         year = 2013,
        month = oct,
       volume = {776},
       number = {2},
          eid = {67},
        pages = {67},
          doi = {10.1088/0004-637X/776/2/67},
archivePrefix = {arXiv},
       eprint = {1306.3701},
 primaryClass = {astro-ph.SR}
}

@ARTICLE{Herbst2001,
       author = {{Herbst}, W. and {Bailer-Jones}, C.~A.~L. and {Mundt}, R.},
        title = "{The Mass Dependence of Stellar Rotation in the Orion Nebula Cluster}",
      journal = {\apjl},
         year = 2001,
        month = jun,
       volume = {554},
       number = {2},
        pages = {L197-L200},
          doi = {10.1086/321701},
archivePrefix = {arXiv},
       eprint = {astro-ph/0104438}
}

@ARTICLE{Somers2017,
       author = {{Somers}, Garrett and {Stauffer}, John and {Rebull}, Luisa and {Cody}, Ann Marie and {Pinsonneault}, Marc H.},
        title = "{M Dwarf Rotation from the K2 Young Clusters to the Field. I. A Mass-Rotation Correlation at 10 Myr}",
      journal = {\apj},
         year = 2017,
        month = nov,
       volume = {850},
       number = {2},
          eid = {134},
        pages = {134},
          doi = {10.3847/1538-4357/aa93ed},
archivePrefix = {arXiv},
       eprint = {1710.07638},
 primaryClass = {astro-ph.SR}
}

@ARTICLE{Matt2008,
       author = {{Matt}, Sean and {Pudritz}, Ralph E.},
        title = "{Accretion-powered Stellar Winds. II. Numerical Solutions for Stellar Wind Torques}",
      journal = {\apj},
         year = 2008,
        month = may,
       volume = {678},
       number = {2},
        pages = {1109-1118},
          doi = {10.1086/533428},
archivePrefix = {arXiv},
       eprint = {0801.0436},
 primaryClass = {astro-ph},
       adsurl = {https://ui.adsabs.harvard.edu/abs/2008ApJ...678.1109M}
}

@ARTICLE{Asplund2009,
       author = {{Asplund}, Martin and {Grevesse}, Nicolas and {Sauval}, A. Jacques and {Scott}, Pat},
        title = "{The Chemical Composition of the Sun}",
      journal = {\araa},
         year = 2009,
        month = sep,
       volume = {47},
       number = {1},
        pages = {481-522},
          doi = {10.1146/annurev.astro.46.060407.145222},
archivePrefix = {arXiv},
       eprint = {0909.0948},
 primaryClass = {astro-ph.SR},
       adsurl = {https://ui.adsabs.harvard.edu/abs/2009ARA&A..47..481A}
}

@ARTICLE{Bellinger2016,
       author = {{Bellinger}, Earl P. and {Angelou}, George C. and {Hekker}, Saskia and {Basu}, Sarbani and {Ball}, Warrick H. and {Guggenberger}, Elisabeth},
        title = "{Fundamental Parameters of Main-Sequence Stars in an Instant with Machine Learning}",
      journal = {\apj},
         year = 2016,
        month = oct,
       volume = {830},
       number = {1},
          eid = {31},
        pages = {31},
          doi = {10.3847/0004-637X/830/1/31},
archivePrefix = {arXiv},
       eprint = {1607.02137},
 primaryClass = {astro-ph.SR},
       adsurl = {https://ui.adsabs.harvard.edu/abs/2016ApJ...830...31B}
}

@ARTICLE{Choi2016,
       author = {{Choi}, Jieun and {Dotter}, Aaron and {Conroy}, Charlie and {Cantiello}, Matteo and {Paxton}, Bill and {Johnson}, Benjamin D.},
        title = "{Mesa Isochrones and Stellar Tracks (MIST). I. Solar-scaled Models}",
      journal = {\apj},
         year = 2016,
        month = jun,
       volume = {823},
       number = {2},
          eid = {102},
        pages = {102},
          doi = {10.3847/0004-637X/823/2/102},
archivePrefix = {arXiv},
       eprint = {1604.08592},
 primaryClass = {astro-ph.SR},
       adsurl = {https://ui.adsabs.harvard.edu/abs/2016ApJ...823..102C}
}

@ARTICLE{Henyey1965,
       author = {{Henyey}, Louis and {Vardya}, M.~S. and {Bodenheimer}, Peter},
        title = "{Studies in Stellar Evolution. III. The Calculation of Model Envelopes.}",
      journal = {\apj},
         year = 1965,
        month = oct,
       volume = {142},
        pages = {841},
          doi = {10.1086/148357},
       adsurl = {https://ui.adsabs.harvard.edu/abs/1965ApJ...142..841H}
}

@ARTICLE{Huber2017,
       author = {{Huber}, Daniel and {Zinn}, Joel and {Bojsen-Hansen}, Mathias and {Pinsonneault}, Marc and {Sahlholdt}, Christian and {Serenelli}, Aldo and {Silva Aguirre}, Victor and {Stassun}, Keivan and {Stello}, Dennis and {Tayar}, Jamie and {Bastien}, Fabienne and {Bedding}, Timothy R. and {Buchhave}, Lars A. and {Chaplin}, William J. and {Davies}, Guy R. and {Garc{\'\i}a}, Rafael A. and {Latham}, David W. and {Mathur}, Savita and {Mosser}, Benoit and {Sharma}, Sanjib},
        title = "{Asteroseismology and Gaia: Testing Scaling Relations Using 2200 Kepler  Stars with TGAS Parallaxes}",
      journal = {\apj},
         year = 2017,
        month = aug,
       volume = {844},
       number = {2},
          eid = {102},
        pages = {102},
          doi = {10.3847/1538-4357/aa75ca},
archivePrefix = {arXiv},
       eprint = {1705.04697},
 primaryClass = {astro-ph.SR},
       adsurl = {https://ui.adsabs.harvard.edu/abs/2017ApJ...844..102H}
}

@ARTICLE{Jermyn2023,
       author = {{Jermyn}, Adam S. and {Bauer}, Evan B. and {Schwab}, Josiah and {Farmer}, R. and {Ball}, Warrick H. and {Bellinger}, Earl P. and {Dotter}, Aaron and {Joyce}, Meridith and {Marchant}, Pablo and {Mombarg}, Joey S.~G. and {Wolf}, William M. and {Sunny Wong}, Tin Long and {Cinquegrana}, Giulia C. and {Farrell}, Eoin and {Smolec}, R. and {Thoul}, Anne and {Cantiello}, Matteo and {Herwig}, Falk and {Toloza}, Odette and {Bildsten}, Lars and {Townsend}, Richard H.~D. and {Timmes}, F.~X.},
        title = "{Modules for Experiments in Stellar Astrophysics (MESA): Time-dependent Convection, Energy Conservation, Automatic Differentiation, and Infrastructure}",
      journal = {\apjs},
         year = 2023,
        month = mar,
       volume = {265},
       number = {1},
          eid = {15},
        pages = {15},
          doi = {10.3847/1538-4365/acae8d},
archivePrefix = {arXiv},
       eprint = {2208.03651},
 primaryClass = {astro-ph.SR},
       adsurl = {https://ui.adsabs.harvard.edu/abs/2023ApJS..265...15J}
}

@ARTICLE{Paxton2011,
       author = {{Paxton}, Bill and {Bildsten}, Lars and {Dotter}, Aaron and {Herwig}, Falk and {Lesaffre}, Pierre and {Timmes}, Frank},
        title = "{Modules for Experiments in Stellar Astrophysics (MESA)}",
      journal = {\apjs},
         year = 2011,
        month = jan,
       volume = {192},
       number = {1},
          eid = {3},
        pages = {3},
          doi = {10.1088/0067-0049/192/1/3},
archivePrefix = {arXiv},
       eprint = {1009.1622},
 primaryClass = {astro-ph.SR},
       adsurl = {https://ui.adsabs.harvard.edu/abs/2011ApJS..192....3P}
}

@ARTICLE{Paxton2015,
       author = {{Paxton}, Bill and {Marchant}, Pablo and {Schwab}, Josiah and {Bauer}, Evan B. and {Bildsten}, Lars and {Cantiello}, Matteo and {Dessart}, Luc and {Farmer}, R. and {Hu}, H. and {Langer}, N. and {Townsend}, R.~H.~D. and {Townsley}, Dean M. and {Timmes}, F.~X.},
        title = "{Modules for Experiments in Stellar Astrophysics (MESA): Binaries, Pulsations, and Explosions}",
      journal = {\apjs},
         year = 2015,
        month = sep,
       volume = {220},
       number = {1},
          eid = {15},
        pages = {15},
          doi = {10.1088/0067-0049/220/1/15},
archivePrefix = {arXiv},
       eprint = {1506.03146},
 primaryClass = {astro-ph.SR},
       adsurl = {https://ui.adsabs.harvard.edu/abs/2015ApJS..220...15P}
}

@ARTICLE{Paxton2013,
       author = {{Paxton}, Bill and {Cantiello}, Matteo and {Arras}, Phil and {Bildsten}, Lars and {Brown}, Edward F. and {Dotter}, Aaron and {Mankovich}, Christopher and {Montgomery}, M.~H. and {Stello}, Dennis and {Timmes}, F.~X. and {Townsend}, Richard},
        title = "{Modules for Experiments in Stellar Astrophysics (MESA): Planets, Oscillations, Rotation, and Massive Stars}",
      journal = {\apjs},
         year = 2013,
        month = sep,
       volume = {208},
       number = {1},
          eid = {4},
        pages = {4},
          doi = {10.1088/0067-0049/208/1/4},
archivePrefix = {arXiv},
       eprint = {1301.0319},
 primaryClass = {astro-ph.SR},
       adsurl = {https://ui.adsabs.harvard.edu/abs/2013ApJS..208....4P}
}

@ARTICLE{Paxton2018,
       author = {{Paxton}, Bill and {Schwab}, Josiah and {Bauer}, Evan B. and {Bildsten}, Lars and {Blinnikov}, Sergei and {Duffell}, Paul and {Farmer}, R. and {Goldberg}, Jared A. and {Marchant}, Pablo and {Sorokina}, Elena and {Thoul}, Anne and {Townsend}, Richard H.~D. and {Timmes}, F.~X.},
        title = "{Modules for Experiments in Stellar Astrophysics (MESA): Convective Boundaries, Element Diffusion, and Massive Star Explosions}",
      journal = {\apjs},
         year = 2018,
        month = feb,
       volume = {234},
       number = {2},
          eid = {34},
        pages = {34},
          doi = {10.3847/1538-4365/aaa5a8},
archivePrefix = {arXiv},
       eprint = {1710.08424},
 primaryClass = {astro-ph.SR},
       adsurl = {https://ui.adsabs.harvard.edu/abs/2018ApJS..234...34P}
}

@ARTICLE{Paxton2019,
       author = {{Paxton}, Bill and {Smolec}, R. and {Schwab}, Josiah and {Gautschy}, A. and {Bildsten}, Lars and {Cantiello}, Matteo and {Dotter}, Aaron and {Farmer}, R. and {Goldberg}, Jared A. and {Jermyn}, Adam S. and {Kanbur}, S.~M. and {Marchant}, Pablo and {Thoul}, Anne and {Townsend}, Richard H.~D. and {Wolf}, William M. and {Zhang}, Michael and {Timmes}, F.~X.},
        title = "{Modules for Experiments in Stellar Astrophysics (MESA): Pulsating Variable Stars, Rotation, Convective Boundaries, and Energy Conservation}",
      journal = {\apjs},
         year = 2019,
        month = jul,
       volume = {243},
       number = {1},
          eid = {10},
        pages = {10},
          doi = {10.3847/1538-4365/ab2241},
archivePrefix = {arXiv},
       eprint = {1903.01426},
 primaryClass = {astro-ph.SR},
       adsurl = {https://ui.adsabs.harvard.edu/abs/2019ApJS..243...10P}
}

@ARTICLE{Thoul1994,
       author = {{Thoul}, Anne A. and {Bahcall}, John N. and {Loeb}, Abraham},
        title = "{Element Diffusion in the Solar Interior}",
      journal = {\apj},
         year = 1994,
        month = feb,
       volume = {421},
        pages = {828},
          doi = {10.1086/173695},
archivePrefix = {arXiv},
       eprint = {astro-ph/9304005},
 primaryClass = {astro-ph},
       adsurl = {https://ui.adsabs.harvard.edu/abs/1994ApJ...421..828T}
}

@article{Sobol1967,
    title = {On the distribution of points in a cube and the approximate evaluation of integrals},
    journal = {USSR Computational Mathematics and Mathematical Physics},
    volume = {7},
    number = {4},
    pages = {86-112},
    year = {1967},
    issn = {0041-5553},
    doi = {https://doi.org/10.1016/0041-5553(67)90144-9},
    url = {https://www.sciencedirect.com/science/article/pii/0041555367901449},
    author = {I.M Sobol'}
}

@book{Eddington1926,
  author    = {Eddington, Arthur S.},
  title     = {The Internal Constitution of the Stars},
  year       = {1926},
  publisher  = {Cambridge University Press},
  address    = {Cambridge}
}

@ARTICLE{vanSaders2016,
       author = {{van Saders}, Jennifer L. and {Ceillier}, Tugdual and {Metcalfe}, Travis S. and {Silva Aguirre}, Victor and {Pinsonneault}, Marc H. and {Garc{\'\i}a}, Rafael A. and {Mathur}, Savita and {Davies}, Guy R.},
        title = "{Weakened magnetic braking as the origin of anomalously rapid rotation in old field stars}",
      journal = {\nat},
         year = 2016,
        month = jan,
       volume = {529},
       number = {7585},
        pages = {181-184},
          doi = {10.1038/nature16168},
archivePrefix = {arXiv},
       eprint = {1601.02631},
 primaryClass = {astro-ph.SR},
       adsurl = {https://ui.adsabs.harvard.edu/abs/2016Natur.529..181V}
}

@ARTICLE{Ong2025,
       author = {{Ong}, J.~M. Joel},
        title = "{Signatures of Core-envelope Rotational Misalignment in the Mixed-mode Asteroseismology of Kepler-56}",
      journal = {\apj},
         year = 2025,
        month = feb,
       volume = {980},
       number = {1},
          eid = {40},
        pages = {40},
          doi = {10.3847/1538-4357/ada4a3},
archivePrefix = {arXiv},
       eprint = {2412.19451},
 primaryClass = {astro-ph.SR},
       adsurl = {https://ui.adsabs.harvard.edu/abs/2025ApJ...980...40O}
}

@ARTICLE{Biswas2026,
       author = {{Biswas}, Shraddha and {Jiang}, Ing-Guey and {Yeh}, Li-Chin and {Liu}, Hsin-Min and {Parthasarathy}, Kaviya and {Bisht}, D. and {Chakrabarti}, Sandip K. and {Bhowmick}, D. and {Bisht}, Mohit Singh and {Raj}, A. and {Martin}, Bryan E. and {Yadav}, R.~K.~S. and {Rangwal}, Geeta},
        title = "{Revisiting the Orbital Dynamics of the Hot Jupiter WASP-12 b with New Transit Times}",
      journal = {\aj},
         year = 2026,
        month = apr,
       volume = {171},
       number = {4},
          eid = {246},
        pages = {246},
          doi = {10.3847/1538-3881/ae4c86},
archivePrefix = {arXiv},
       eprint = {2510.06589},
 primaryClass = {astro-ph.EP},
       adsurl = {https://ui.adsabs.harvard.edu/abs/2026AJ....171..246B}
}

@ARTICLE{Ricker2015,
       author = {{Ricker}, George R. and {Winn}, Joshua N. and {Vanderspek}, Roland and {Latham}, David W. and {Bakos}, G{\'a}sp{\'a}r {\'A}. and {Bean}, Jacob L. and {Berta-Thompson}, Zachory K. and {Brown}, Timothy M. and {Buchhave}, Lars and {Butler}, Nathaniel R. and {Butler}, R. Paul and {Chaplin}, William J. and {Charbonneau}, David and {Christensen-Dalsgaard}, J{\o}rgen and {Clampin}, Mark and {Deming}, Drake and {Doty}, John and {De Lee}, Nathan and {Dressing}, Courtney and {Dunham}, Edward W. and {Endl}, Michael and {Fressin}, Francois and {Ge}, Jian and {Henning}, Thomas and {Holman}, Matthew J. and {Howard}, Andrew W. and {Ida}, Shigeru and {Jenkins}, Jon M. and {Jernigan}, Garrett and {Johnson}, John Asher and {Kaltenegger}, Lisa and {Kawai}, Nobuyuki and {Kjeldsen}, Hans and {Laughlin}, Gregory and {Levine}, Alan M. and {Lin}, Douglas and {Lissauer}, Jack J. and {MacQueen}, Phillip and {Marcy}, Geoffrey and {McCullough}, Peter R. and {Morton}, Timothy D. and {Narita}, Norio and {Paegert}, Martin and {Palle}, Enric and {Pepe}, Francesco and {Pepper}, Joshua and {Quirrenbach}, Andreas and {Rinehart}, Stephen A. and {Sasselov}, Dimitar and {Sato}, Bun'ei and {Seager}, Sara and {Sozzetti}, Alessandro and {Stassun}, Keivan G. and {Sullivan}, Peter and {Szentgyorgyi}, Andrew and {Torres}, Guillermo and {Udry}, Stephane and {Villasenor}, Joel},
        title = "{Transiting Exoplanet Survey Satellite (TESS)}",
      journal = {Journal of Astronomical Telescopes, Instruments, and Systems},
         year = 2015,
        month = jan,
       volume = {1},
          eid = {014003},
        pages = {014003},
          doi = {10.1117/1.JATIS.1.1.014003},
archivePrefix = {arXiv},
       eprint = {1406.0151},
 primaryClass = {astro-ph.EP},
       adsurl = {https://ui.adsabs.harvard.edu/abs/2015JATIS...1a4003R}
}

@ARTICLE{Yang2025,
       author = {{Yang}, Huiqin and {Liu}, Jifeng and {Soria}, Roberto and {Spada}, Federico and {Wang}, Song and {Fang}, Xiangsong and {Li}, Xue},
        title = "{Four ages of rotating stars in the rotation─activity relationship and gyrochronology}",
      journal = {\aap},
         year = 2025,
        month = jul,
       volume = {699},
          eid = {A251},
        pages = {A251},
          doi = {10.1051/0004-6361/202554379},
archivePrefix = {arXiv},
       eprint = {2506.07383},
 primaryClass = {astro-ph.SR},
       adsurl = {https://ui.adsabs.harvard.edu/abs/2025A&A...699A.251Y}
}

@article{Bergfors2012,
   title={Stellar companions to exoplanet host stars: Lucky Imaging of transiting planet hosts★},
   volume={428},
   ISSN={0035-8711},
   url={http://dx.doi.org/10.1093/mnras/sts019},
   DOI={10.1093/mnras/sts019},
   number={1},
   journal={Monthly Notices of the Royal Astronomical Society},
   publisher={Oxford University Press (OUP)},
   author={Bergfors, C. and Brandner, W. and Daemgen, S. and Biller, B. and Hippler, S. and Janson, M. and Kudryavtseva, N. and Geißler, K. and Henning, T. and Köhler, R.},
   year={2012},
   month=Oct, pages={182–189} 
}

@article{Bechter2014,
   title={WASP-12b AND HAT-P-8b ARE MEMBERS OF TRIPLE STAR SYSTEMS},
   volume={788},
   ISSN={1538-4357},
   url={http://dx.doi.org/10.1088/0004-637X/788/1/2},
   DOI={10.1088/0004-637x/788/1/2},
   number={1},
   journal={The Astrophysical Journal},
   publisher={American Astronomical Society},
   author={Bechter, Eric B. and Crepp, Justin R. and Ngo, Henry and Knutson, Heather A. and Batygin, Konstantin and Hinkley, Sasha and Muirhead, Philip S. and Johnson, John Asher and Howard, Andrew W. and Montet, Benjamin T. and Matthews, Christopher T. and Morton, Timothy D.},
   year={2014},
   month=May, pages={2} 
}

@ARTICLE{Barker2019,
       author = {{Barker}, A.~J. and {Jones}, C.~A. and {Tobias}, S.~M.},
        title = "{Angular momentum transport by the GSF instability: non-linear simulations at the equator}",
      journal = {\mnras},
         year = 2019,
        month = aug,
       volume = {487},
       number = {2},
        pages = {1777-1794},
          doi = {10.1093/mnras/stz1386},
archivePrefix = {arXiv},
       eprint = {1905.06962},
 primaryClass = {astro-ph.SR},
       adsurl = {https://ui.adsabs.harvard.edu/abs/2019MNRAS.487.1777B}
}

@ARTICLE{GoldreichSchubert1967,
       author = {{Goldreich}, Peter and {Schubert}, Gerald},
        title = "{Differential Rotation in Stars}",
      journal = {\apj},
         year = 1967,
        month = nov,
       volume = {150},
        pages = {571},
          doi = {10.1086/149360},
       adsurl = {https://ui.adsabs.harvard.edu/abs/1967ApJ...150..571G}
}

@ARTICLE{Fricke1968,
       author = {{Fricke}, K.},
        title = "{Instabilit{\"a}t station{\"a}rer Rotation in Sternen}",
      journal = {\zap},
         year = 1968,
        month = jan,
       volume = {68},
        pages = {317},
       adsurl = {https://ui.adsabs.harvard.edu/abs/1968ZA.....68..317F}
}

@unpublished{Lammers2026,
  author = {Lammers, Caleb and Su, Yubo and Winn, Joshua N.},
  title  = {The Mysterious Inspiral of {WASP-12 b}: Why Obliquity Tides
            Cannot Drive Orbital Decay},
  note   = {Manuscript in preparation},
  year   = {2026}
}

@misc{wang_2026_21792490,
       author = {{Wang}, Xian-Yu},
        title = "{Allesfitters: a modified version of allesfitter for advanced Rossiter-McLaughlin effect modeling}",
         year = 2026,
        month = aug,
          eid = {10.5281/zenodo.21792490},
          doi = {10.5281/zenodo.21792490},
      version = {1.0.0},
    publisher = {Zenodo},
       adsurl = {https://ui.adsabs.harvard.edu/abs/2026zndo..21792490W}
}
\bibliographystyle{aasjournalv7.1}

\end{document}